\documentclass[prx,twocolumn,secnumarabic,amssymb,showpacs, superscriptaddress]{revtex4-1}
\usepackage{longtable}

\newcommand{\be}{\begin{eqnarray}}
\newcommand{\ee}{\end{eqnarray}}

\newcommand{\bra}[1]{\langle{#1}|}
\newcommand{\ket}[1]{|{#1}\rangle}

\newcommand{\sz}[1]{\hat{Z}_{#1}}
\newcommand{\szz}{\hat{Z}}
\newcommand{\sxx}[1]{\hat{X}_{#1}}
\newcommand{\syy}[1]{\hat{Y}_{#1}}
\newcommand{\lam}[1]{\lambda_\mathrm{#1}}

\newcommand{\catpm}{\ket{\mathcal{C}^\pm_\alpha}}
\newcommand{\catmp}{\ket{\mathcal{C}^\mp_\alpha}}

\newcommand{\catpmb}{\bra{\mathcal{C}^\pm_\alpha}}

\newcommand{\catp}{\ket{\mathcal{C}^+_\alpha}}
\newcommand{\catm}{\ket{\mathcal{C}^-_\alpha}}
\newcommand{\catpb}{\bra{\mathcal{C}^+_\alpha}}
\newcommand{\catmb}{\bra{\mathcal{C}^-_\alpha}}

\newcommand{\sx}{\hat{X}}
\newcommand{\sy}{\hat{Y}}

\newcommand{\catpr}{\ket{\mathcal{C}^+_{\alpha e^{i\phi}}}}
\newcommand{\catmr}{\ket{\mathcal{C}^-_{\alpha e^{i\phi}}}}
\newcommand{\catpmr}{\ket{\mathcal{C}^\pm_{\alpha e^{i\phi}}}}

\newcommand{\catprt}{\ket{\mathcal{C}^+_{\alpha e^{i\phi(t)}}}}
\newcommand{\catmrt}{\ket{\mathcal{C}^-_{\alpha e^{i\phi(t)}}}}
\newcommand{\catpmrt}{\ket{\mathcal{C}^\pm_{\alpha e^{i\phi(t)}}}}

\newcommand{\catprtau}{\ket{\mathcal{C}^+_{\alpha e^{i\phi(\tau)}}}}
\newcommand{\catmrtau}{\ket{\mathcal{C}^-_{\alpha e^{i\phi(\tau)}}}}

\newcommand{\catprtt}{\ket{\mathcal{C}^+_{\alpha e^{i\phi(\tau)}}}}
\newcommand{\catmrtt}{\ket{\mathcal{C}^-_{\alpha e^{i\phi(\tau)}}}}

\newcommand{\hta}{\hat{a}}

\usepackage{mathtools}
\usepackage{dsfont}
\renewcommand{\thesubsection}{\Alph{subsection}}
\usepackage[table]{xcolor}    
\usepackage{tabularx}

\newcolumntype{C}{>{$}c<{$}}
\AtBeginDocument{
\heavyrulewidth=.08em
\lightrulewidth=.05em
\cmidrulewidth=.03em
\belowrulesep=.65ex
\belowbottomsep=0pt
\aboverulesep=.4ex
\abovetopsep=0pt
\cmidrulesep=\doublerulesep
\cmidrulekern=.5em
\defaultaddspace=.5em
\tabcolsep=7pt
}
\usepackage{booktabs}

\usepackage{hyperref}
\definecolor{darkblue}{RGB}{0,0,127} 
\definecolor{darkgreen}{RGB}{0,150,0}
\hypersetup{breaklinks, colorlinks, linkcolor=darkblue, citecolor=darkgreen, filecolor=red, urlcolor=darkblue}

\begin{document}

\title{{Bias-preserving gates with stabilized cat qubits }}

\author{S.~Puri}
\affiliation{Department of Physics, Yale University, New Haven, CT 06520, USA}
\affiliation{Yale Quantum Institute, Yale University, New Haven, CT 06520, USA}
\author{L.~St-Jean}
\affiliation{Institut Quantique and D\'epartment de Physique, Universit\'e de Sherbrooke, Qu\'ebec J1K 2R1, Canada}
\author{J.~A.~Gross}
\affiliation{Institut Quantique and D\'epartment de Physique, Universit\'e de Sherbrooke, Qu\'ebec J1K 2R1, Canada}
\author{A.~Grimm}
\affiliation{Department of Applied Physics, Yale University, New Haven, CT 06511, USA}
\affiliation{Yale Quantum Institute, Yale University, New Haven, CT 06520, USA}
\author{N.~E.~Frattini}
\affiliation{Department of Applied Physics, Yale University, New Haven, CT 06511, USA}
\affiliation{Yale Quantum Institute, Yale University, New Haven, CT 06520, USA}
\author{P.~S.~Iyer}
\affiliation{Institute of Quantum Computing, 200 University Of Waterloo, Ontario, Canada}
\author{A.~Krishna}
\affiliation{Institut Quantique and D\'epartment de Physique, Universit\'e de Sherbrooke, Qu\'ebec J1K 2R1, Canada}
\author{S.~Touzard}
\affiliation{Department of Applied Physics, Yale University, New Haven, CT 06511, USA}
\affiliation{Yale Quantum Institute, Yale University, New Haven, CT 06520, USA}
\author{L.~Jiang}
\affiliation{Department of Applied Physics, Yale University, New Haven, CT 06511, USA}
\affiliation{Yale Quantum Institute, Yale University, New Haven, CT 06520, USA}
\author{A.~Blais}
\affiliation{Institut Quantique and D\'epartment de Physique, Universit\'e de Sherbrooke, Qu\'ebec J1K 2R1, Canada}
\affiliation{Canadian Institute for Advanced Research, Toronto, Canada}
\author{S.~T.~Flammia}
\affiliation{Centre for Engineered Quantum Systems, School of Physics, University of Sydney, Sydney, NSW 2006 Australia}
\affiliation{Yale Quantum Institute, Yale University, New Haven, CT 06520, USA}
\author{S.~M.~Girvin}
\affiliation{Department of Physics, Yale University, New Haven, CT 06520, USA}
\affiliation{Yale Quantum Institute, Yale University, New Haven, CT 06520, USA}

\date{\today}
\begin{abstract}
The code capacity threshold for error correction using qubits which exhibit {\it asymmetric} or {\it biased} noise channels is known to be much higher than with qubits without such structured noise. 
However, it is unclear how much this improvement persists when realistic circuit level noise is taken into account. 
This is because implementations of gates which do not commute with the dominant error {\it un-bias} the noise channel. 
In particular, a native bias-preserving controlled-NOT (CX) gate, which is an essential ingredient of stabilizer codes, is not possible in strictly two-level systems. 
Here we overcome the challenge of implementing a bias-preserving CX gate by using stabilized cat qubits in driven nonlinear oscillators. 
The physical noise channel of this qubit is biased towards phase-flips, which increase linearly with the size of the cat, while bit-flips are exponentially suppressed with cat size. 
Remarkably, the error channel of this native CX gate between two such cat qubits is also dominated by phase-flips, while bit-flips remain exponentially suppressed. 
This CX gate relies on the topological phase that arises from the rotation of the cat qubit in phase space. 
The availability of bias-preserving CX gates opens a path towards fault-tolerant codes tailored to biased-noise cat qubits with high threshold and low overhead. 
As an example, we analyze a scheme for concatenated error correction using cat qubits. 
We find that the availability of CX gates with moderately sized cat qubits, having mean photon number $<10$, improves a rigorous lower bound on the fault-tolerance threshold by a factor of two and decreases the overhead in logical Clifford operations by a factor of 5. 
We expect these estimates to improve significantly with further optimization and with direct use of other codes such as topological codes tailored to biased noise.
\end{abstract}


\maketitle
\section{Introduction}
\label{intro}

With fault-tolerant quantum error-correction, it is possible to perform arbitrarily long quantum computations provided the error rate per physical gate or time step is below some constant threshold value and the correlations in the noise remain weak~\cite{aliferis2008p}. 
However, codes which exhibit high thresholds, such as surface codes, come at the cost of prohibitively large overheads in the number of physical qubits and gates~\cite{knill2005quantum,raussendorf2007topological}. 
Current efforts in quantum error correction (QEC) are largely devoted to recovery from generic noise which lacks any special structure. 
For example, in the widely studied {\it depolarizing} noise model, errors are represented with the stochastic action of the Pauli operators $\hat{X}$, $\hat{Y}$, $\hat{Z}$, and the probability of these errors is assumed to be (roughly) equal. 
However, several types of physical qubits have a {\it biased} noise channel, that is, one type of error dominates over all the others. 
Some examples of such biased-noise qubits are superconducting fluxonium qubits~\cite{pop2014coherent}, quantum-dot spin qubits~\cite{shulman2012demonstration,watson2018programmable}, nuclear spins in diamond~\cite{waldherr2014quantum}, and many others. 
It is therefore natural to consider whether the threshold and overhead requirements for fault-tolerant QEC can be improved by exploiting the structure of the noise.

Some efforts have been made towards designing QEC codes for biased-noise qubits~\cite{aliferis2008fault,aliferis2009fault,webster2015reducing,robertson2017tailored,tuckett2018ultrahigh,tuckett2018tailoring}. 
In particular, recent studies have shown ultra-high code-capacity thresholds for surface codes tailored to biased noise~\cite{tuckett2018ultrahigh,tuckett2018tailoring}. 
The code capacity is calculated by assuming noisy data qubits and noiseless syndrome-extraction circuits. 
However, errors during gate operations or {\it circuit-level noise} must be taken into account in order to estimate the fault-tolerance threshold. 
Importantly, in the case of qubits with biased noise, operations which do not commute with the dominant error can {\it un-bias} or {\it  depolarize} the noise channel, reducing or eliminating any advantages conferred by the original biased noise. 

To illustrate this point, consider first a system that preserves the noise bias.
Suppose we have a gate 
\begin{align}
    ZZ(\theta)=\exp(i\theta\sz{1}\sz{2}/2)
\end{align}
between two qubits suffering only from phase-flip errors with a tuneable phase angle $\theta$. 
When $\theta = \pi/2$, we recover the usual controlled-PHASE gate, CZ, up to local Pauli rotations and an overall phase. 
The $ZZ(\theta)$ gate can be implemented with an interaction of the form $\hat{H}_\mathrm{ZZ}=-V\sz{1}\sz{2}$ with the evolution unitary $\hat{U}(t)=\exp(iVt\sz{1}\sz{2})$. 
A $ZZ(\theta)$ gate is realized at time $T=\theta/2V$. 
Suppose a phase-flip error occurs in either of the two qubits at time $0\leq\tau\leq T$, in which case the evolution is modified into $\hat{U}_\mathrm{e}(T)=\hat{U}(T-\tau)\sz{1/2}\hat{U}(\tau)=\sz{1/2}\hat{U}(T)$. 
In other words, an erroneous gate operation $\hat{U}_\mathrm{e}(T)$ is equivalent to an error-free gate followed by a phase-flip and therefore the $ZZ(\theta)$ gate preserves the error bias. 

Now consider a CX gate between the two qubits, implemented with an interaction of the form
\be
\hat{H}_\mathrm{CX}=V\left[\left(\frac{\hat{I}_1+\sz{1}}{2}\right)\otimes\hat{I}_{2}+\left(\frac{\hat{I}_1-\sz{1}}{2}\right)\otimes\sxx{2}\right]\nonumber
\ee
with the evolution unitary $\hat{U}(t)=\exp(-i\hat{H}_\mathrm{CX}t)$.
Here the qubits labeled 1 and 2 are the control and target respectively. 
A CX gate is realized at time $T$ when $VT=\pi/2$ and 
\be
\hat{U}(T)=\left[\left(\frac{\hat{I}_1+\sz{1}}{2}\right)\otimes\hat{I}_{2}+\left(\frac{\hat{I}_1-\sz{1}}{2}\right)\otimes\sxx{2}\right],\nonumber
\ee
where we have ignored an overall phase.
In this case, a phase-flip error in the target qubit at time $0\leq\tau\leq T$ modifies the evolution to
\begin{align}
\hat{U}_\mathrm{e}(T)&=\hat{U}(T-\tau)\hat{I}_1\otimes \sz{2}\hat{U}(\tau)\nonumber\\
&=\hat{I}_1\otimes \sz{2}e^{iV(T-\tau)(\hat{I}_1-\sz{1})\otimes \sxx{2}}\hat{U}(T). 
\end{align}
Consequently, a phase-flip error is introduced in the control qubit depending on when the phase error on the target occurred. 
But more importantly, the phase-flip of the target qubit during the gate propagates as a combination of phase-flip and bit-flip in the same qubit (for $\tau\neq 0,T$). 
Application of the CX gate therefore reduces the bias of the noise channel by introducing bit-flips in the target qubit. 
In the same way, coherent errors in the gate operation arising from any uncertainty in $V$ and $T$ will also give rise to bit-flip errors in the target qubit. 
As a result, a native bias-preserving CX gate seems to be unphysical~\cite{aliferis2008fault,guillaud2019repetition}.
This is a serious drawback because the CX is a standard gate required to extract error syndromes in many error-correcting codes, including codes tailored to biased noise~\cite{tuckett2018ultrahigh,tuckett2018tailoring}. 
In the absence of a bias-preserving CX, alternate circuits are required for syndrome extraction. 
This was achieved in Ref~\cite{aliferis2008fault}, for example, by using teleportation schemes which require several CZ gates, measurements and state preparations. 
The added complexity, however, limits the potential gains in fault-tolerance thresholds for error correction with biased-noise qubits. 

\begin{figure}
\begin{centering}
 \includegraphics[width=\columnwidth]{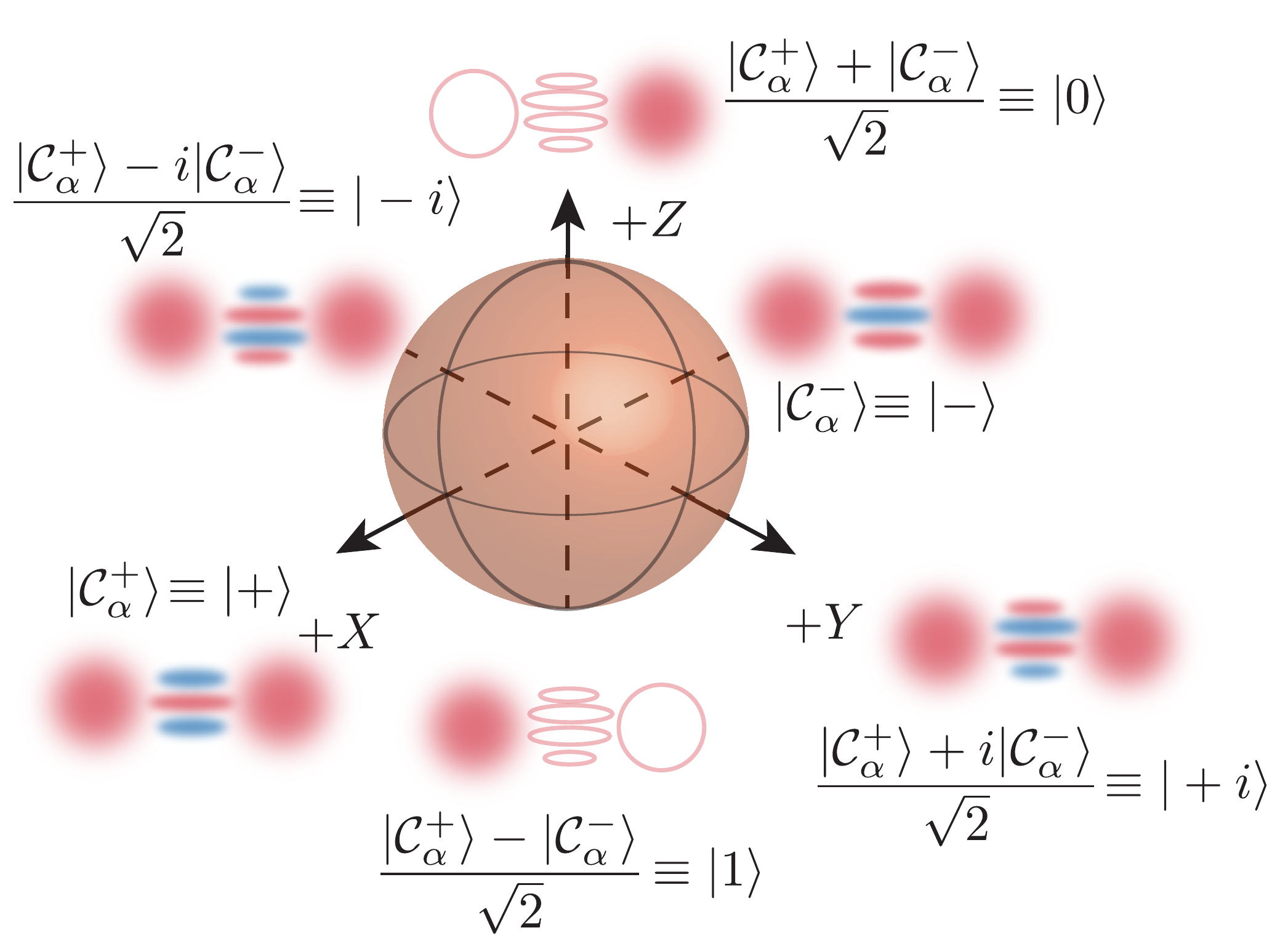}
 \caption{Bloch sphere of the cat qubit. The figure also shows cartoons of the Wigner functions corresponding to the eigenstates of $\hat{X},\hat{Y}$ and $\hat{Z}$ Pauli operators. }
 \label{bloch}
 \end{centering}
 \end{figure}

In this paper, we show that a radical solution to the problem of implementing a bias-preserving CX exists with two-component cat-qubits realized in a parametrically driven nonlinear oscillator~\cite{puri2017engineering}. 
We choose to work in a basis in which the cat states $\catpm=\mathcal{N}_\pm(\ket{\alpha}\pm\ket{-\alpha})$ define the $X$-axis of the qubit Bloch sphere shown in Fig~\ref{bloch} (that is, $\ket{\pm}\equiv\catpm)$. 
Here $\ket{\pm\alpha}$ are coherent states, which have the same amplitude but differ in phase by $\pi$ and $\mathcal{N}_\pm=1/\sqrt{2(1\pm e^{-2|\alpha|^2})}$ are the normalization constants. 
Note that the cat states are orthogonal, $\langle\mathcal{C}^-_\alpha|\mathcal{C}^+_\alpha\rangle=0$. 
For simplicity, we assume that the qubit is defined with real and positive $\alpha$. 
The $Z$-axis of the Bloch sphere, or the computational basis, is defined as,
\begin{align}
\ket{0}=\frac{\catp+\catm}{\sqrt{2}},\quad
\ket{1}=\frac{\catp-\catm}{\sqrt{2}}.
\end{align}
Note that, in the limit of large $\alpha$, the states $(\catp\pm\catm)/\sqrt{2}$ are exponentially close to the coherent states $\ket{\pm\alpha}$. 

The cat states, or equivalently their superpositions, $\ket{{0}}$ and $\ket{{1}}$, are the degenerate eigenstates of a parametrically driven Kerr-nonlinear oscillator~\cite{puri2017engineering}. 
Compared to schemes based on harmonic oscillators~\cite{ofek2016extending,leghtas2015confining,touzard2018coherent}, the advantage of the realization considered here is that the intrinsic Kerr nonlinearity, required to realize the cat qubit, also provides the ability to perform fast gates~\cite{GrimmStabilizing}. 
Additionally, it has been theoretically shown that although phase-flips increase linearly with the size of the cat $\alpha^2$, bit-flips are exponentially suppressed~\cite{puri2017engineering,puri2018stabilized}. 
As a result, this cat-qubit exhibits a strongly biased noise channel.
Remarkably, with such cats we show that it is possible to perform a native CX gate while preserving error bias. 
This gate is based on the topological phase that arises from the rotation of the cats in phase space generated by continuously changing the phase of the parametric drive. 
Because of the topological construction, the proposed CX gate preserves the error bias. 
Moreover, the noise channel of the gate also remains biased in the presence of coherent control errors. 
The ability to realize a bias-preserving CX gate differentiates the cat-qubit from strictly two-level systems with biased noise and demonstrates the advantage of continuous-variable systems for fault-tolerant quantum computing.   

This paper is organized as follows. 
We first describe the preparation of the driven cat-qubit and present its error channel. 
We also discuss the implementation of trivially biased $Z(\theta)$- and $ZZ(\theta)$-gates. The $ZZ(\theta)$-gate can be used to reduce the overhead for magic-state distillation~\cite{webster2015reducing}.
We then show how the bias-preserving CX gate is implemented and provide the $\chi$-matrix representation of the noisy gate. 
Finally, in order to demonstrate the advantage of having physical bias-preserving CX gates, we analyze the scheme for concatenated error correction tailored to biased noise in Ref.~\cite{aliferis2008fault}. 
The scheme first uses a repetition code to correct for the dominant phase-flip errors. 
The overall noise strength after the first encoding is reduced compared to the unencoded qubits and the effective noise strength is more symmetric. 
The repetition code is then concatenated with a CSS code. 
We find that the availability of a bias-preserving CX considerably simplifies the gadgets needed to implement fault-tolerant logical gates. 
Consequently, we are able to achieve an increase in the threshold by a factor of $\gtrsim 2$ and a reduction in the overhead by a factor of $\gtrsim 5$ for the repetition code gadgets. 
 


\section{Stabilized cat-qubit}\label{sec_intro}
\begin{figure}
\begin{centering}
 \includegraphics[width=.9\columnwidth]{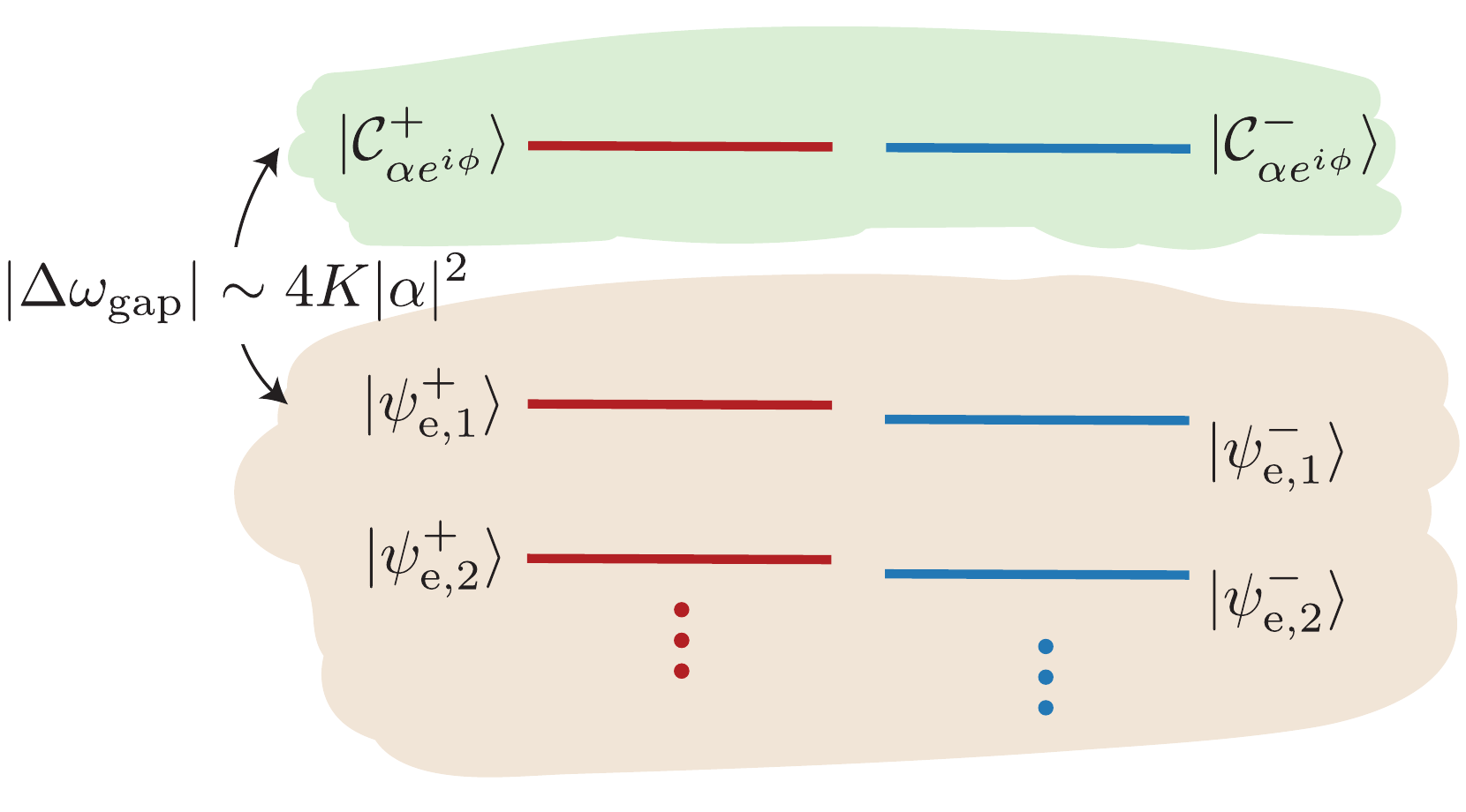}
 \caption{Eigenspectrum of the two-photon driven nonlinear oscillator in the rotating frame. The Hamiltonian in the rotating wave approximation is given in Eq.~\eqref{H0}. 
 The cat states $\catpmr$ with $\alpha=\sqrt{P/K}$ are exactly degenerate. The eigenspectrum can be divided into an even and an odd parity manifold. The cat subspace, highlighted in green, is separated from the first excited state by an energy gap $|\Delta\omega_\mathrm{gap}|
 \sim4K\alpha^2$. In the rotating frame, the excited states appear at a lower energy. This is because the Kerr nonlinearity is negative and implies that transitions out of the cat manifold occur at a lower frequency compared to transitions within the cat subspace. 
 The energy difference between the first $n\sim\alpha^2/4$ pairs of excited states (highlighted in orange) $\ket{\psi^\pm_{\mathrm{e},n}}$ decreases exponentially with $P$ or equivalently with $\alpha^2$. 
 These excited state pairs are consequently referred to as quasi-degenerate states.}
 \label{eigen}
 \end{centering}
\end{figure}

\subsection{Two-photon driven nonlinear oscillator}
The Hamiltonian of a two-photon driven Kerr-non\-linear oscillator in a frame rotating at the oscillator frequency $\omega_\mathrm{r}$ is given by
\be
\hat{H}_0(\phi)&=&-K\hta^{\dag 2}\hta^2+P(\hta^{\dag 2}e^{2i\phi}+\hta^2e^{-2i\phi}),\\
\label{H0}
&=&-K\left(\hta^{\dag 2}-\alpha^2 e^{-2i\phi}\right)\left(\hta^{2}-\alpha^2 e^{2i\phi}\right)+\frac{P^2}{K}.
\label{H0_2}
\ee
Here $K$ is the strength of the nonlinearity while $P$ and $\phi$ are respectively the amplitude and phase of the drive and $\alpha=\sqrt{P/K}$. 
The second line makes it clear that the even- and odd-parity cat states $\catpmr=\mathcal{N}_\pm(\ket{\alpha e^{i\phi}}\pm\ket{-\alpha e^{i\phi}})$ are the degenerate eigenstates of this Hamiltonian~\cite{puri2017engineering,goto2016universal}. 
Figure~\ref{eigen} shows the eigenstates of the oscillator in the rotating frame. 

Since Eq.~\eqref{H0} commutes with the photon number parity operator, its eigenspace can be divided into even- and odd-parity subspaces, labeled in the figure by the red and blue levels, respectively. 
The degenerate cat-subspace $\mathcal{C}$ (green) is separated from the rest of the Hilbert space $\mathcal{C}_\perp$ (orange) by a large energy gap, which in the rotating frame and in the limit of large $\alpha$ is well approximated as $\Delta\omega_\mathrm{gap}\sim-4K\alpha^2$. The negative energy gap implies that in the lab frame, transitions out of the cat manifold occur at a lower frequency compared to $\omega_\mathrm{r}$, the transition frequency within it. 
For large $\alpha$, the energy gap between pairs of even and odd excited states $\ket{\psi^\pm_{\mathrm{e},n}}$ decreases exponentially for $n\lessapprox \alpha^2/4$. 
As a result, the eigenspace of the two-photon driven oscillator reduces to $\alpha^2/4$ pairs of quasi-degenerate states (recall that the cat-subspace is exactly degenerate). 
This Hilbert space symmetry is important for the exponential suppression of bit-flip errors. 
Moreover, observe that in the limit $P\rightarrow 0$, the even and odd parity cat states continuously approach the vacuum and single-photon Fock states respectively. 
Consequently, starting from an undriven oscillator in vacuum (or single-photon Fock state) it is possible to adiabatically prepare the state $\catpr$ (or $\catmr$) by increasing the amplitude of a resonant two-photon drive at a rate $\ll 1/|\Delta\omega_\mathrm{gap}|$~\cite{puri2017engineering}. 

The phase $\phi$ of the two-photon drive is a continuous parameter that specifies the orientation of the cat in phase space. 
We define the cat-qubit with the phase $\phi=0$ (see Fig.~\ref{bloch}) and for the discussion of the following two sections we will fix this phase. 
This phase degree of freedom is however crucial for the implementation of the CX gate and we will return to it in section~\ref{sec-CX}. 

\subsection{Dynamics in the qubit subspace}
Suppose a single-photon drive is applied to the oscillator at the resonance frequency $\omega_\mathrm{r}$. 
In the rotating frame, the resulting Hamiltonian is $\hat{H}_1=\hat{H}_0+J(\hta e^{-i\theta}+\hta^\dag e^{i\theta})$. 
Since coherent states are eigenstates of $\hta$ it is easy to see that $\hta\catpm=\alpha r^{\pm 1}\catmp$, where
\be
r=\mathcal{N}_+/\mathcal{N}_- = \frac{\sqrt{1-e^{-2\alpha^2}}}{\sqrt{1+e^{-2\alpha^2}}} \sim 1-e^{-2\alpha^2},
\ee
and the last expression is taken in the limit of large $\alpha$. 
Unlike for $\hta$, coherent states are not eigenstates of $\hta^\dag$. 
The action of $\hta^\dag$ on a state in the cat-subspace causes transitions to the excited states $\hta^\dag\catpm\sim\alpha\catmp+\ket{\psi^\mp_{\mathrm{e},1}}$.
Recall that the cat-subspace is separated from the rest of the Hilbert space by an energy gap. 
The applied drive however is at frequency $\omega_\mathrm{r}$ and therefore the probability of excitation to the states $\ket{\psi^\mp_{\mathrm{e},1}}$ is suppressed by $\sim (J/\Delta\omega_\mathrm{gap})^2$. 
On the other hand these excitations would be permitted if the external drive had a frequency close to $\omega_\mathrm{r}+\Delta\omega_\mathrm{gap}\sim\omega_\mathrm{r}-4K\alpha^2$. 
Since the on-resonance drive only causes transitions within the cat-subspace, the Hamiltonian projected onto $\mathcal{C}$ is 
\begin{align}
\hat{P}_\mathcal{C}\hat{H}_{1}\hat{P}_\mathcal{C}&= \alpha\cos(\theta)J\left(r+r^{-1}\right)\hat{Z}\nonumber\\
& \ \ \ + \alpha\sin(\theta)J\left(r-r^{-1}\right)\hat{Y},\nonumber
\end{align}
where $\hat{P}_\mathcal{C}=\catp\catpb+\catm\catmb$ is the projection operator onto the cat subspace. 
In the limit of large $\alpha$ (or equivalently $P$), the above equation reduces to
\be
\hat{P}_\mathcal{C}\hat{H}_{1}\hat{P}_\mathcal{C}\sim 2\alpha\cos(\theta)J\hat{Z}- 2\alpha\sin(\theta)Je^{-2\alpha^2}\hat{Y}.
\label{single_a}
\ee
This expression shows that it is possible to implement an arbitrary rotation around the $Z$-axis of the Bloch sphere using a single-photon drive by choosing $\theta=0$~\cite{puri2017engineering,puri2018stabilized}. 
The angle of rotation is determined by the strength $J$ and duration of the single-photon drive, and the amplitude $\alpha$ of the cat state. 
Importantly, equation~\eqref{single_a} also shows that, unlike rotation around the $Z$-axis, rotation around the $Y$-axis is suppressed exponentially with $\alpha^2$. 
In other words, the external drive couples predominantly to $\szz$. 
This observation implies that control errors (such as errors in the amplitude, frequency, phase and duration of the single-photon drive) will lead to over-rotation or under-rotation around the $Z$-axis, but only cause exponentially small angle rotations around the $Y$-axis. 
Recall that for virtual excitations out of the cat-subspace to be small we require $J\ll |\Delta\omega_\mathrm{gap}|$. 
That is, the energy gap governs rate of gate operations. 
It is easy to achieve $|\Delta\omega_\mathrm{gap}|/2\pi\sim 200$ MHz in superconducting cavities and therefore fast rotations $\lesssim 100$ ns are possible~\cite{GrimmStabilizing}. 
This is to be contrasted with cat states produced in a harmonic oscillator by means of two-photon drive and dissipation~\cite{mirrahimi2014dynamically,cohen2017degeneracy,guillaud2019repetition}. 
The ``dissipative gap" which defines the cat qubit subspace is significantly smaller ($\lesssim 1$ MHz)~\cite{leghtas2015confining,touzard2018coherent} and therefore the gates are slower ($\gtrsim 1$ $\mu$s.) 

It is easy to extend the analysis above to realize a $ZZ(\theta)=\exp{(i\theta\sz{1}\sz{2})/2}$ gate. 
This gate is implemented between two driven nonlinear oscillators coupled via a resonant beam-splitter type interaction~\cite{puri2017engineering,gao2018programmable}, $\hat{H}_{ZZ}=\hat{H}_{0,1}+\hat{H}_{0,2}+J_{12}(\hta^\dag_1\hta_2+\hta^\dag_2\hta_1)$, with $\hat{H}_{0,i}=-K\hta^{\dag 2}_i\hta^2_i+P(\hta^{\dag 2}_i+\hta^2_i)$, and $i=1,2$. 
For small $J_{12}$ the evolution under the Hamiltonian $\hat{H}_{ZZ}$ is confined within the cat subspace and 
\begin{align}
\hat{P}_\mathcal{C}\hat{H}_{ZZ}\hat{P}_\mathcal{C}&= J_{12}\alpha^2(r^2+r^{-2})\sz{1} \sz{2}\nonumber\\
&\ \ \ +J_{12}\alpha^2(r^2-r^{-2})\hat{Y}_1 \hat{Y}_2\nonumber\\
& \sim 2J_{12}\alpha^2\sz{1} \sz{2}-4J_{12}\alpha^2e^{-2\alpha^2}\hat{Y}_1\hat{Y}_2.
\end{align}
The last line in the above equation is written in the limit of large $\alpha$. 
In this limit, the term $\propto \hat{Y}_1\hat{Y}_2 $ is negligibly small. 
Therefore, the unitary evolution under $\hat{H}_{ZZ}$ realizes a $ZZ(\theta)$ gate with $\theta=4J_{12}\alpha^2 t_\mathrm{gate}$, where $t_\mathrm{gate}$ is the duration for which the beam-splitter coupling is turned on. 
Following the previous arguments, the control errors during this gate only lead to over- or under-rotation around the $Z$-axis, or correlated $\sz{1} \sz{2}$-errors. 
On the other hand, errors involving $\hat{X}_i$ or $\hat{Y}_i$ are exponentially suppressed with $\alpha^2$.

\begin{table*}[ht]
\centering
\begin{tabular}[c]{c c c}
\toprule
Noise type & Error Channel  & Coefficients\\
\midrule
& &  \\
Single-photon dissipation  & $
(\lam{I}\hat{I}+\lam{X}\sx)\hat{\rho}(\lam{I}\hat{I}+\lam{X}\sx)+$ &  $\lam{I,X}=(\sqrt{1-pr^{-2}}\pm \sqrt{1-pr^2})/{2}$\\
(bath at zero temperature) & $(\lam{Z}\szz+i\lam{Y}\sy)\hat{\rho}(\lam{Z}\szz-i\lam{Y}\sy)$ & $\lam{Z,Y}=\sqrt{p}\left(r\pm r^{-1}\right)/2$, $p=\kappa t \alpha^2$\\
 & & \\
& & \\
Thermal bath  & $
(\lam{I}\hat{I}+\lam{X}\sx)\hat{\rho}(\lam{I}\hat{I}+\lam{X}\sx)+$ & $\lam{I,X}=(\sqrt{1-p_1r^{-2}-p_2 r^2}\pm \sqrt{1-p_1r^2-p_2r^{-2}})/{2}$ \\
(narrow spectral density) & $(\lam{Z}\szz+i\lam{Y}\sy)\hat{\rho}(\lam{Z}\szz-i\lam{Y}\sy)$ & $\lam{Z,Y}=\sqrt{p_1}\left(r\pm r^{-1}\right)/2$, $\lam{Z,Y}'=\sqrt{p_2}\left(r\pm r^{-1}\right)/2$\\
 &$(\lam{Z}'\szz-i\lam{Y}'\sy)\hat{\rho}(\lam{Z}'\szz+i\lam{Y}'\sy)$ & $p_1=\kappa t \alpha^2 [1+n_\mathrm{th}(\omega_\mathrm{r})]$, $p_2=\kappa t \alpha^2 n_\mathrm{th}(\omega_\mathrm{r})$  \\
 & &  \\
 & & \\
Pure-dephasing  & $(\lam{I}\hat{I}+\lam{X}\sx)\hat{\rho}(\lam{I}\hat{I}+\lam{X}\sx)+$ & $\lam{I,X}=(\sqrt{1-pr^{-4}}\pm \sqrt{1-pr^4})/{2}$ \\
(narrow spectral density) & $(\lam{I}'\hat{I}-\lam{X}'\sx)\hat{\rho}(\lam{I}'\hat{I}-\lam{X}'\sx)+$ & $\lam{I,X}'=\sqrt{p}(r^2\pm r^{-2})/2$, $p=\kappa_\phi t\alpha^4$\\
& & \\
& & \\
Thermal bath \&  & $\lambda_\mathrm{II}\hat{I}\hat{\rho}\hat{I}+\lambda_{\mathrm{IX}}\hat{I}\hat{\rho}\hat{X}+\lambda^*_\mathrm{IX}\hat{X}\hat{\rho}\hat{I}+$ & \\
two-photon dissipation & $\lambda_\mathrm{XX}\hat{X}\hat{\rho}\hat{X}+\lambda_\mathrm{YY}\hat{Y}\hat{\rho}\hat{Y}+\lambda_\mathrm{YZ}\hat{Y}\hat{\rho}\hat{Z}+$ & See Eq.~\eqref{ch_therm} \& Fig.~\ref{thermal_i}\\
(white spectral density) & $\lambda^*_\mathrm{YZ}\hat{Z}\hat{\rho}\hat{Y}+\lambda_\mathrm{ZZ}\hat{Z}\hat{\rho}\hat{Z}$ & \\
& & \\
& & \\
Pure-dephasing \&  &  &  \\
two-photon dissipation & $\lambda_\mathrm{II}\hat{I}\hat{\rho}\hat{I}+\lambda_{\mathrm{IX}}\hat{I}\hat{\rho}\hat{X}+\lambda^*_\mathrm{IX}\hat{X}\hat{\rho}\hat{I}+\lambda_\mathrm{XX}\hat{X}\hat{\rho}\sx$ & See Eq.~\eqref{ch_dephase} \& Fig.~\ref{dephase_i} \\
(white spectral density) & & \\
& & \\
\bottomrule
\end{tabular}
\caption{The table lists the error channel of the cat-qubit for different sources of decoherence. 
For the first three error sources, single-photon dissipation, thermal noise with narrow spectral density and pure-dephasing with narrow spectral density, it is possible to obtain an analytical expression for the channel. 
The expressions for the coefficients in the limit when the product of rate of decoherence and time is small (i.e., $\kappa t\alpha^2<1$ and $\kappa_\phi t\alpha^4 <1$) are given in the third column. 
Recall that $r=\sqrt{1-e^{-2\alpha^2}}/\sqrt{1+e^{-2\alpha^2}}$ approaches 1 in the limit of large $\alpha$. 
Consequently, we find that all the coefficients involving the matrices $\sx$ and $\sy$ are suppressed exponentially in $\alpha^2$.}
\label{table}
\end{table*}

\section{Error Channel of the cat-qubit}\label{sec_err}
We will now show that irrespective of the nature of the coupling between the cat qubit and the bath, its error channel is biased towards dephasing error.

\subsection{Noise with narrow-band spectral density}
Suppose the oscillator couples to the environment with a general operator $\hat{O}=\sum_{m,n}\chi_{m,n}(t)\hta^{\dag m}\hta^n+\mathrm{h.c.}$.
From the analysis in the previous section, we see that $\hta^{\dag m}$ will cause excitations out of the cat subspace. 
In fact, in the limit of large $\alpha$, $\hta^{\dag m}$ will excite the $m^\mathrm{th}$ excited manifold. 
However, when the frequency spectrum of $\chi_{m,n}(t)$ is narrow and centered around $(n-m)\omega_\mathrm{r}$ and $\mathrm{max}[|\chi_{m,n}(t)|]\alpha^{m+n-1}\ll |\Delta\omega_\mathrm{gap}|$, then resonant (or {\it real}) and non-resonant (or {\it virtual}) excitations out of the cat-manifold are negligible. 
The effect of the coupling in the cat manifold is then described by $\hat{P}_\mathcal{C}\hat{O}_{m,n}\hat{P}_\mathcal{C}= g_{m,n}f^{*m}(t)f^{n}(t)\hta^{\dag m}_\mathcal{C}\hta^{n}_\mathcal{C}$. Here,
\begin{align} 
\hta_\mathcal{C}&=\hat{P}_\mathcal{C}\hta\hat{P}_\mathcal{C}=\alpha\left(\frac{r+r^{-1}}{2}\right)\hat{Z}+i \alpha\left(\frac{r-r^{-1}}{2}\right)\hat{Y},\nonumber\\
\hta^\dag_\mathcal{C}&=\hat{P}_\mathcal{C}\hta^\dag\hat{P}_\mathcal{C}=\alpha\left(\frac{r+r^{-1}}{2}\right)\hat{Z}-i \alpha\left(\frac{r-r^{-1}}{2}\right)\hat{Y},
\label{proj}
\end{align}
are the annihilation and creation operators projected onto the cat manifold. 
Note that for large $\alpha$, we have $\hta_\mathcal{C}, \hta^\dag_\mathcal{C}\approx \alpha\hat{Z}\mp i \alpha e^{-2\alpha^2}\hat{Y}$. 
Hence, we find that the oscillator-environment interaction is dominant along the $Z$-axis ($\propto \alpha^{m+n}$), while suppressed along $X,Y$-axis ($\propto \alpha^{m+n} e^{-2\alpha^2}$) and the resulting noise channel is biased. 
We now list the error channels for a few sources of narrow-bandwidth noise.

\subsubsection{Thermal bath with narrow spectral density}
\label{sec_narrow_therm}
By far, the main source of noise in oscillators is single-photon loss. In the cat subspace, one photon at a time is lost to the environment at frequency $\omega_\mathrm{r}$. 
However, it is also possible for the oscillator to gain photons if the bath is at non-zero temperature. 
If the spectral density of thermal photons is narrow, but smooth and centered around $\omega_\mathrm{r}$ then addition of a single photon to the oscillator (i.e., action of $\hta^\dag$) cannot cause transitions out of the cat subspace. 
The Born-Markov approximation in this limit leads to the Lindbladian $\mathcal{D}[\hat{O}_1]\hat{\rho}+\mathcal{D}[\hat{O}_2]\hat{\rho}$~\cite{puri2018stabilized} with,
\begin{align}
\label{narrow_therm_loss_op}
\hat{O}_1&=\sqrt{\kappa [1+n_\mathrm{th}(\omega_\mathrm{r})]}\alpha\left[\left(\tfrac{r+r^{-1}}{2}\right)\hat{Z}+i\left(\tfrac{r-r^{-1}}{2}\right)\hat{Y}\right],\\
&\sim \sqrt{\kappa [1+n_\mathrm{th}(\omega_\mathrm{r})]}\alpha[\hat{Z}-ie^{-2\alpha^2}\hat{Y}] \hspace{0.2cm}(\alpha \rightarrow \mathrm{large}),\\
\label{narrow_therm_gain_op}
\hat{O}_2&=\sqrt{\kappa n_\mathrm{th}(\omega_\mathrm{r})}\alpha\left[\left(\tfrac{r+r^{-1}}{2}\right)\hat{Z}-i\left(\tfrac{r-r^{-1}}{2}\right)\hat{Y}\right],\\
&\sim \sqrt{\kappa n_\mathrm{th}(\omega_\mathrm{r})}\alpha[\hat{Z}+ie^{-2\alpha^2}\hat{Y}] \hspace{0.2cm}(\alpha \rightarrow \mathrm{large}).
\end{align}
In the above expressions $n_\mathrm{th}(\omega_\mathrm{r})$ is the thermal photon number at $\omega_\mathrm{r}$. When $n_\mathrm{th}=0$, the above equation reduces to the master equation of the cat qubit coupled to zero temperature bath. 
Table~\ref{table} shows the error-channel corresponding to the above Lindbladian in the operator-sum representation, in the limit of small $\kappa\alpha^2 t$ for both $n_\mathrm{th}=0$ and $n_\mathrm{th}\neq 0$.

\subsubsection{Narrow spectral density frequency noise}
Apart from gain and loss of photons, it is possible that coupling with the environment causes the frequency of the oscillator to fluctuate. 
This noise channel is often referred to as {\it pure dephasing}. 
However, if these fluctuations are slow and of small amplitude compared to the energy gap, such as in the case of flux noise in superconducting circuits~\cite{bialczak2007flux,sank2012flux}, then the out-of-cat excitations are suppressed. 
Consequently, in the Born-Markov approximation the Lindbladian is given by $\mathcal{D}[\hat{O}]\hat{\rho}$~\cite{puri2018stabilized}  with,
\begin{align}
\hat{O}&=\sqrt{\kappa_\phi}\alpha^2\left[\left(\frac{r^2+r^{-2}}{2}\right)\hat{I}+\left(\frac{r^2-r^{-2}}{2}\right)\hat{X}\right],\\
&\sim \sqrt{\kappa_\phi}\alpha^2[\hat{I}-2e^{-2\alpha^2}\hat{X}] \hspace{0.2cm}(\alpha \rightarrow \mathrm{large}).
\end{align}
Table~\ref{table} presents the corresponding error channel in the limit of small $\kappa_\phi\alpha^4 t$.  


\subsection{Noise with wide-band spectral density}
The previous section described the noise channel of the cat qubit coupled to a bath with narrow-band spectral density so that leakage is avoided. But what if the spectrum of the environment-oscillator coupling is such that leakage out of the cat subspace becomes non-negligible? Firstly, we will show that the leakage errors can be autonomously corrected by addition of photon dissipation. Secondly, we find that the amount of non-dephasing errors introduced due to the autonomous correction process depends on the energy difference between the even and odd parity states of the $m^\mathrm{th}$ excited manifold $\ket{\psi^\pm_{\mathrm{e},m}}$. However, since this energy difference decreases exponentially with $\alpha^2$ for $m<\alpha^2$, the non-dephasing errors also remain exponentially suppressed with $\alpha^2$. It is important to emphasize that for the exponential suppression of non-dephasing errors, the weight $m$ must be smaller than the number of quasi-degenerate pairs of excited states $\alpha^2/4$. Therefore,
it becomes possible to think of the driven nonlinear oscillator as a code which protects against non-dephasing errors in the cat-qubit. Moreover, the distance of this protection is $\sim\alpha^2/4$, which increases with the strength of the drive $P$.
We explain these results further using explicit examples in the following sections.

\subsubsection{Two-photon dissipation channel}~\label{diss_2ph}
In the presence of two-photon dissipation, the oscillator loses pairs of photons to the environment. The master equation of the parametrically driven oscillator in presence of a white two-photon dissipation channel is given by,
\be
\dot{\hat{\rho}}&=-i[\hat{H}_0(\phi),\hat{\rho}]+\kappa_{2}\mathcal{D}[\hta^{ 2}]\hat{\rho},
\ee
where $\kappa_{2}$ is the rate of two-photon dissipation. Superconducting cavities with $\kappa_\mathrm{2}/2\pi\sim 200$ kHz have been engineered ~\cite{touzard2018coherent}. 
The dissipative dynamics can be understood in the quantum-jump approach in which the deterministic evolution governed by the non-Hermitian effective Hamiltonian $\hat{H}={H}_0-i\kappa_{2}\hta^{\dag 2}\hta^2/2$ is interrupted by two-photon jump events. 
The non-Hermitian Hamiltonian is analogous to Eq.~\eqref{H0} with the Kerr-nonlinearity $K$ replaced by a complex quantity $K+i\kappa_\mathrm{2}/2$. 
The nature of the eigenspectrum of the non-Hermitian Hamiltonian is therefore the same as the actual Hamiltonian of Eq.~\eqref{H0}. 
However, unlike Eq.~\eqref{H0}, the eigenenergies of $\hat{H}$ become complex implying linewidth broadening. 

The cat states $\ket{\mathcal{C}^\pm_\beta}$ are degenerate eigenstates of the non-Hermitian Hamiltonian $\hat{H}\ket{\mathcal{C}^\pm_\beta}=E\ket{\mathcal{C}^\pm_\beta}$ where $E$ is a complex quantity $E=P^2/(K+i\kappa_{2}/2)$ and $\beta=e^{i\phi}\sqrt{P/(K+i\kappa_{2}/2)}$. 
Moreover, the cat states are also eigenstates of the two-photon jump operator $\hta^2\ket{\mathcal{C}^\pm_\beta}=\beta^2\ket{\mathcal{C}^\pm_\beta}$. 
Therefore, the states $\ket{\mathcal{C}^\pm_\beta}$ are invariant to two-photon dissipation. 
We have defined the cat qubit $\catpm$ using real and positive coherent state amplitude $\alpha$. 
For this qubit to be stabilized in the presence of two-photon dissipation, the phase and amplitude of the required two-photon drive are $2\phi_0=\tan^{-1}(\kappa_{2}/2K)$ and $P=\alpha^2 \sqrt{K^2+\kappa^2_{2}/4}$, respectively.

\begin{figure}
\begin{centering}
 \includegraphics[width=.8\columnwidth]{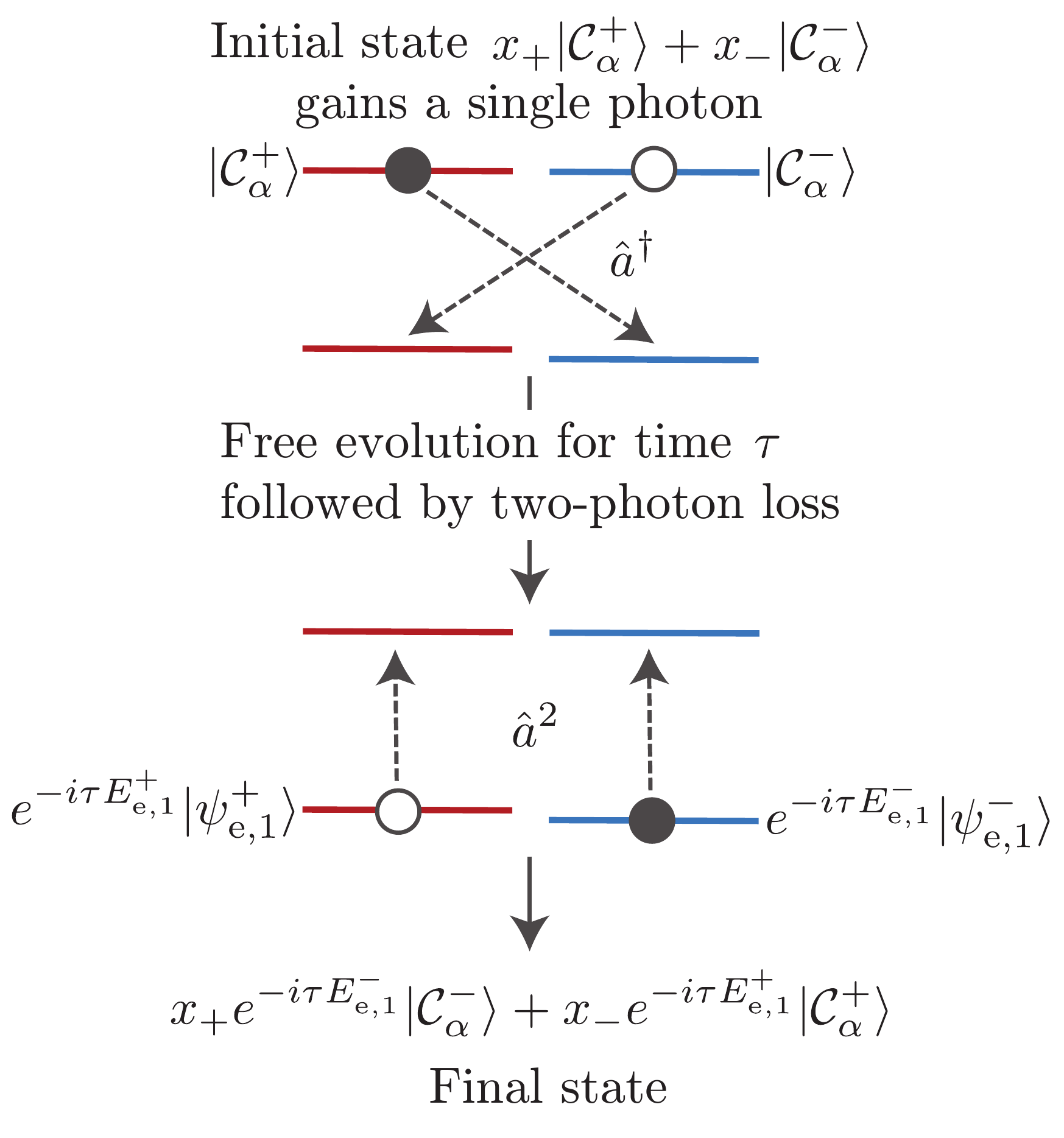}
 \caption{As shown in the top panel, addition of single photon at frequency $\omega_\mathrm{r}+\Delta\omega_\mathrm{gap}\sim \omega_\mathrm{r}-4K\alpha^2$ excites the 
 $\ket{\psi_0}=x_+\catp+x_-\catm$ to $x_+\ket{\psi^-_\mathrm{e,1}}+x_-\ket{\psi^+_\mathrm{e,1}}$. The state then evolves freely for some time $\tau$ before a two-photon loss event brings it back to the cat subspace. This transition is shown in the bottom panel. During the time $\tau$, $\ket{\psi^\mp_\mathrm{e,1}}$ acquire phases $\sim \tau E^\mp_\mathrm{e,1}$ (assuming that the linewidth is small.) That is, just before the loss of two photons, the state of the system is $x_+e^{-i\tau E^-_\mathrm{e,1}}\ket{\psi^-_\mathrm{e,1}}+x_-e^{-i\tau E^-_\mathrm{e,1}}\ket{\psi^+_\mathrm{e,1}}$. Loss of two photons causes transition within the same parity subspace, and hence the final state is $x_+e^{-i\tau E^-_\mathrm{e,1}}\catm+x_-e^{-i\tau E^+_\mathrm{e,1}}\catp\equiv \szz e^{i(E^-_\mathrm{e,1}-E^+_\mathrm{e,1})\tau\sx/2}\ket{\psi_0} $. Therefore, the autonomous correction of leakage leads to both dephasing and non-dephasing error. However, $E^-_\mathrm{e,1}-E^+_\mathrm{e,1}$ decreases exponentially with $\alpha^2$ and hence the non-dephasing error is exponentially suppressed. }
 \label{therm_cartoon}
 \end{centering}
 \end{figure}

\subsubsection{Thermal bath with white-noise spectrum}\label{sec_th}

White thermal noise leads to the Lindbladian master equation,
\begin{align}
\dot{\hat{\rho}}=-i[\hat{H}_0(\phi),\hat{\rho}]+\kappa (n_\mathrm{th}+1)\mathcal{D}[\hta]\hat{\rho}+\kappa n_\mathrm{th}\mathcal{D}[\hta^\dag]\hat{\rho},
\end{align}
where $n_\mathrm{th}$ is the number of thermal photons. Again following the quantum-jump approach, the dynamics of the oscillator can be described by evolution under a non-Hermitian Hamiltonian $\hat{{H}}=\hat{H}_0(\phi)-i\kappa(1+n_\mathrm{th})\hta^\dag\hta/2-i\kappa n_\mathrm{th}\hta\hta^\dag/2$ which is interrupted by stochastic quantum jumps corresponding to the operators $\hta$, $\hta^\dag$ ~\cite{walls2007quantum}. When $\kappa\ll |\Delta\omega_\mathrm{gap}|$, it is possible to replace $\hta,\hta^\dag$ with their projections in the cat basis $\hta_\mathcal{C},\hta^\dag_\mathcal{C}$ given in Eq.~\eqref{proj}. As a result, the dominant effect of the non-Hermitian terms in $\hat{{H}}$ is to broaden the linewidths of the cat states. A stochastic jump corresponding to the action of $\hta$ on a state in the cat-qubit subspace does not cause leakage. 
More importantly however, the action of $\hta^\dag$ on a state in the cat subspace causes leakage, $\hta^\dag\catpm\sim\alpha\catmp+\ket{\psi^\mp_{\mathrm{e},1}}$ (note the change in parity). In other words, $\bra{\psi^\mp_{\mathrm{e},1}}\hta^\dag\catpm\sim 1$ so that a single photon gain event excites the first excited subspace at a rate $\sim\kappa n_\mathrm{th}$. This transition to the first excited state is illustrated in Fig.~\ref{therm_cartoon}. In fact, $m$ photon gain events excite the $m^\mathrm{th}$ excited subspace (with opposite parity if $m$ is odd, or same parity if $m$ is even). Suppose a single-photon loss event followed a gain event. In this case, $\catpmb \hta \ket{\psi^\mp_{\mathrm{e},1}}\sim 1$ and hence a single-photon loss event corrects the leakage at a rate $\kappa (n_\mathrm{th}+1)$. As a result, at steady state, the amount of leakage is $\sim \kappa n_\mathrm{th}/\kappa(n_\mathrm{th}+1)\sim n_\mathrm{th}$ (for $n_\mathrm{th}\ll 1$). Now suppose a two-photon dissipation channel is introduced such that the rate of two-photon loss is $\kappa_\mathrm{2}$. In this case $\catpmb \hta^2 \ket{\psi^\pm_{\mathrm{e},1}}\sim 2\alpha$ and hence a two-photon loss event will correct the leakage at a rate $4\kappa_\mathrm{2ph}\alpha^2$. As a result, the residual leakage at steady state, is given by $\sim \kappa n_\mathrm{th}/4\kappa_\mathrm{2ph}\alpha^2<n_\mathrm{th}$ for $4\kappa_\mathrm{2ph}\alpha^2>\kappa$. Typically in superconducting circuits $\kappa/2\pi\sim 10$ KHz, $n_\mathrm{th}=1\%$ so that even with a moderately sized cat $\alpha=2$ and small amount of two-photon dissipation $\kappa_2/2\pi=200$ kHz, the residual leakage is reduced to $\sim 3\times 10^{-3}\%$. 

Observe that the loss of two-photons causes transitions within the same parity subspace. Therefore, as illustrated in Fig.~\ref{therm_cartoon}, two-photon loss immediately after a single-photon gain event does result in phase-flips. But phase-flips are already the dominant error channel in the system and therefore this effect does not change the structure of noise. However, the process of correcting leakage can also introduce bit-flips. Before a two-photon jump event brings the population back to the cat manifold, the states $\ket{\psi^\pm_{\mathrm{e},1}}$ accumulate a phase proportional to their energies $E^\pm_{\mathrm{e},1}$, shown in the second panel in Fig.~\ref{therm_cartoon}. As a result, when the population in the state $\ket{\psi^\pm_{\mathrm{e},1}}$ returns to the cat manifold, the two states $\catp$ and $\catm$ accumulate a phase difference $\propto (E^+_{\mathrm{e},1}-E^-_{\mathrm{e},1})$ (see Fig.~\ref{therm_cartoon}). In the cat-qubit's computational basis $\ket{{0},{1}}=(\catp\pm\catm)/\sqrt{2}$, this corresponds to a bit-flip. However, recall from section~\ref{sec_intro} that $(E^+_{\mathrm{e},1}-E^-_{\mathrm{e},1})$ decreases exponentially with $\alpha^2$ and the exited state manifold is quasi-degenerate. Consequently, the probability of a bit-flip error due to leakage also decreases exponentially with $\alpha^2$ and the noise bias is preserved. 

\begin{figure}
\begin{centering}
 \includegraphics[width=\columnwidth]{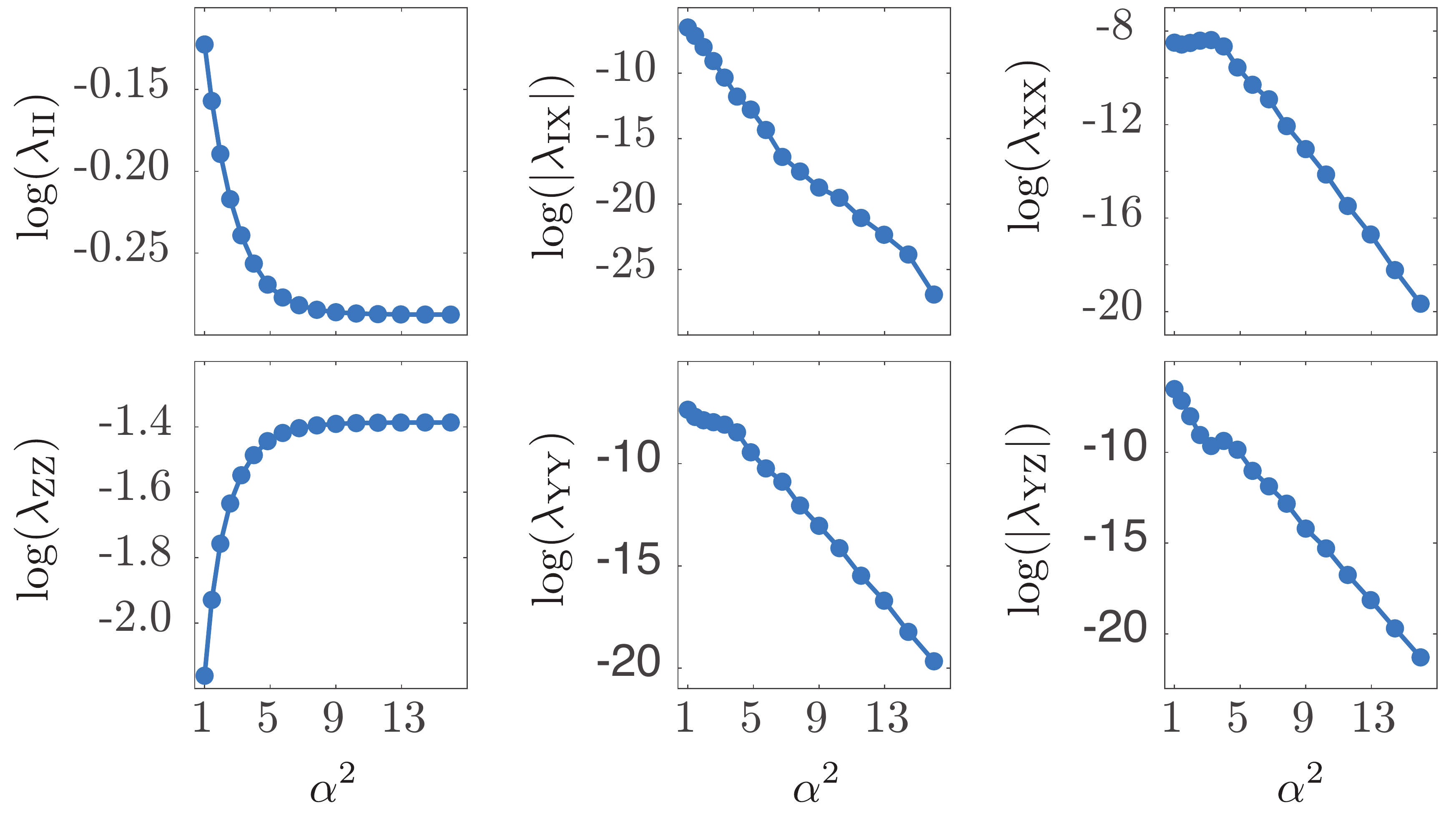}
 \caption{Natural logarithm of the coefficients of the error channel  Eq.~\eqref{ch_therm} of an idle cat qubit in the presence of white thermal noise and two-photon dissipation. As expected the amount of non-phase errors decreases exponentially with $\alpha^2$. The parameters for the simulations are $n_\mathrm{th}=0.01$, $\kappa=K/400$ and $\kappa_\mathrm{2ph}=K/10$ . The coefficients are evaluated at $t=50/K$. }
 \label{thermal_i}
 \end{centering}
 \end{figure}

\begin{figure}
\begin{centering}
 \includegraphics[width=\columnwidth]{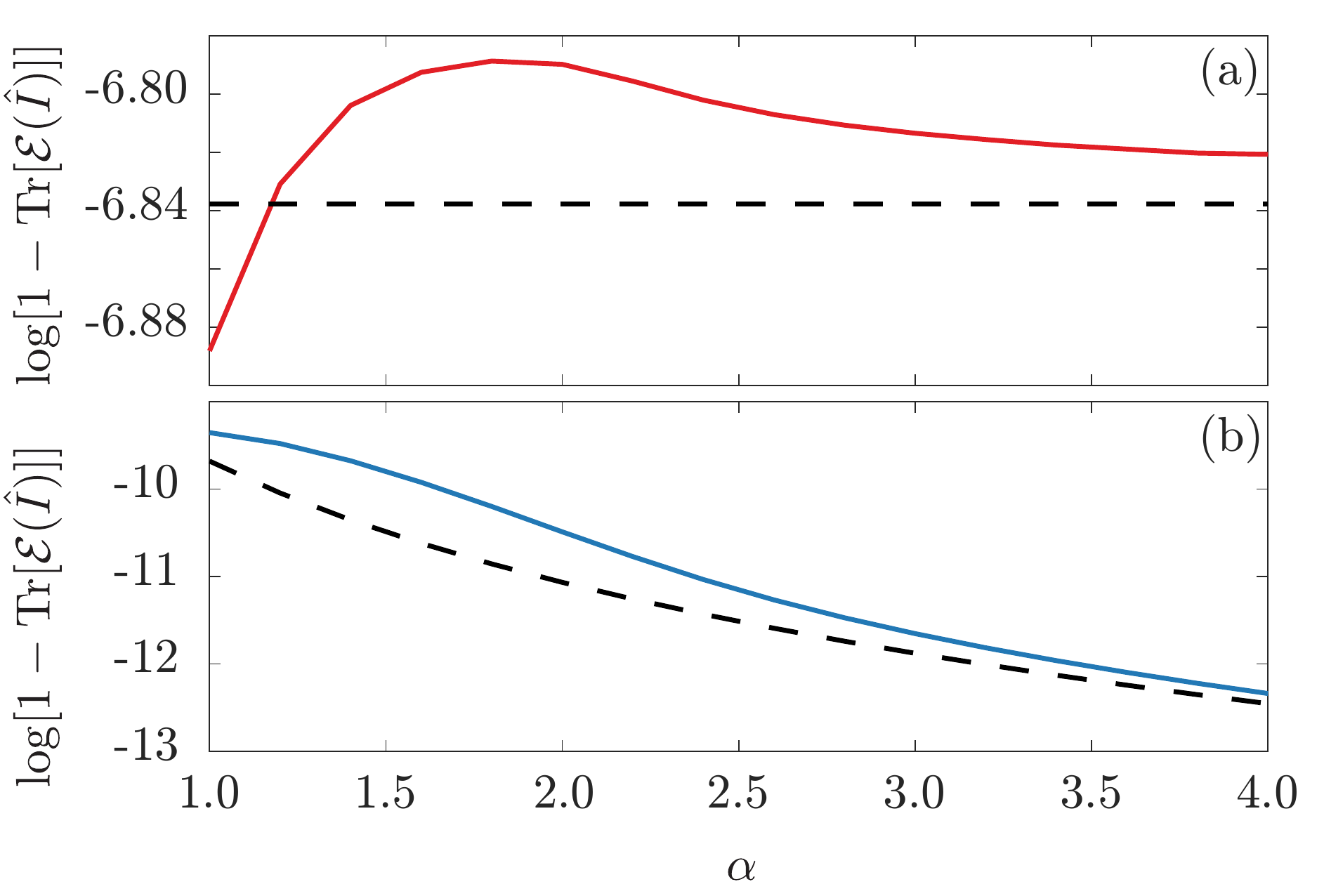}
 \caption{Natural logarithm of the amount of leakage, $\log[1-\mathrm{Tr}[\mathcal{E}(\hat{I})]]$ in the presence of white thermal noise without two-photon dissipation (red solid line in (a)) and with it (blue solid line in (b)). As expected, the two-photon dissipation autonomously corrects for leakage. The dashed black line shows the leakage predicted by the theoretical expressions for the rates of out-of-subspace excitations ($\sim\kappa n_\mathrm{th}$) and correction due to single-photon loss $\sim\kappa(1+n_\mathrm{th})$ and two-photon loss $\sim4\kappa_\mathrm{2ph}\alpha^2$. These expressions are only approximations which become more and more exact as $\alpha$ increases. As we also see from the figure, the numerically estimated leakage converges to the theoretically predicted value for large $\alpha$. 
 }
 \label{thermal_oos}
 \end{centering}
 \end{figure}

In order to confirm the analysis above, we numerically evaluate the error channel of the cat-qubit as a function of $\alpha^2$ by simulating the master equation,
\begin{align}
\dot{\hat{\rho}}=&-i[\hat{H}_0(\phi_0),\hat{\rho}]+\kappa (1+n_\mathrm{th})\mathcal{D}[\hta]\hat{\rho}+\kappa n_\mathrm{th}\mathcal{D}[\hta^\dag]\hat{\rho}\nonumber\\
&+\kappa_\mathrm{2ph} \mathcal{D}[\hta^2]\hat{\rho}.
\end{align}
The Hamiltonian $\hat{H}_0(\phi_0)$ stabilizes a cat qubit of real and positive amplitude $\alpha$. Following the results in section~\ref{sec_intro}.\ref{diss_2ph},
\begin{align}
\hat{H}_0(\phi_0)&=-K\hta^{\dag 2}\hta^2+P(\hta^{\dag 2} e^{2i\phi_0}+\mathrm{h.c.}), \\
 2\phi_0&=\tan^{-1}(\kappa_{2\mathrm{ph}}/2K), \quad P=\alpha^2\sqrt{K^2+\frac{\kappa^2_2}{4}}.
 \label{H0_2ph}
\end{align} 
From the simulations, we find that the  error channel takes the form,
\begin{align}
\mathcal{E}(\hat{\rho})=&\lambda_\mathrm{II}\hat{I}\hat{\rho}\hat{I}+\lambda_{\mathrm{IX}}\hat{I}\hat{\rho}\hat{X}+\lambda^*_\mathrm{IX}\hat{X}\hat{\rho}\hat{I}+\lambda_\mathrm{XX}\hat{X}\hat{\rho}\hat{X}\nonumber\\
&+\lambda_\mathrm{YY}\hat{Y}\hat{\rho}\hat{Y}+\lambda_\mathrm{YZ}\hat{Y}\hat{\rho}\hat{Z}+\lambda^*_\mathrm{YZ}\hat{Z}\hat{\rho}\hat{Y}+\lambda_\mathrm{ZZ}\hat{Z}\hat{\rho}\hat{Z}.
\label{ch_therm}
\end{align}
The coefficients $\lambda_\mathrm{II}$, $\lambda_{\mathrm{IX}}$, etc. are shown in Fig.~\ref{thermal_i} at time $t=50/K$ as a function of $\alpha$ for $n_\mathrm{th}=0.01$, $\kappa=K/400$ and $\kappa_\mathrm{2ph}=K/10$. Appendix~\ref{app}.~\ref{app_ch} discusses how the error channel is extracted from master equation simulations. The time $50/K$ is chosen because it is the typical gate time on the stabilized cat qubit. As expected, for large $\alpha$ $|\lambda_\mathrm{IX}|$, $\lambda_\mathrm{XX}$, $|\lambda_\mathrm{YZ}|$ and $\lambda_\mathrm{YY}$ decreases exponentially with $\alpha^2$. The amount of leakage is quantified by $1-\mathrm{Tr}[\mathcal{E}(\hat{I})]$ which is shown in Fig.~\ref{thermal_oos}(a) for $\kappa_\mathrm{2ph}=0$ (solid red line) and Fig~\ref{thermal_oos}(b) for $\kappa_\mathrm{2ph}=K/10$ (solid blue line). As expected leakage decreases in the presence of two-photon dissipation. The simple theoretical model predicts that for large $\alpha$, the leakage rate out of the cat manifold is $\sim\kappa n_\mathrm{th}$. The rate at which the excited state population decays back to the cat manifold due to single- and two-photon dissipation is $\sim\kappa(1+n_\mathrm{th})$ and $\sim4\kappa_\mathrm{2ph}\alpha^2$ respectively. Using these rates, it is possible to analytically estimate the amount of leakage, which is shown by the dashed black lines in Fig.~\ref{thermal_oos}(a,b). Indeed, the agreement between the numerical results and approximate analytical expressions is very good at large $\alpha$. 

Just like thermal noise, frequency fluctuations of the oscillator can also have a white spectral density. In Appendix~\ref{app}.~\ref{sec_fr}, we discuss the error channel for white frequency noise and provide numerical estimate for the corresponding error channel. As expected, we find that the non-dephasing errors are suppressed exponentially with $\alpha^2$.

The analysis in this section can easily be extended to any form of incoherent and coherent (or control) errors. We can now summarize the results for a general environment-oscillator interaction. Suppose, the system operator that enters in the interaction Hamiltonian is of the form $\sum_{m,n} \chi_{m,n}\hta^{\dag m}\hta^n+\mathrm{h.c.}$. The $\hta^{\dag m}$ term excites $\catpm$ to the $m^\mathrm{th}$ excited manifold $\ket{\psi^\pm_\mathrm{e,m}}$. Addition of two-photon dissipation autonomously corrects for this leakage error. Moreover, if the order of $\hta^\dag$ in the interaction, is smaller than the number of pairs of quasi-degenerate excited states, $\alpha^2/4$, the dominant error is of the form $f(\alpha)\szz$, while the non-dephasing errors are exponentially suppressed. Here $f(\alpha)$ is a polynomial function which depends on the details of the interaction and amount of two-photon dissipation added to correct for leakage. In other words, the two-photon driven nonlinear oscillator effectively results in an inherent quantum code to correct for up to $\alpha^2/4$ bit-flip errors.


\section{Bias-preserving CX gate}
\label{sec-CX}
\subsection{Overview}
\label{subsec-CX_overview}

As discussed in section~\ref{intro}, for the noise channel to remain biased, the time-dependent unitary describing the system evolution during the gate must not explicitly contain a $\sx$ operator. 
How can we then implement a controlled-$\sx$ (CX) gate? 
In order to build intuition about on how to address this problem, it is useful to note that $\sx\catpm=\pm\catpm$. 
Now, recall from Eq.~\eqref{H0} that the orientation of the cat state in phase-space is defined by the phase $\phi$ of the two-photon drive. 
If this phase changes adiabatically from $0$ to $\pi$, then the cat states $\catpm$ transform to $\ket{\mathcal{C}^\pm_{-\alpha}}=\pm\catpm$. 
Therefore, rotating the phase of the two-photon drive by $\pi$ is equivalent to a $\sx$ operation. 
Our proposal for a two-qubit bias-preserving CX gate is based on this phase-space rotation of a target cat-qubit conditioned on the state of a control cat-qubit. 
In this section, we first describe the desired evolution of the system under a CX gate and show that this evolution preserves the bias. 
Subsequently, in the next section we describe the underlying Hamiltonian achieving this evolution. 

Consider two cat qubits each stabilized in a two-photon driven Kerr nonlinear oscillator. 
The initial state of the system is
\begin{align}
\ket{\psi(0)}&=(c_0\ket{{0}} +c_1\ket{{1}})\otimes (d_0\ket{{0}} +d_1\ket{{1}})\nonumber\\
&=(c_0\ket{{0}} +c_1\ket{{1}})\otimes [(d_0+d_1)\catp+(d_0-d_1)\catm],\nonumber
\end{align}
where the first and second terms in the tensor product refer to the control and target qubits, respectively. Now suppose that the phase of the two-photon drive applied to the target oscillator is conditioned on the state of the control cat-qubit so that at time $t$ the state of the system is
\begin{align}
\ket{\psi(t)}=& c_0\ket{{0}} \otimes [(d_0+d_1)\catp +(d_0-d_1)\catm]\nonumber\\
&+c_1\ket{{1}} \otimes [(d_0+d_1)\catprt +(d_0-d_1)\catmrt].
\label{evol}
\end{align}
If the phase $\phi(t)$ is such that $\phi(0)=0$ and $\phi(T)=\pi$, then at time $T$

\begin{align}
\ket{\psi(T)}=&c_0\ket{{0}} \otimes  \{(d_0+d_1)\catp +(d_0-d_1)\catm\}\nonumber\\
&+c_1\ket{{1}} \otimes  \{(d_0+d_1)\ket{\mathcal{C}^+_{\alpha e^{i\pi}}} +(d_0-d_1)\ket{\mathcal{C}^-_{\alpha e^{i\pi}}}\} \nonumber\\
=&c_0\ket{{0}} \otimes \{(d_0+d_1)\catp +(d_0-d_1)\catm\}\nonumber\\
&+c_1\ket{{1}} \otimes \{(d_0+d_1)\catp -(d_0-d_1)\catm \}\nonumber\\
=&c_0\ket{{0}} \otimes (d_0\ket{{0}} +d_1\ket{{1}})+c_1\ket{{1}} \otimes (d_0\ket{{1}} +d_1\ket{{0}})\nonumber\\
=&\hat{U}_\mathrm{CX}\ket{\psi(0)}.
\label{CX_x}
\end{align}
As expected from the above discussion, a CX gate is realized by rotating the phase of the cat in the target oscillator by $\pi$ conditioned on the control cat. The CX operation is based on the fact that during this rotation, the $\catm$ state acquires a $\pi$ phase relative to $\catp$. This is a topological phase as it does not depend on energy like a dynamic phase, or the geometry of the path like a geometric phase. This phase will arise as long as the states $\ket{\pm\alpha}$ move along a loop in phase space that doesn't come too close to the origin (see further discussion in the next section and Appendix~\ref{appen_phases}.)
If the number of times that the states $\ket{\pm\alpha}$ go around the origin to $\ket{\mp\alpha}$ is given by $u$, then the phase acquired by $\catm$ is $e^{iu\pi}$. In other words, $u$ is the winding number.

Coupling with the environment during this evolution leads to errors in both the control and target cats. From the analysis in section~\ref{sec_err}, the predominant stochastic errors are of the form $\hat{O}_\mathrm{c}=f(\alpha)\sz{\mathrm{c}}$ in the control cat and $\hat{O}^\tau_\mathrm{t}=f'(\alpha e^{i\phi (\tau)}) \sz{\mathrm{t}}^{\tau}$ in the target cat where the superscript $\tau$ refers to the operator in the instantaneous basis $\sz{\mathrm{t}}^{\tau}=\ket{\mathcal{C}^+_{\alpha e^{i\phi(\tau)}}}\bra{\mathcal{C}^-_{\alpha e^{i\phi(\tau)}}}+\ket{\mathcal{C}^-_{\alpha e^{i\phi(\tau)}}}\bra{\mathcal{C}^+_{\alpha e^{i\phi(\tau)}}}$. We now show that these dominant phase errors during the CX evolution propagate as phase errors. To see this, assume that a phase error occurred in the control qubit at time $\tau$. Consequently, immediately after this error has occured the state of the system is
\begin{widetext}
\begin{align}
\ket{\psi(\tau)}^\mathrm{phase-flip}_\mathrm{control}&=\hat{O}_\mathrm{c}\otimes \hat{I}^\tau_{\mathrm{t}}\left\{c_0\ket{{0}} \otimes [(d_0+d_1)\catp +(d_0-d_1)\catm]+c_1\ket{{1}} \otimes  [(d_0+d_1)\catprtau +(d_0-d_1)\catmrtau]\right\}\nonumber\\
&=c_0\ket{{0}} \otimes [(d_0+d_1)\catp +(d_0-d_1)\catm]-c_1\ket{{1}}\otimes [(d_0+d_1)\catprtau +(d_0-d_1)\catmrtau].
\end{align}
After this phase-flip event, the conditional phase continues to evolve and at time $T$,
\begin{align}
\ket{\psi(T)}^\mathrm{phase-flip}_\mathrm{control}&=c_0\ket{{0}} \otimes [(d_0+d_1)\catp +(d_0-d_1)\catm]-c_1\ket{{1}} \otimes [(d_0+d_1)\catp -(d_0-d_1)\catm]\nonumber\\
&=\sz{\mathrm{c}}\otimes\hat{I}^\tau_{\mathrm{t}}\left\{c_0\ket{{0}} \otimes [(d_0+d_1)\catp +(d_0-d_1)\catm]+c_1\ket{{1}} \otimes [(d_0+d_1)\catp -(d_0-d_1)\catm]\right\}\nonumber\\
&=(\sz{\mathrm{c}}\otimes\hat{I}^\tau_{\mathrm{t}})\hat{U}_\mathrm{CX}\ket{\psi(0)}.
\end{align}
Therefore, a phase error on the control cat qubit at any time during the implementation of the CX is equivalent to a phase-flip on the control qubit after an ideal CX. 

Now, assume that a phase error occurred on the target at time $\tau$. Immediately after this error, the state is
\begin{align}
\ket{\psi(\tau)}^\mathrm{phase-flip}_\mathrm{target}=&\hat{I}_{\mathrm{c}}\otimes\hat{O}^\tau_\mathrm{t}\left\{c_0\ket{{0}} \otimes  [(d_0+d_1)\catp +(d_0-d_1)\catm]+c_1\ket{{1}} \otimes  [(d_0+d_1)\catprtau +(d_0-d_1)\catmrtau\} \right]\nonumber\\
=&f(\alpha)\left\{ c_0\ket{{0}} \otimes [(d_0+d_1)\catm +(d_0-d_1)\catp]\right]\nonumber\\
&+f(\alpha e^{i\phi(\tau)})\left[ c_1\ket{{1}} \otimes [(d_0+d_1)\catmrtt +(d_0-d_1)\catprtt]\right\}.
\end{align}
As before, after this phase-flip event the conditional phase continues to evolve, and at time $T$,
\begin{align}
\ket{\psi(T)}^\mathrm{phase-flip}_\mathrm{target}=&f(\alpha)\left\{ c_0\ket{{0}} \otimes [(d_0+d_1)\catm +(d_0-d_1)\catp]\right\}\nonumber\\
&+f(\alpha e^{i\phi(\tau)})\left\{ c_1\ket{{1}} \otimes [-(d_0+d_1)\catm +(d_0-d_1)\catp]\right\}\nonumber\\
=&\hat{I}_\mathrm{c}\otimes\sz{\mathrm{t}}\left\{ f(\alpha )c_0\ket{{0}} \otimes [ d_0\ket{{0}}+d_1\ket{{1}} ]-f(\alpha e^{i\phi(\tau)})c_1\ket{{1}} \otimes [d_0\ket{{1}}+d_1\ket{{0}}]\right\}\nonumber\\
=&\left[\sz{\mathrm{c}}f(\alpha e^{i\phi(\tau)(1-\sz{\mathrm{c}})/2})\otimes \sz{\mathrm{t}}\right]\hat{U}_\mathrm{CX}\ket{\psi(0)}.
\label{CX_et}
\end{align}
\end{widetext}

Remarkably, the above equations show that a phase-flip error on the target qubit at any time during the CX evolution is equivalent to phase errors on the control and target qubits after the ideal CX gate. 
In other words, this CX gate based on rotation of the target cat-qubit in phase space does not un-bias the noise channel. 
This is in stark contrast with the CX gate implementation between two strictly two-level qubits and shows the advantage of using the larger Hilbert space of an oscillator. 
Although we have only explicitly showed the bias preserving nature of the CX with respect to one phase-flip in either the control or target cats, it is easy to extend the analysis above to multiple phase-flips to see that the bias remains preserved. 
Moreover, note that any control errors in the target or control qubit can be expanded in the form $\sum_{m,n,p,q} \chi_{m,n,p,q}\hta^{\dag m}_\mathrm{c}\hta^n_\mathrm{c}\hta^{\dag p}_\mathrm{t}\hta^q_\mathrm{t}$, where $\hta_\mathrm{c}$ and $\hta_\mathrm{t}$ are the annihilation operators for control and target oscillators respectively. 
Of course the terms $\hta^{\dag m}_\mathrm{c},\hta^{\dag p}_\mathrm{t}$ will excite the control and target oscillators out of the cat qubit subspace. 
As we have already seen, addition of photon dissipation will autonomously correct this leakage while keeping bit-flips exponentially suppressed as long as the weights $p,m<\alpha^2/4$. 
In fact, small amounts of control error will only lead to low weight terms in the expansion above and therefore the bias will be maintained. We will now explain this more in detail with an example. 

Suppose the control error was such that at the end of the gate, $\phi(T)=\pi+\Delta$ (instead of $\phi(T)=\pi$). That is,
\begin{align}
\ket{\psi(T)}=&c_0\ket{{0}} \otimes [(d_0+d_1)\catp +(d_0-d_1)\catm]\nonumber\\
&+c_1\ket{{1}} \otimes   [(d_0+d_1)\ket{\mathcal{C}^+_{-\alpha e^{i\Delta}}}
 +(d_0-d_1)\ket{\mathcal{C}^-_{-\alpha e^{i\Delta}}}
 ].
 \label{cont_e}
\end{align}
Now, $\ket{\mathcal{C}^\pm_{-\alpha e^{i\Delta}}}=\pm e^{i\Delta\hta^\dag\hta}\catpm=\pm(1+i\Delta\hta^\dag\hta-\Delta^2\hta^\dag\hta\hta^\dag\hta/2+...)\catpm$ and, for small $\Delta$, only a few terms in the expansion are important. 
In fact, below a threshold error $\Delta<\Delta_\mathrm{th}$ the high weight $(>\alpha^2/4)$ terms exponentially decrease. 
The control error in this case only causes excitation of states in the pair-wise quasi-degenerate manifold, which are subsequently corrected by two-photon dissipation. 
Note that during this autonomous correction, the cat states pick up an overall phase depending on when the photon jump events happened, $\ket{\mathcal{C}^\pm_{-\alpha e^{i\Delta}}}\rightarrow \pm e^{i\chi}\catpm $. 
Similar to Eq.~\eqref{CX_et}, this extra phase leads to dephasing of the control cat-qubit. 
In general, the threshold $\Delta_\mathrm{th}$ depends on the strength of the Kerr nonlinearity and rate of two-photon dissipation. 
However, numerical and analytical estimates predict that in the experimentally relevant limit $K\gg \kappa_\mathrm{2}$, the threshold is as large as $\Delta_\mathrm{th}\sim \pi/6$ (see Appendix~\ref{app}.~\ref{app_thresh}). 
The large threshold shows the robustness of the gate to rotation errors.  

Note that there is another source of rotation errors in the target cat. Indeed, any non-dephasing error in the control qubit during the CX gate will cause leakage in the target oscillator. 
For example, a bit-flip error in the control cat at $t=T/2$ causes a phase-space rotation error in the target cat by $\pi/2$. 
That is, at the end of the gate the target cat states are $\ket{\mathcal{C}^\pm_{i\alpha}}$ rather than $\catpm$. 
This can, however, be corrected by two-photon dissipation. 
Moreover, since the non-dephasing errors in the control cat are exponentially suppressed, so is the leakage and the non-dephasing faults from subsequent correction of leakage.

\subsection{Hamiltonian of the bias-preserving CX gate}~\label{sec_hamil}
Having seen that the evolution in Eq.~\eqref{evol} results in a CX gate with biased-noise error channel, we will now present the physical interaction Hamiltonian required to implement it. 
In general we assume that the amplitude of the cats in the target and control oscillators, $\alpha$ and $\beta$ respectively, are different.
The following time-dependent interaction Hamiltonian implements the bias-preserving CX between the two oscillators,
\begin{align}
\hat{H}_\mathrm{CX}=&-K\left(\hta_\mathrm{c}^{\dag 2}-\beta^2\right)\left(\hta_\mathrm{c}^{ 2}-\beta^2\right)
\nonumber\\
&-K\left[\hta^{\dag 2}_\mathrm{t}-\alpha^2e^{-2i\phi(t)}\left(\frac{\beta-\hta^\dag_\mathrm{c}}{2\beta}\right)-\alpha^2\left(\frac{\beta+\hta^\dag_\mathrm{c}}{2\beta}\right)\right]\nonumber\\
&\times\left[\hta^2_\mathrm{t}-\alpha^2e^{2i\phi(t)}\left(\frac{\beta-\hta_\mathrm{c}}{2\beta}\right)-\alpha^2\left(\frac{\beta+\hta_\mathrm{c}}{2\beta}\right)\right]\nonumber\\
&-\frac{\dot{\phi}(t)}{4\beta}\hta^\dag_\mathrm{t}\hta_\mathrm{t}(2\beta-\hta^\dag_\mathrm{c}-\hta_\mathrm{c}).
\label{CX_h}
\end{align}
The first line in the above expression is the Hamiltonian of the parametrically driven nonlinear oscillator stabilizing the control cat-qubit. 
The phase of the drive to this oscillator is fixed $\phi=0$. 
To understand the other two lines, recall that $\hta^\dag_\mathrm{c},\hta_\mathrm{c}\sim\beta\sz{c}\pm i\beta e^{-2\beta^2}\syy{c}$. 
Therefore, if the control qubit is in the state $\ket{{0}}$ ($\sim \ket{\beta}$, for large $\beta$) and we ignore the exponentially small contribution from the term $\propto \syy{c}$, then the above Hamiltonian is equivalent to
\begin{align}
\hat{H}_\mathrm{CX}^{\ket{{0}}_\mathrm{c}}\equiv&-K\left(\hta_\mathrm{c}^{\dag 2}-\beta^2\right)\left(\hta_\mathrm{c}^{ 2}-\beta^2\right)\nonumber\\
&-K\left(\hta^{\dag 2}_\mathrm{t}-\alpha^2\right)\left(\hta^2_\mathrm{t}-\alpha^2\right).
\end{align}
Consequently, when the control qubit is in the state $\ket{{0}}$ the state of the target oscillator remains unchanged. 
On the other hand, if the control qubit is in the state $\ket{{1}}$ ($\sim\ket{-\beta}$, for large $\beta$) then Eq.~\eqref{CX_h} is equivalent to
\begin{align}
\hat{H}_\mathrm{CX}^{\ket{{1}}_\mathrm{c}}\equiv &-K\left(\hta_\mathrm{c}^{\dag 2}-\beta^2\right)\left(\hta_\mathrm{c}^{ 2}-\beta^2\right)\nonumber\\
&-K\left(\hta^{\dag 2}_\mathrm{t}-\alpha^2 e^{-2i\phi(t)}\right)\left(\hta^2_\mathrm{t}-\alpha^2e^{2i\phi(t)}\right)\nonumber\\
&-\dot{\phi}(t)\hta^\dag_\mathrm{t}\hta_\mathrm{t}.
\label{CX_g}
\end{align}
From the second line of this expression we see that the cat states $\catpmrt$ are the instantaneous eigenstates in the target oscillator. 
As a result, if the phase $\phi(t)$ changes adiabatically, respecting $\dot{\phi}(t)\ll |\Delta\omega_\mathrm{gap}|$ then the orientation of the target cats follow $\phi(t)$ and $\alpha$ evolves in time to $\alpha e^{i\phi(t)}$. 
During this rotation in phase space the target cat also acquires a geometric phase $\Phi^\pm_\mathrm{g}(t)$ proportional to the area under the phase space path, $e^{i\Phi^\pm_\mathrm{g}(t)}\catpmrt$ where $\Phi^\pm_\mathrm{g}(t)=\phi(t)\alpha^2r^{\mp 2}$. 
The difference in the two geometric phases, $\Phi^+_\mathrm{g}$ and $\Phi^-_\mathrm{g}$, reflects the fact that the mean photon numbers are different for the two states $\catpmrt$ and the area of the path followed by $\catmrt$ in phase space is larger than that followed by $\catprt$. 
This geometric phase has some interesting properties which are discussed in Appendix~\ref{appen_phases}.  
In the limit of large $\alpha$, the difference in the two decreases exponentially in $\alpha^2$, $\Phi^-_\mathrm{g}-\Phi^+_\mathrm{g}=4\phi(t)\alpha^2 e^{-2\alpha^2}/(1-e^{-4\alpha^2})$. 
Consequently, for large $\alpha$, the state $\ket{1}\otimes d_0\catp+d_1\catm$ evolves in time to $e^{i\Phi_\mathrm{g}(t)}\ket{1}\otimes d_0\catprt+d_1\catmrt$ where $\Phi_g(t)=\Phi^-_\mathrm{g}(t)\sim\Phi^+_\mathrm{g}(t)$. 
In other words, the geometric phase, effectively, is only an overall phase which results in an additional $Z_\mathrm{c}(\Phi_g)$ rotation on the control qubit. 
This rotation can be accounted for in software or by an application of $Z_\mathrm{c}(-\Phi_g)$ operation. 
Or it can be directly cancelled during the CX gate itself by the addition of an additional interaction, given by the last term in Eq.~\eqref{CX_g}. 
The projection of this term in the cat basis is given by
\begin{align}
\dot{\phi}(t)\hta^\dag_\mathrm{t}\hta_\mathrm{t}\equiv&\dot{\phi}(t)\alpha^2 \left[r^2\catprt\bra{\mathcal{C}^+_{\alpha e^{i\phi(t)}}}\right.\nonumber\\
&+\left.{r^{-2}}\catmrt\bra{\mathcal{C}^-_{\alpha e^{i\phi(t)}}}\right].
\end{align}

The above equation shows that the last term of Eq.~\eqref{CX_g} leads to a dynamic phase which exactly cancels the geometric phase. 
As a result, we find that when the control cat is in state $\ket{{1}}$, an arbitrary state of the target qubit $d_0\catp+d_1\catm$ evolves in time to $d_0\catprt+d_1\catmrt$. 
Consequently, the Hamiltonian in Eq.~\eqref{CX_h} leads to the evolution desired to implement the bias-preserving CX gate.

\subsection{Numerical simulations}~\label{CX_num}
\begin{figure}
\begin{centering}
 \includegraphics[width=\columnwidth]{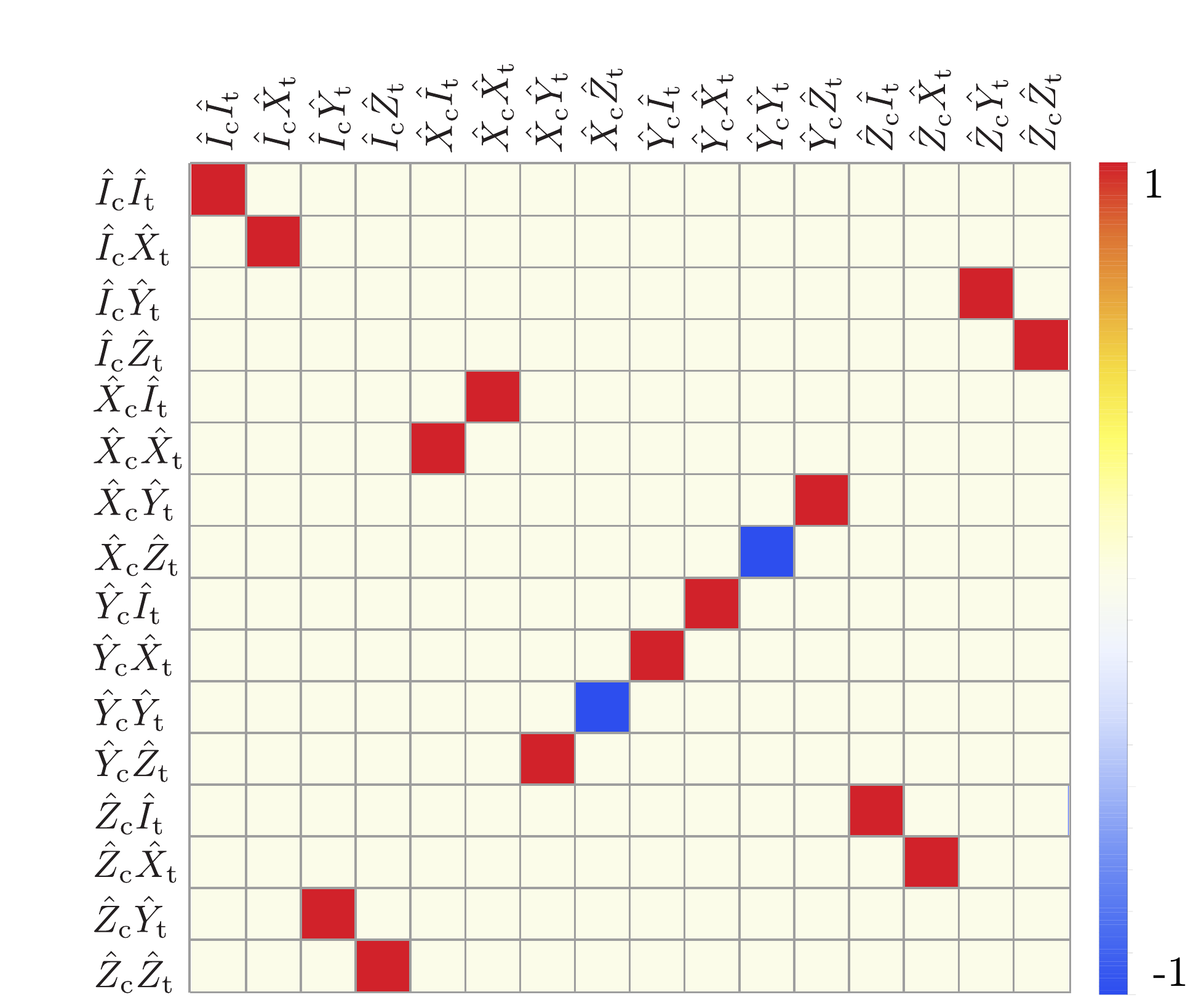}
 \caption{Pauli transfer matrix of the CX gate obtained by simulating the Hamiltonian in Eq.~\eqref{CX_h} with $\alpha=\beta=2$, $\phi(t)=\pi t/T$ and $T=10/K$. The infidelity of this CX operation w.r.t. an ideal two-level CX is $9.3\times 10^{-7}$ and results from non-adiabatic transitions due to finite $KT$. }
 \label{CX_i}
 \end{centering}
 \end{figure}

To show that the Hamiltonian of Eq.~\eqref{CX_h} does result in a bias-preserving CX, we first simulate Eq.~\eqref{CX_h} without noise in the oscillators. 
We chose $\alpha=\beta=2$, $\phi(t)=\pi t/T$ and $T=10/K$. 
Figure~\eqref{CX_i} shows the Pauli transfer matrix obtained in this way. 
The infidelity between the CX resulting from the evolution under Eq.~\eqref{CX_h} and an ideal CX is as small as $\sim 9.3\times 10^{-7}$. 
This small infidelity, primarily resulting from non-adiabatic transitions due to finite $KT$, clearly shows that the Hamiltonian of Eq.~\eqref{CX_h} implements an ideal CX gate with an extremely high degree of accuracy.

Next, in order to account for losses we numerically simulate evolution under the master equation 
\begin{align}
\dot{\hat{\rho}}=&-i[\hat{H}_\mathrm{CX},\hat{\rho}]+\kappa(n_\mathrm{th}+1)\sum_{i=\mathrm{c},\mathrm{t}}\mathcal{D}[\hta_i]\hat{\rho}\nonumber\\
&+\kappa n_\mathrm{th}\sum_{i=\mathrm{c},\mathrm{t}} \mathcal{D}[\hta^\dag_i]\hat{\rho}.
\end{align}
From this, we obtain the Pauli transfer matrix of the noisy CX, $R^\mathrm{CX}_\mathrm{noisy}$. The transfer matrix of the error channel is evaluated as $R_\mathrm{noise}=R^\mathrm{CX}_\mathrm{noisy}(R^\mathrm{CX}_\mathrm{ideal})^{-1}$. Finally, the error channel in the operator sum form is obtained from this transfer matrix. Instead of listing all the 256 matrix entries of the channel, we present its dominant terms. Moreover, to quantify the asymmetry in the noise channel of the CX gate, we introduce a quantity $\eta$ referred to as the bias. The bias, $\eta_\mathrm{CX}$, is defined as the ratio of probability of dephasing and non-dephasing faults. The probability of dephasing errors is obtained from the error channel as the sum of the coefficients corresponding to the terms $\hat{I}_\mathrm{c}\sz{\mathrm{t}}\hat{\rho}\hat{I}_\mathrm{c}\sz{\mathrm{t}}$, $\sz{\mathrm{c}}\hat{I}_\mathrm{t}\hat{\rho}\sz{\mathrm{c}}\hat{I}_\mathrm{t}$ and $\sz{\mathrm{c}}\sz{\mathrm{t}}\hat{\rho}\sz{\mathrm{c}}\sz{\mathrm{t}}$. In the same way, the probability of non-dephasing error is the sum of the coefficients corresponding to the remaining diagonal terms (except for $\hat{I}_\mathrm{c}\hat{I}_\mathrm{t}\hat{\rho}\hat{I}_\mathrm{c}\hat{I}_\mathrm{t}$). The coefficient corresponding to $\hat{I}_\mathrm{c}\hat{I}_\mathrm{t}\hat{\rho}\hat{I}_\mathrm{c}\hat{I}_\mathrm{t}$ yields the gate fidelity.

For $n_\mathrm{th}=0$, we find that the error channel is dominantly given by
\begin{align}
\mathcal{E}(\hat{\rho})\sim &\lambda_\mathrm{I_cI_t,I_cI_t}\hat{I}_\mathrm{c}\hat{I}_\mathrm{t}\hat{\rho}\hat{I}_\mathrm{c}\hat{I}_\mathrm{t}+\lambda_\mathrm{Z_cZ_t,Z_cZ_t}\sz{\mathrm{c}}\sz{\mathrm{t}}\hat{\rho}\sz{\mathrm{c}}\sz{\mathrm{t}}
\nonumber\\
&+\lambda_\mathrm{Z_c I_t,Z_c I_t}\sz{\mathrm{c}}\hat{I}_\mathrm{t}\hat{\rho}\sz{\mathrm{c}}\hat{I}_\mathrm{t}+
\lambda_\mathrm{I_cZ_t,I_cZ_t}\hat{I}_\mathrm{c}\sz{\mathrm{t}}\hat{\rho}\hat{I}_\mathrm{c}\sz{\mathrm{t}}\nonumber\\
&+\left(i\lambda_\mathrm{I_cZ_t,Z_cZ_t}\hat{I}_\mathrm{c}\sz{\mathrm{t}}\hat{\rho}\sz{\mathrm{c}}\sz{\mathrm{t}}+\mathrm{h.c.}\right).
\end{align}
For $\kappa=K/4000$, $T=10/K$ and $\alpha=\beta=2$, $\lambda_\mathrm{I_cI_t,I_cI_t}\sim 0.94$, $\lambda_\mathrm{Z_cI_t,Z_cI_t}\sim 0.029$, $\lambda_\mathrm{I_cZ_t,I_cZ_t}\sim 0.015$, $\lambda_\mathrm{Z_cZ_t,Z_cZ_t}\sim 0.015$, $\lambda_\mathrm{I_cZ_t,Z_cZ_t}\sim  -0.009$ and the gate fidelity is $94\%$.
The leakage is $9.6\times 10^{-7}$ which does not significantly increase from the case when losses are absent and the bias is $\eta\sim 10^{7}$.

Next, we obtain the error channel for $n_\mathrm{th}=1\%$. In order to correct for leakage two-photon dissipation $\kappa_2\mathcal{D}[\hta^2]\hat{\rho}$ is added after the gate operation (see Appendix~\ref{appen_thm} for details).
In the absence of the two-photon dissipation $\kappa_2=0$, the amount of leakage due to thermal photons is $\sim 3\times 10^{-5}$. With $\kappa_2=K/5$, leakage is reduced by almost two-orders of magnitude to $\sim 5\times 10^{-6}$. The gate fidelity in this case is reduced to $\sim 89\%$ and the error channel is dominantly given by
\begin{align}
\mathcal{E}(\hat{\rho})\sim &\lambda_\mathrm{I_cI_t,I_cI_t}\hat{I}_\mathrm{c}\hat{I}_\mathrm{t}\hat{\rho}\hat{I}_\mathrm{c}\hat{I}_\mathrm{t}+\lambda_\mathrm{Z_cZ_t,Z_cZ_t}\sz{\mathrm{c}}\sz{\mathrm{t}}\hat{\rho}\sz{\mathrm{c}}\sz{\mathrm{t}}
\nonumber\\
&+\lambda_\mathrm{Z_cI_t,Z_cI_t}\sz{\mathrm{c}}\hat{I}_\mathrm{t}\hat{\rho}\sz{\mathrm{c}}\hat{I}_\mathrm{t}+
\lambda_\mathrm{I_cZ_t,I_cZ_t}\hat{I}_\mathrm{c}\sz{\mathrm{t}}\hat{\rho}\hat{I}_\mathrm{c}\sz{\mathrm{t}}\nonumber\\
&+\left(i\lambda_\mathrm{I_cI_t,Z_cI_t}\hat{I}_\mathrm{c}\hat{I}_\mathrm{t}\hat{\rho}\sz{\mathrm{c}}\hat{I}_\mathrm{t}+\mathrm{h.c.}\right)\nonumber\\
&+\left(i\lambda_\mathrm{I_cZ_t,Z_cZ_t}\hat{I}_\mathrm{c}\sz{\mathrm{t}}\hat{\rho}\sz{\mathrm{c}}\sz{\mathrm{t}}+\mathrm{h.c.}\right),
\end{align}
with $\lambda_\mathrm{I_cI_t,I_cI_t}\sim 0.89$, $\lambda_\mathrm{Z_cI_t,Z_cI_t}\sim 0.052$, $\lambda_\mathrm{I_cZ_t,I_cZ_t}\sim 0.016$, $\lambda_\mathrm{Z_cZ_t,Z_cZ_t}\sim 0.038$,
$\lambda_\mathrm{I_cI_t,Z_cI_t}\sim  -0.0002$ and $\lambda_\mathrm{I_cZ_t,Z_cZ_t}\sim -  0.008$. The order of magnitude of the other terms in the error channel is $\leq 10^{-5}$ and the bias is $\eta\sim 732$. When the size of the cats is increased to $\alpha=\beta=2.2$ and $\alpha=\beta=2.5$, the bias increases to $\eta\sim 902$ and $\eta\sim 3000$ respectively.

Finally, we numerically estimate the error channel in case of over-rotation. This can happen, for example, when control errors lead to the gate being implemented for slightly longer time $T'=T+\delta(T)$. For the simulation we choose $\pi\delta(T)=0.01 T$ corresponding to an over-rotation of the target cat by an angle $\Delta=0.01$ (see Eq.~\eqref{cont_e}). In this case, we simulate the master equation $\dot{\hat{\rho}}=-i[\hat{H}_\mathrm{CX},\hat{\rho}]+\kappa\mathcal{D}[\hta_\mathrm{c}]\hat{\rho}+\kappa\mathcal{D}[\hta_\mathrm{t}]\hat{\rho}$ for time $T'$ and then add two-photon dissipation $\kappa_2\mathrm{D}[\hta_\mathrm{t}^2]+\kappa_2\mathrm{D}[\hta_\mathrm{c}^2]$ to correct for over-rotation. The dominant terms of the resulting error-channel are
\begin{align}
\mathcal{E}(\hat{\rho})\sim &\lambda_\mathrm{I_cI_t,I_cI_t}\hat{I}_\mathrm{c}\hat{I}_\mathrm{t}\hat{\rho}\hat{I}_\mathrm{c}\hat{I}_\mathrm{t}+\lambda_\mathrm{Z_cZ_t,Z_cZ_t}\sz{\mathrm{c}}\sz{\mathrm{t}}\hat{\rho}\sz{\mathrm{c}}\sz{\mathrm{t}}
\nonumber\\
&+\lambda_\mathrm{Z_cI_t,Z_cI_t}\sz{\mathrm{c}}\hat{I}_\mathrm{t}\hat{\rho}\sz{\mathrm{c}}\hat{I}_\mathrm{t}+
\lambda_\mathrm{I_cZ_t,I_cZ_t}\hat{I}_\mathrm{c}\sz{\mathrm{t}}\hat{\rho}\hat{I}_\mathrm{c}\sz{\mathrm{t}}\nonumber\\
&+\left(i\lambda_\mathrm{I_cZ_t,Z_cZ_t}\hat{I}_\mathrm{c}\sz{\mathrm{t}}\hat{\rho}\sz{\mathrm{c}}\sz{\mathrm{t}}+\mathrm{h.c.}\right).
\end{align}
For $\kappa=K/4000$, $\kappa_2=K/5$, and $\alpha=\beta=2$, $\lambda_\mathrm{I_cI_t,I_cI_t}\sim 0.97$, $\lambda_\mathrm{Z_cI_t,Z_cI_t}\sim 0.038$, $\lambda_\mathrm{I_cZ_t,I_cZ_t}\sim 0.015$, $\lambda_\mathrm{Z_cZ_t,Z_cZ_t}\sim 0.024$, $\lambda_\mathrm{I_cZ_t,Z_cZ_t}\sim -0.009$ and the bias is $\eta_\mathrm{CX}\sim 1955$. For $\alpha=\beta=2.2$, the bias increases to $\eta_\mathrm{CX}\sim 2796$. The above examples confirm that the noise channel of the CX gate is biased and the bias increases with the size of the cat. Because of the large Hilbert space size, it becomes difficult to perform numerical simulations for larger $\alpha$. However, using the insights from single oscillator simulations in section~\ref{sec_err}.\ref{sec_th} and appendix~\ref{sec_fr} we expect to achieve a bias of $\sim 10^4$ for $\alpha^2<10$ with experimentally reasonable experimental parameters.


\section{Threshold and overhead for concatenation-based codes}
\label{sec_thresh}
In section~\ref{sec_intro} we have described the adiabatic preparation of the cat states $\catpm$, $\mathcal{P}_{\ket{\pm}}$. We have also outlined the implementation of arbitrary rotations about the $Z$-axis and implement $ZZ(\theta)$-gates. In addition, measurements along $Z$-axis, $\mathcal{M}_{\hat{Z}}$, can be performed using homodyne detection, while measurements along $X$-axis, $\mathcal{M}_{\hat{X}}$, require intermediary gates or ancilla~\cite{puri2018stabilized,ofek2016extending}. The preparation operation, measurements and the gates $Z(\theta)$, and $ZZ(\theta)$ are trivially biased. More importantly however, we have shown that it is also possible to implement a biased-noise CX gate between two cat-qubits. 
 Observe that, the bias-preserving set of unitaries $\{\mathrm{CX},Z(\theta),ZZ(\theta)\}$ is not universal. In fact, as shown in Appendix~\ref{app}.~\ref{appen_univ}, no matter how the Hamiltonian evolution is constructed a native, universal set of bias preserving unitaries is impossible. However the unitaries $\{\mathrm{CX},Z(\theta),ZZ(\theta)\}$, in combination with state preparation, $\mathcal{P}_{\ket{\pm}}$, and measurements $\mathcal{M}_{\hat{X},\hat{Z}}$~\cite{puri2018stabilized}, are sufficient to implement universal fault-tolerant quantum computation~\cite{aliferis2009fibonacci}. 
 
 In this section, we will use the physical bias preserving set of operations $\{\mathrm{CX},Z(\theta),ZZ(\theta), \mathcal{P}_{\ket{\pm}},\mathcal{M}_{\hat{X}},\mathcal{M}_{\hat{Z}}\}$ to realize efficient and compact circuits for fault-tolerant error correction based on concatenation~\cite{aliferis2008fault} (Ref.~\cite{guillaud2019repetition} also discusses the repetition code using the idea of CX gates described here adapted to dissipative cats.)  
 For the following analysis we will consider the error-channel in the Pauli-twirling approximation. 
 That is, we ignore the off-diagonal elements in the error-channel. 
 This approximation can always be enforced by actively randomizing the Pauli frame at each step of a computation~\cite{knill2005quantum, Wallman2015}
 The resulting channel can then be understood in the stochastic noise model by assigning a probability to each fault-path. 
 In this approximation, for example, the error channel of the two-qubit CX is dominantly of the form 
 $\mathcal{E}(\hat{\rho})\sim\lambda_\mathrm{I_tI_c,I_tI_c}(\hat{I}_\mathrm{t}\hat{I}_\mathrm{c}\hat{\rho}\hat{I}_\mathrm{t}\hat{I}_\mathrm{c})+\lambda_\mathrm{Z_tZ_c,Z_tZ_c}(\sz{\mathrm{t}}\sz{\mathrm{c}}\hat{\rho}\sz{\mathrm{t}}\sz{\mathrm{c}})+\lambda_\mathrm{I_tZ_c,I_tZ_c}(\hat{I}_\mathrm{t}\sz{\mathrm{c}}\hat{\rho}\hat{I}_\mathrm{t}\sz{\mathrm{c}})+
\lambda_\mathrm{Z_tI_c,Z_tI_c}(\sz{\mathrm{t}}\hat{I}_\mathrm{c}\hat{\rho}\sz{\mathrm{t}}\hat{I}_\mathrm{c})$. 
This noise channel effectively introduces dephasing errors in the target and control cat qubits with probability $\lambda_\mathrm{Z_tI_c,Z_tI_c}+\lambda_\mathrm{Z_tZ_c,Z_tZ_c}$ and $\lambda_\mathrm{I_tZ_c,I_tZ_c}+\lambda_\mathrm{Z_tZ_c,Z_tZ_c}$ respectively. 
For simplicity, we will denote by $\varepsilon$ the upper bound on the probability of a dephasing error in a cat-qubit resulting from the noise during a single-qubit gate, two-qubit gate, state preparation or measurement. For the example of the CX gate, this means that $\lambda_\mathrm{Z_tI_c,Z_tI_c}+\lambda_\mathrm{Z_tZ_c,Z_tZ_c}, \lambda_\mathrm{I_tZ_c,I_tZ_c}+\lambda_\mathrm{Z_tZ_c,Z_tZ_c}\leq\varepsilon$. 
We define a bias $\eta$ so that the probability of a $\sx$ or $\sy$ error is $\varepsilon/\eta$. 
\begin{figure*}[ht]
\begin{centering}
 \includegraphics[width=2\columnwidth]{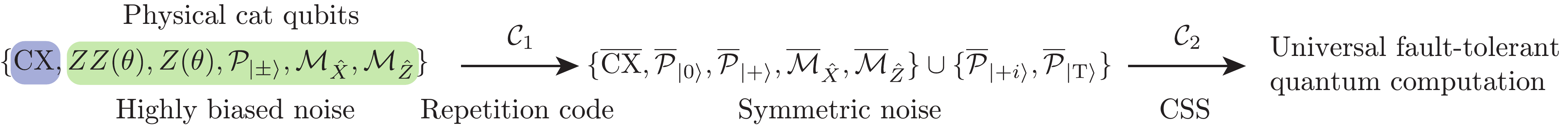}
 \caption{Overview of the scheme for concatenated biased-noise error correction introduced in~\cite{aliferis2008fault}. 
 The set of physical operations which have biased noise channel which is used in~\cite{aliferis2008fault} is highlighted in green. 
 As we have shown, when cat states in parametrically driven non-linear oscillators are used as qubits, a CX gate (highlighted in blue) respecting the bias becomes possible. 
 In the first step, the physical qubits are encoded in a repetition code $\mathcal{C}_1$ to correct for phase-errors. 
 Next fault-tolerant $\mathcal{C}_1$ protected gadgets are constructed for the next level of concatenation with a CSS code. 
 As shown in the main text, the addition of a physical CX gate with biased noise simplifies the construction of $\mathcal{C}_1$ protected $\mathcal{C}_2$ gadgets.}
 \label{overview}
 \end{centering}
\end{figure*}
 
The idea introduced in~\cite{aliferis2008fault} is to first encode the physical biased-noise qubits in a repetition code $\mathcal{C}_1$ and correct for dominant errors, in this case phase-flips. 
A repetition code with $n$ qubits can correct $(n-1)/2$ phase-flip errors. The codewords are $\ket{0}_\mathrm{L}=(\ket{+}_\mathrm{L}+\ket{-}_\mathrm{L})/\sqrt{2}$ and $\ket{1}_\mathrm{L}=(\ket{+}_\mathrm{L}-\ket{-}_\mathrm{L})/\sqrt{2}$, where $\ket{+}_\mathrm{L}=\catp\catp\catp...$ and $\ket{-}_\mathrm{L}=\catm\catm\catm...$. 
The result of the first encoding is a more symmetric noise channel with reduced noise strength. 
The repetition code with errors below a threshold can then be concatenated to a CSS code $\mathcal{C}_2$ to further reduce the errors. Figure~\ref{overview} summarizes this scheme. 
In~\cite{aliferis2008fault}, error correction and $\mathcal{C}_2$-compatible logical operations at $\mathcal{C}_1$ are implemented using only trivially biased CZ gates, preparations and measurements (see Fig.~\ref{overview}). 
The $\mathcal{C}_1$ protected $\mathcal{C}_2$-gadgets considered in~\cite{aliferis2008fault} are $\{\overline{\mathrm{CX}},\overline{P}_{\ket{0}},\overline{P}_{\ket{+}},\overline{\mathcal{M}}_{\hat{X}},\overline{\mathcal{M}}_{\hat{Z}}\}$. 
These Clifford operations are supplemented with preparation of magic-states $\overline{P}_{\ket{+i}}$ and $\overline{P}_{\ket{\mathrm{T}}}$. 
The error strengths at $\mathcal{C}_1$ is upper-bounded by the $\overline{\mathrm{CX}}$ gadget. 
In this section, we simplify the circuit for the $\overline{\mathrm{CX}}$ and error-correction gadgets by exploiting the availability of the physical biased-noise CX gate between the cat qubits. 
We show that the error rate and volume of this $\overline{\mathrm{CX}}$-gate is lower than that proposed in~\cite{aliferis2008fault}. 
Implementation of the other $\mathcal{C}_1$-protected $\mathcal{C}_2$ Clifford operations is the same as in~\cite{aliferis2008fault} and is outlined in the Appendix. 
We also complete the analysis by outlining the preparation of magic-states using the trivially bias-preserving physical $ZZ(\theta)$-gates~\cite{webster2015reducing}.

\subsection{Error correction in the repetition code}
\label{subsec_rep}

\begin{figure}
\begin{centering}
 \includegraphics[width=.6\columnwidth]{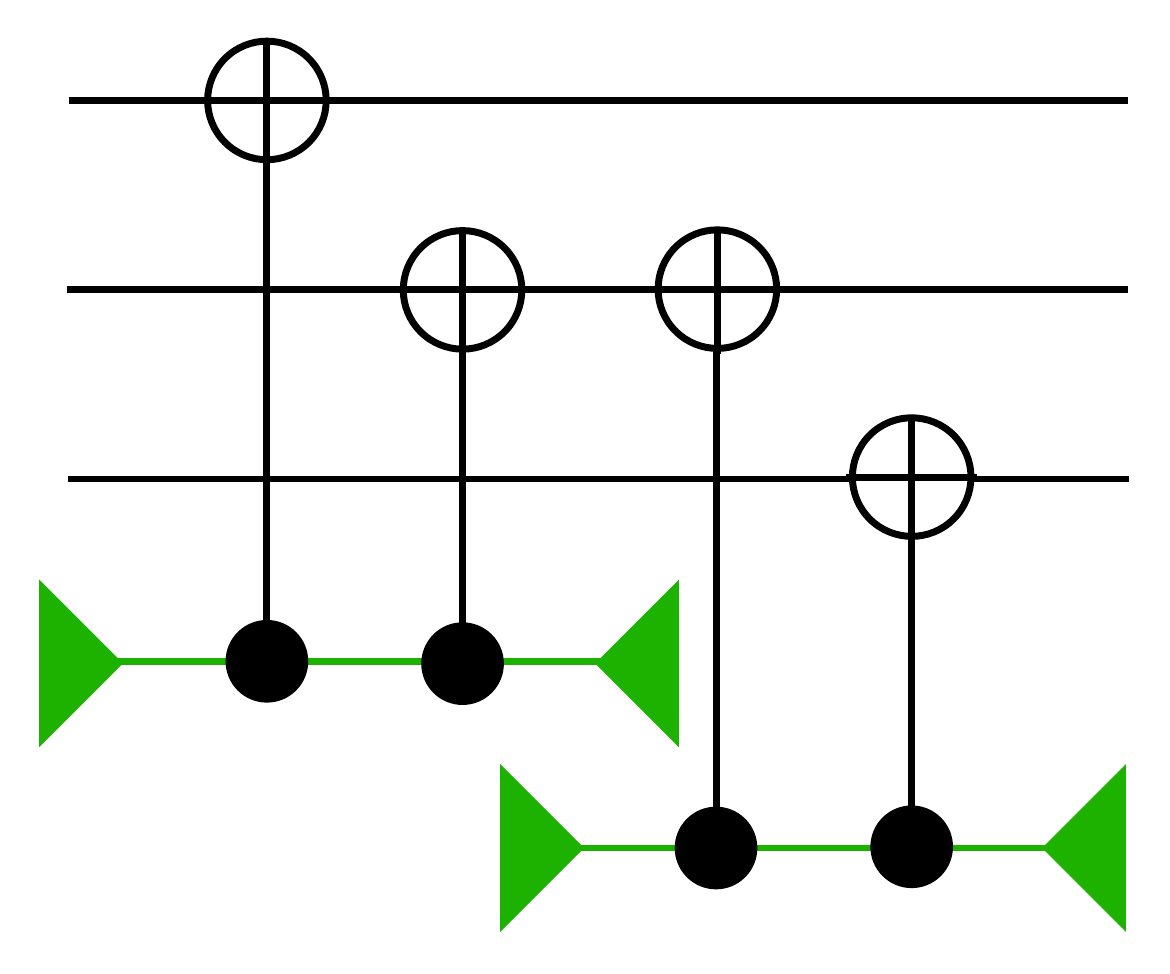}
 \caption{Error-correction gadget for a repetition code with $n=3$. The black and green lines indicate code and ancilla qubits respectively. The green triangles facing the left and right represent preparation and measurement of the ancilla respectively. In the naive scheme, $(n-1)$ stabilizer generators for the repetition code are measured using CX gates between pairs of data qubits and ancilla.}
 \label{EC_g}
 \end{centering}
 \end{figure}
 
The $(n-1)$ stabilizer generators for the repetition code are $\sxx{1}\otimes\sxx{2}\otimes\hat{I}_3\otimes\hat{I}_4...$, $\hat{I}_1\otimes\sxx{2}\otimes\sxx{3}\otimes\hat{I}_4...$, etc. 
The most naive way to detect errors is to measure each stabilizer generator using an ancilla as shown in Fig~\ref{EC_g}. 
Each ancilla is initialized in the state $\catp$. 
Then two CX gates are implemented between the ancilla and qubits $j$, $j+1$. 
Finally, the $(n-1)$ ancillas are measured along the $X$ axis $\mathcal{M}_{\hat{X}}$. 
To be fault-tolerant, each of the stabilizer generator is measured $r$ times and the syndrome bit is determined with a majority vote on the measurement outcomes. 
A syndrome bit is incorrect if $m\geq (r+1)/2$ of the measurements are faulty.

This decoding scheme is equivalent to constructing an $r$-bit repetition code for each of the $(n-1)$ stabilizer generators of the repetition code. 
Thus, each bit of syndrome from the inner code is itself encoded in an $[r,1,r]$ repetition code so that decoding can proceed by first decoding the syndrome bits and then decoding the resulting syndrome. 
As we will see shortly, this naive way to decode the syndrome results in a simple analytic expressions for the logical error rates. 
However, it is by no means an ideal approach to decode and one can imagine that the two-stage decoder above could be replaced by one that directly infers the most likely error on the $n$-qubit repetition code given $s$ measured syndrome bits. 
Below, we introduce the notion of a {\it measurement code} that exploits these insights to improve on the naive scheme by constructing a block code that can directly correct the bit-flip errors on the $n$ data qubits in a single decoding step. 
This will be discussed in section~\ref{sec_thresh}.\ref{subsec_optimal}.

\subsection{\texorpdfstring{$\overline{\mathrm{CX}}$}{Logical CX}-gate with naive decoding}
 
\begin{figure}
\begin{centering}
 \includegraphics[width=\columnwidth]{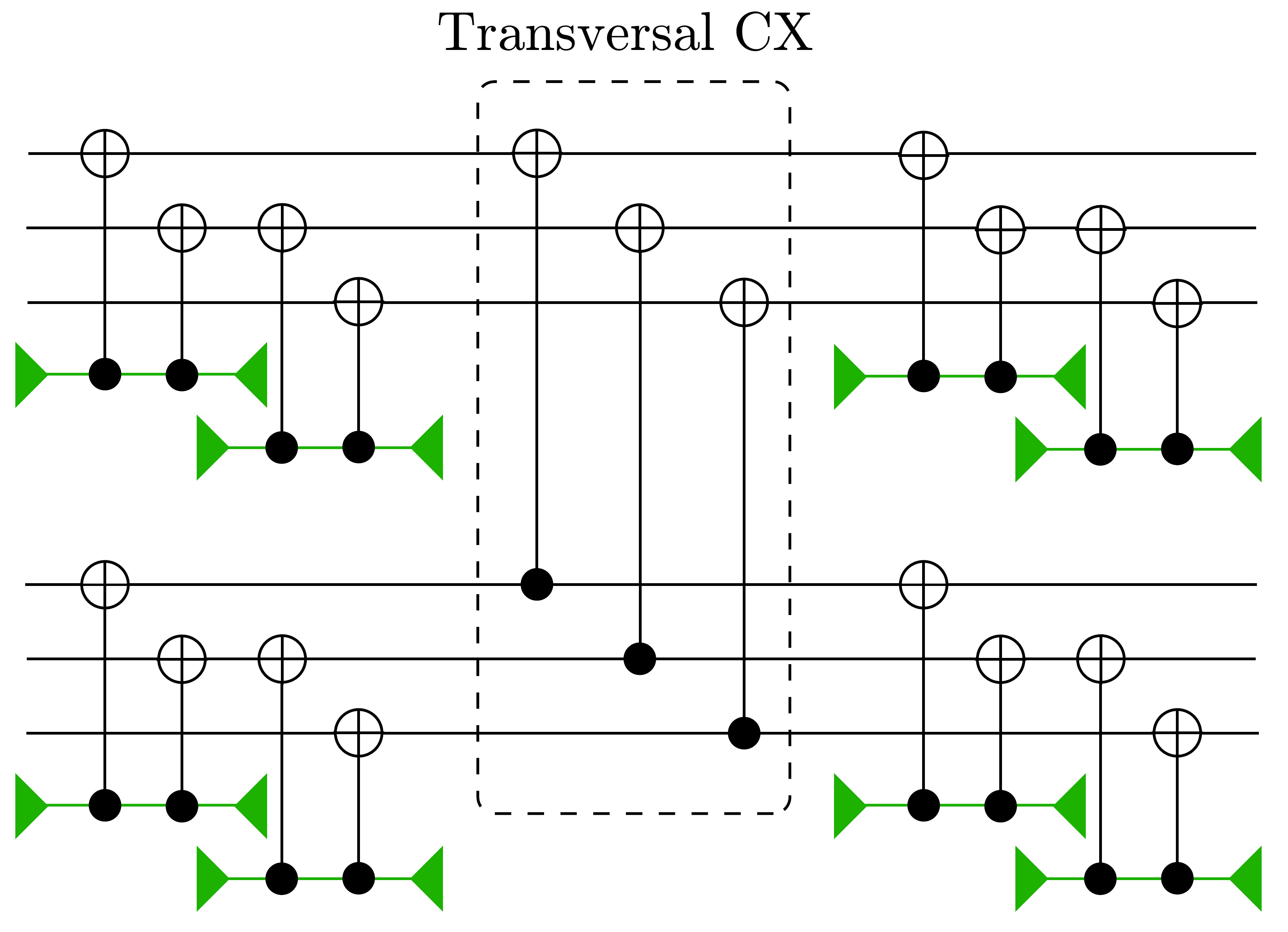}
 \caption{ $\overline{\mathrm{CX}}_\mathrm{cat}$ gadget based on tranversal biased-noise CX gates between the codewords. The codewords are error-corrected at the input and output.}
 \label{CX_gadget}
 \end{centering}
\end{figure}
Since a physical CX with error channel biased towards dephasing errors is available, the $\overline{\mathrm{CX}}$-gadget can be implemented with transversal CXs between two code blocks, as shown in Fig.~\ref{CX_gadget}. 
We will refer to this as $\overline{\mathrm{CX}}_\mathrm{cat}$-gadget, because the biased-noise CX gates are realized using cat-qubits. 
We will now estimate an upper bound for the logical error rate of the $\overline{\mathrm{CX}}_\mathrm{cat}$-gadget. 

Each data qubit coming into the target and control blocks of the $\overline{\mathrm{CX}}_\mathrm{cat}$-gadget is subject to $2r$ CX gates during the previous error-correction step. 
The probability of a dephasing fault in each data qubit is therefore $2r\varepsilon$.
Next, each data qubit in the target and control block is subject to one CX gate. 
Note however that phase errors from the target can spread to phase errors on the control. 
Therefore, the probability of a dephasing fault in each qubit in the target and control blocks is $2r\varepsilon+\varepsilon$ and $4r\varepsilon+\varepsilon$ respectively. 
A logical error will occur if $m\geq (n+1)/2$ qubits in the target or control blocks are faulty. 
Therefore, the probability of a logical error in the control and target blocks before they are input into the error correction gadgets are
\begin{align}
\varepsilon_\mathrm{target}&\leq\binom{n}{\frac{n+1}{2}}(2r\varepsilon+\varepsilon)^{(n+1)/2}\\
\varepsilon_\mathrm{control}&\leq\binom{n}{\frac{n+1}{2}}(4r\varepsilon+\varepsilon)^{(n+1)/2}.
\end{align}
Each of the error correction gadgets now measure $(n-1)$ syndromes and each syndrome bit must be read correctly for successful decoding. 
Each syndrome bit is measured $r$ times and requires 2 CX gates between a pair of code qubits and an ancilla. 
A syndrome measurement can be incorrect if the preparation or measurement of the ancilla was incorrect, or if there was a dephasing error on the ancilla during the CXs. 
Therefore, an upper bound on the probability of error due to failure of the error correction in the target and control blocks is
\begin{align}
\varepsilon_\mathrm{ec}\leq2(n-1)\binom{r}{\frac{r+1}{2}}(4\varepsilon)^{(r+1)/2}.
\end{align}

 In the worst case, a single non-dephasing error occurring with probability $\epsilon/\eta$ anywhere in the circuit will cause the failure of the gadget.  
 There are $4(n-1)r$ CX gates in each of the error-correction gadgets at the input and output and $n$ transversal CX gates. 
 As a result, the probability of an error due to a non-dephasing fault is
 \be
 \varepsilon'\leq (8(n-1)r+n)\frac{\varepsilon}{\eta}.
 \ee
Finally, the probability of a logical error in the $\mathrm{\overline{CX}}$-gadget is given by
 \begin{align}
 \varepsilon_\mathrm{cat}=&\varepsilon_\mathrm{target}+\varepsilon_\mathrm{control}+\varepsilon_\mathrm{ec}+ \varepsilon'\\
 =&\binom{n}{\frac{n+1}{2}}(2r\varepsilon+\varepsilon)^{(n+1)/2}+\binom{n}{\frac{n+1}{2}}(4r\varepsilon+\varepsilon)^{(n+1)/2}\nonumber\\
 &+2(n-1)\binom{r}{\frac{r+1}{2}}(4\varepsilon)^{(r+1)/2}+(8(n-1)r+n)\frac{\varepsilon}{\eta}.
 \label{log_P}
 \end{align}

\begin{figure}
\begin{centering}
 \includegraphics[width=\columnwidth]{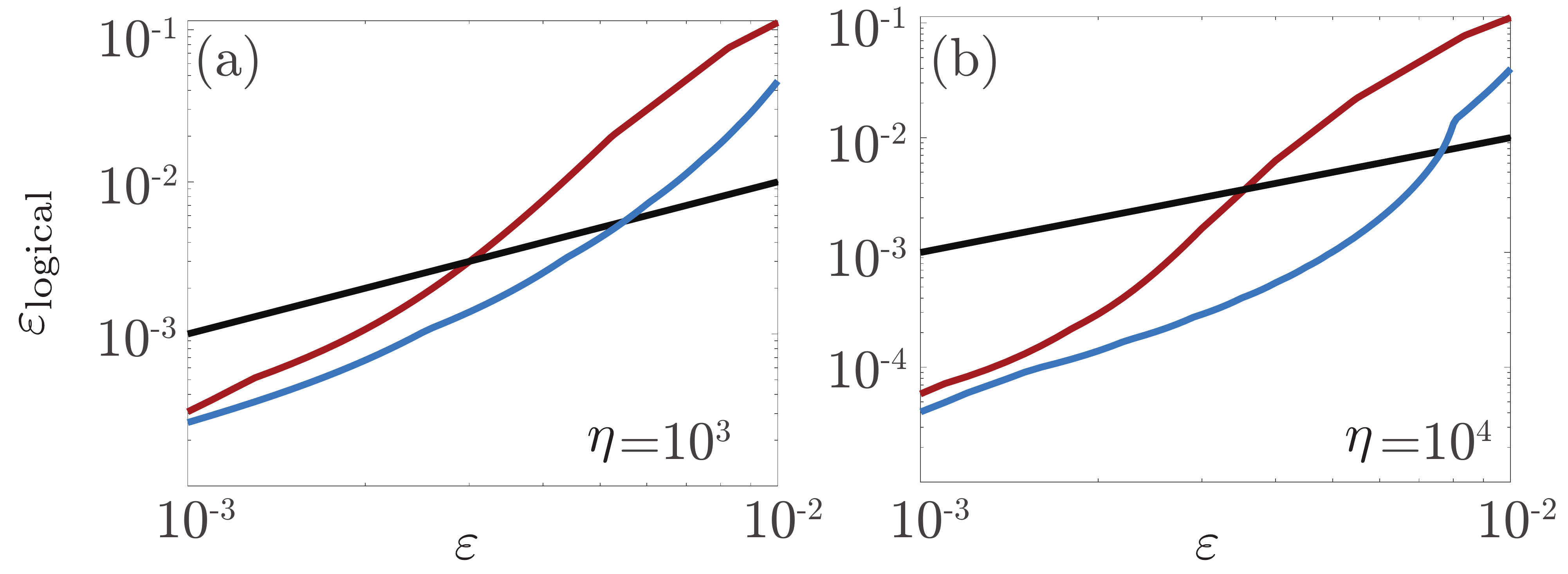}
 \caption{Logical error rate for $\overline{\mathrm{CX}}$ given in Eqs.~\eqref{log_P} (solid blue line) and from~\cite{aliferis2008fault} (solid red line) for different bias $\eta$. The black line with slope=1 is shown for reference. }
 \label{log_comp}
 \end{centering}
 \end{figure}
 
 \begin{figure}
\begin{centering}
 \includegraphics[width=\columnwidth]{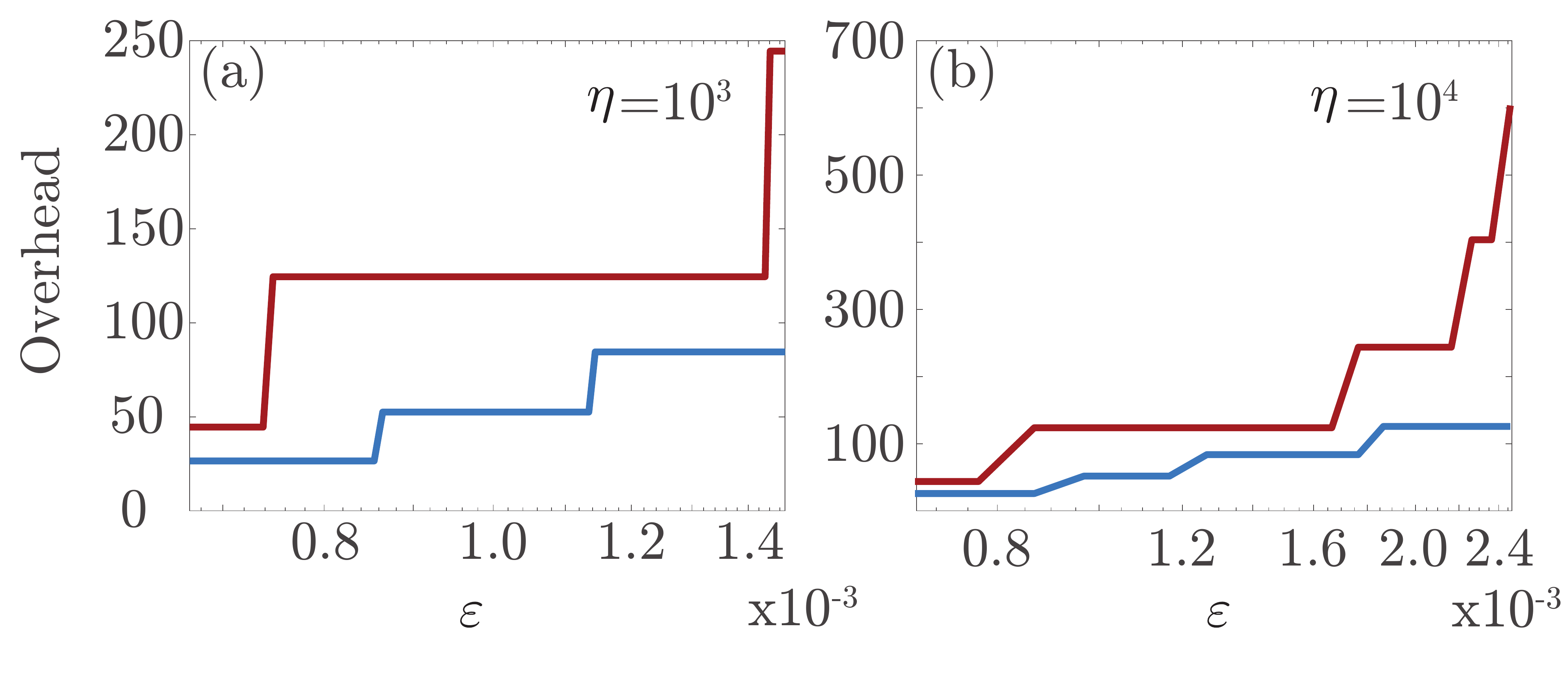}
\caption{Overhead of the $\overline{\mathrm{CX}}_\mathrm{cat}$-gadget (blue line) for a target logical error rate of $0.67\times 10^{-3}$~\cite{aliferis2008fault,aliferis2009fibonacci}. The overhead for the gadget proposed in~\cite{aliferis2008fault} for the same target error rate is shown in red.} 
\label{over_comp}
 \end{centering}
 \end{figure}

 Figure~\ref{log_comp} compares the logical error rates for the $\overline{\mathrm{CX}}_\mathrm{cat}$-gadget in  Eq.~\eqref{log_P} (blue line) and that for the gadget in~\cite{aliferis2008fault} (red line) as a function of the bare error $\varepsilon$ for different bias $\eta$. For reference, a line with slope=1 is also shown (black). The $\overline{\mathrm{CX}}_\mathrm{cat}$-gadget clearly has lower probability for logical errors. In fact, for $\eta=10^4$ the threshold error for the gadget (that is, where the blue curve intersects the black line) is $\varepsilon_\mathrm{cat}=7.5\times 10^{-3}$. This is more than twice the threshold of the $\mathrm{\overline{CX}}$-gadget in~\cite{aliferis2008fault}, 
  $\varepsilon_\mathrm{AP}=3.55\times 10^{-3}$. For smaller bias, the contribution from the non-dephasing term in Eq.~\eqref{log_P} takes over and the performance of $\mathrm{\overline{CX}}$ degrades. 
  
  Moreover, we find that the $\overline{\mathrm{CX}}_\mathrm{cat}$-gadget also requires less overhead to reach the same target logical error rate compared to the gadget in ~\cite{aliferis2008fault}. In order to demonstrate this, we estimate the circuit volume required to reach a target error rate of $0.67\times 10^{-3}$. Using Eqs.~\eqref{log_P} we find the $n$ and $r$ required so that $\varepsilon_\mathrm{cat}\leq 0.67\times 10^{-3}$. The circuit volume for the $\overline{\mathrm{CX}}$ in  Ref~\cite{aliferis2008fault} and that described here are $7nr$ and $8(n-1)r+2n$ respectively. Figure~\ref{over_comp} compares these overheads for $\eta=10^3$ and $\eta=10^4$, as a function of $\varepsilon$. Clearly, the $\mathrm{\overline{CX}}_\mathrm{cat}$ described here has a smaller overhead. For example, with $\varepsilon=2.5\times 10^{-3}$ and $\eta=10^4$, the overhead for $\overline{\mathrm{CX}}_\mathrm{cat}$ is $\sim 5$ times smaller than that for the gadget described in~\cite{aliferis2008fault}.

\begin{figure}
\begin{centering}
 \includegraphics[width=\columnwidth]{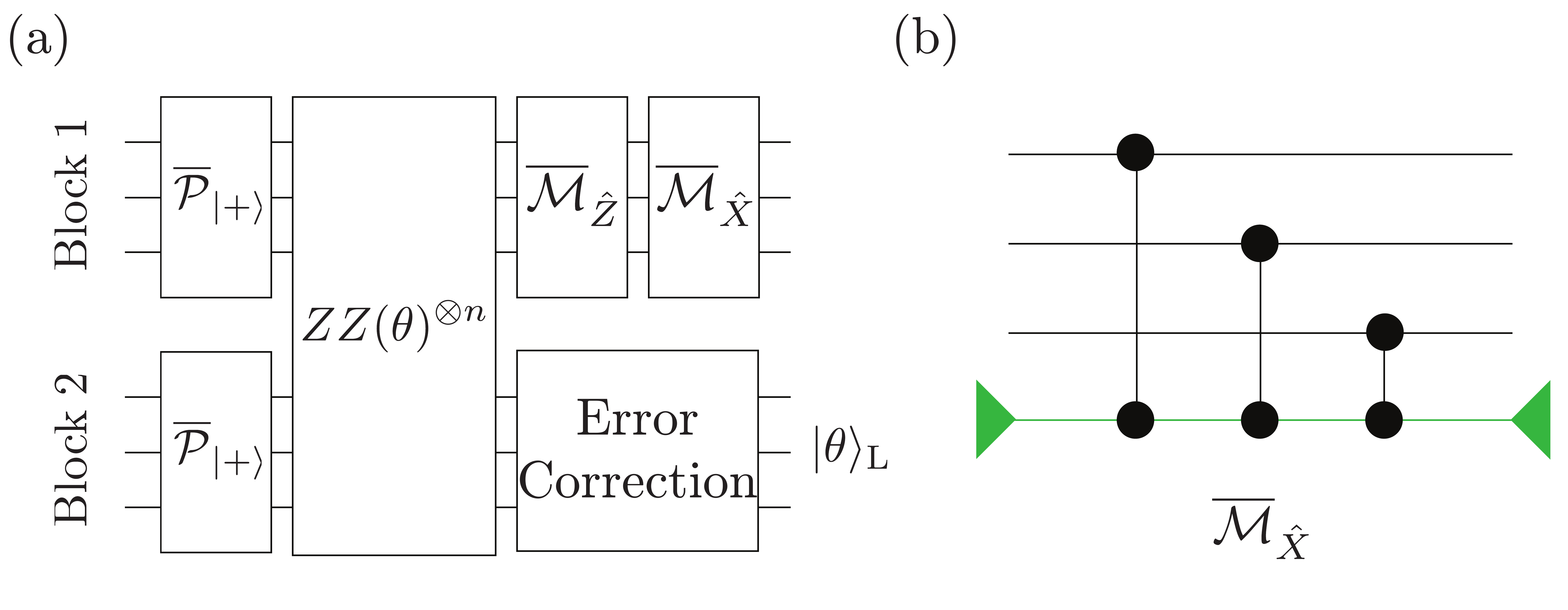}
 \caption{(a) Scheme for magic state preparation adapted from~\cite{webster2015reducing}. The error-correction gadget used is shown in Fig.~\ref{EC_g}. (b) The $\overline{\mathcal{M}}_{\hat{Z}}$ gadget.}
 \label{magic}
 \end{centering}
 \end{figure}
 
\subsection{Magic-state preparation}
In this section, we extend the analysis in~\cite{webster2015reducing} for the preparation of magic states $\bar{\mathcal{P}}_{\ket{i}}$ and $\bar{\mathcal{P}}_{\ket{\mathrm{T}}}$. The gadget for magic state preparation is shown in Fig \ref{magic} (a). The gadget consists of two
repetition code blocks initialized in the state $\ket{+}_\mathrm{L}$. The non-Clifford operation in this gadget is the transversal application of $ZZ(\theta)$ gates between the physical qubits in block 1 and block 2. The transversally applied $ZZ(\theta)$ gates do not preserve the codespace and this operation is not equivalent to a logical $\overline{ZZ}(\theta)$ gate. Following this step, a $\overline{\mathcal{M}}_{\hat{Z}}$ measurement is performed on block 1, followed by $\overline{\mathcal{M}}_{\hat{X}}$ measurement which disentangles block 1 from block 2. Finally, error correction is performed on block 2 to map it into the codespace. We will now elaborate how this gadget leads to a deterministic preparation of $\ket{i}_\mathrm{L}$ state and probabilistic preparation of $\ket{\mathrm{T}}_\mathrm{L}$ with appropriate choice of $\theta$.

After blocks 1 and 2 are each prepared in the $\ket{+}_\mathrm{L}=\ket{+}^{\otimes n}$ state, the pairwise $ZZ(\theta)$ gates entangles them. The state after the application of this operation is
\be
\ket{\psi}=[\ket{0}\ket{\theta}+\sx\sx\ket{0}\ket{\theta}]^{\otimes n},
\ee 
where $\ket{\theta}=\cos(\theta/2)\ket{+}+i\sin(\theta/2)\ket{-}$. The $\overline{\mathcal{M}}_{\hat{Z}}$ measurement on block 1 projects it in a state with even or odd number of $\ket{1}$ depending on the measurement outcome $\overline{\mathcal{M}}_{\hat{Z}}=1$ or $\overline{\mathcal{M}}_{\hat{Z}}=-1$ respectively. 
The state of the system after the $\overline{\mathcal{M}}_{\hat{Z}}$ measurement is 
\begin{align}
\ket{\psi}^{\overline{\mathcal{M}}_{\hat{Z}}=1}&=\sum_{|a|=\mathrm{even}} \bigotimes^n_{i=1}(\sx\sx)^{a_i}\ket{0}\ket{\theta}\nonumber\\
\ket{\psi}^{\overline{\mathcal{M}}_{\hat{Z}}=-1}&=\sum_{|a|=\mathrm{odd}} \bigotimes^n_{i=1}(\sx\sx)^{a_i}\ket{0}\ket{\theta},
\end{align}
where $a_i\in\{0,1\}$. To give an example, if $n=3$, $\ket{\psi} ^{\overline{\mathcal{M}}_{\hat{Z}}=1}=\ket{000}_1\otimes \ket{\theta\theta\theta}_2+\ket{011}_1\otimes \hat{I}\sx\sx\ket{\theta\theta\theta}_2+\ket{101}_1\otimes \sx\hat{I}\sx\ket{\theta\theta\theta}_2+\ket{110}_1\otimes \sx\sx\hat{I}\ket{\theta\theta\theta}_2$ and $\ket{\psi} ^{\overline{\mathcal{M}}_{\hat{Z}}=-1}=\ket{111}_1\otimes \sx\sx\sx\ket{\theta\theta\theta}_2+\ket{100}_1\otimes \sx\hat{I}\hat{I}\ket{\theta\theta\theta}_2+\ket{010}_1\otimes \hat{I}\sx\hat{I}\ket{\theta\theta\theta}_2+\ket{001}_1\otimes \hat{I}\hat{I}\sx\ket{\theta\theta\theta}_2$. Here the subscripts $1,2$ denote the blocks 1 and 2 respectively. 
Next, an $\overline{\mathcal{M}}_{\hat{X}}$ measurement is performed on block 1. 
That is, each qubit in block 1 is measured along the $X$ axis and is projected onto either the $\catp$ or $\catm$ state. 
If $b_i=\pm 1$ represents the result of the measurement on the $i^\mathrm{th}$ qubit, then the state of the qubits in block 2 is
\begin{align}
\ket{\psi}^{\overline{\mathcal{M}}_{\hat{Z}}=1}_b&=\sum_{|a|=\mathrm{even}} \bigotimes^n_{i=1}(b_i\sx)^{a_i}\ket{\theta}=A^1_b\ket{\theta}^{\otimes n}\\
\ket{\psi}^{\overline{\mathcal{M}}_{\hat{Z}}=-1}_b&=\sum_{|a|=\mathrm{odd}} \bigotimes^n_{i=1}(b_i\sx)^{a_i}\ket{\theta}=A^{-1}_b\ket{\theta}^{\otimes n}.
\label{md1}
\end{align}
Here $A^{\pm1}_b=\sum_{|a|=\mathrm{even/odd}} \bigotimes^n_{i=1}(b_i\sx)^{a_i}$ and $A^{-1}_b=\sx^{\otimes n}A^1_b=\sxx{\mathrm{L}}A^1_b$. 
Consequently, the state of block 2 when the result of the $\overline{\mathcal{M}}_{\hat{Z}}$ measurement is $-1$ differs from that when the measurement result is $1$ simply by $\sxx{\mathrm{L}}$. 
Therefore, for fault-tolerance, this measurement is repeated $r_\mathrm{z,L}$ times and a majority vote is taken over the measurement outcomes. The above expressions can be simplified to

\begin{align}
\ket{\psi}^{\overline{\mathcal{M}}_{\hat{Z}}=1}_b&=\left[\cos\left(\frac{\theta}{2}\right)\right]^s\left[i\sin\left(\frac{\theta}{2}\right)\right]^{n-s}\bigotimes_{i=1}^n\ket{\mathcal{C}^{\mathrm{sign}(b_i)}_\alpha}\nonumber\\
&+\left[i\sin\left(\frac{\theta}{2}\right)\right]^s\left[\cos\left(\frac{\theta}{2}\right)\right]^{n-s}\sz{\mathrm{L}}\bigotimes_{i=1}^n\ket{\mathcal{C}^{\mathrm{sign}(b_i)}_\alpha}\\
\ket{\psi}^{\overline{\mathcal{M}}_{\hat{Z}}=-1}_b&=\sxx{\mathrm{L}}\ket{\psi}^{\overline{\mathcal{M}}_{\hat{Z}}=1}_b,
\end{align}
where $s$ is the number of $+1$ in $b$. The above expressions show that block 2 is in a superposition of 
qubit states which are completely correlated and completely anti-correlated to the qubit states in block 1. 
To elaborate with an example, for $n=3$ if the $\overline{\mathcal{M}}_{\hat{Z}}$ measurement yields 1 and if the $\overline{\mathcal{M}}_{\hat{X}}$ measurement yields $b={1,1,1}$ or $b={-1,-1,-1}$ then the state of the qubits in block 2 is $\cos(\theta/2)^3 \catp\catp\catp+(i\sin(\theta/2))^3\catm\catm\catm$. If the $\overline{\mathcal{M}}_{\hat{X}}$ measurement yields $b={1,1,-1}$ or $b={-1,-1,1}$ then the state of the qubits in block 2 is $\cos(\theta/2)^2i\sin(\theta/2) \catp\catp\catm+(i\sin(\theta/2))^2\cos(\theta/2)\catm\catm\catp$.
Finally, the error-correction step on block 2 using the gadget shown in Fig.~\ref{EC_g} projects block 2 back into the codespace,
\begin{align}
\ket{\psi}^{\overline{\mathcal{M}}_{\hat{Z}}=1}_b=&\sz{\mathrm{L}}^{H[s-(n+1)/2]}\times\nonumber\\
&\left\{\left[\cos\left(\frac{\theta}{2}\right)\right]^s\left[i\sin\left(\frac{\theta}{2}\right)\right]^{n-s}\ket{+}_\mathrm{L}\right.   \nonumber\\
&+\left. \left[i\sin\left(\frac{\theta}{2}\right)\right]^s\left[\cos\left(\frac{\theta}{2}\right)\right]^{n-s}\ket{-}_\mathrm{L}\right\}\\
\ket{\psi}^{\overline{\mathcal{M}}_{\hat{Z}}=-1}_b=&\sxx{\mathrm{L}}\ket{\psi}^{\overline{\mathcal{M}}_{\hat{Z}}=1}_b,
\label{md2}
\end{align}
where $H[s-(n+1)/2]$ is the heaviside step function. 
The above expression shows that the resulting code state depends on the choice of $\theta$ and the intermediate measurement results. 
Using $\theta=\pi/2$, the state $\ket{i}_\mathrm{L}$ is prepared (up to correctable Pauli rotations) irrespective of the measurement outcome. 
Moreover, if we find that if the error syndromes obtained from the measurement of the stabilizer generators do not agree with the $\overline{\mathcal{M}}_{\hat{Z}}$ measurement on block 1, then we reject the state on block 2.
Using $\theta=\pi/4$, the state $\ket{\mathrm{T}}_\mathrm{L}$ is prepared (up to correctable Pauli rotations) only when $n-s=s\pm1$ or $s=(n\pm 1)/2$. 
Out of all the $2^n$ possible measurement results, the number of outputs that result in the correct state is $\binom{n}{\frac{n+1}{2}}+\binom{n}{\frac{n-1}{2}}$. 
Therefore, the probability of successful  $\ket{\mathrm{T}}_\mathrm{L}$ state preparation is,
$p_\mathrm{success}=2^{1-n}\binom{n}{\frac{n-1}{2}}$. 
For example, with $n=9$ or $n=21$, $p_\mathrm{success}=0.49$ or $0.34$ respectively. 
The probability of success decreases with increasing $n$ because the $ZZ(\theta)^{\otimes n}$ gates do not preserve the codespace. 
Next, we will examine the probability of logical error in the preparation of the states.

\subsubsection{Upper bound on the probability of a logical non-dephasing error}
A logical non-dephasing fault in block 2 can be due to a physical non-dephasing error during the $2nr$ CX gates in the error correction gadget or during any of the $n$ $ZZ(\theta)$ gates. Moreover, a faulty $\overline{\mathcal{M}}_{\hat{Z}}$ measurement also leads to a non-dephasing error (see Eq.~\eqref{md2}). 
This can happen due to a non-dephasing error in any of the $n$ $ZZ(\theta)$ gates or any of the $nr_\mathrm{z,L}$ CZ gates. 
It can also result if a majority $m=(r_\mathrm{z,L}+1)/2$ of the measurements are faulty (either due to dephasing faults in any one of the $n$ CZ gates or the preparation and measurement of the ancilla.) 
Therefore, the probability of a logical non-dephasing error is upper bounded by
\begin{align}
\begin{split}
\varepsilon_\mathrm{x,L}\leq\ & (2nr+2n+nr_\mathrm{z,L})\frac{\varepsilon}{\eta}\\
&\ +\binom{r_\mathrm{z,L}}{\frac{r_\mathrm{z,L}+1}{2}}(n\varepsilon+2\varepsilon)^{(r_\mathrm{z,L}+1)/2}.
\end{split}
\label{xT}
\end{align}

\subsubsection{Upper bound on the probability of \texorpdfstring{$\sz{\mathrm{L}}$}{logical dephasing} error}
\begin{figure}
\begin{centering}
 \includegraphics[width=.8\columnwidth]{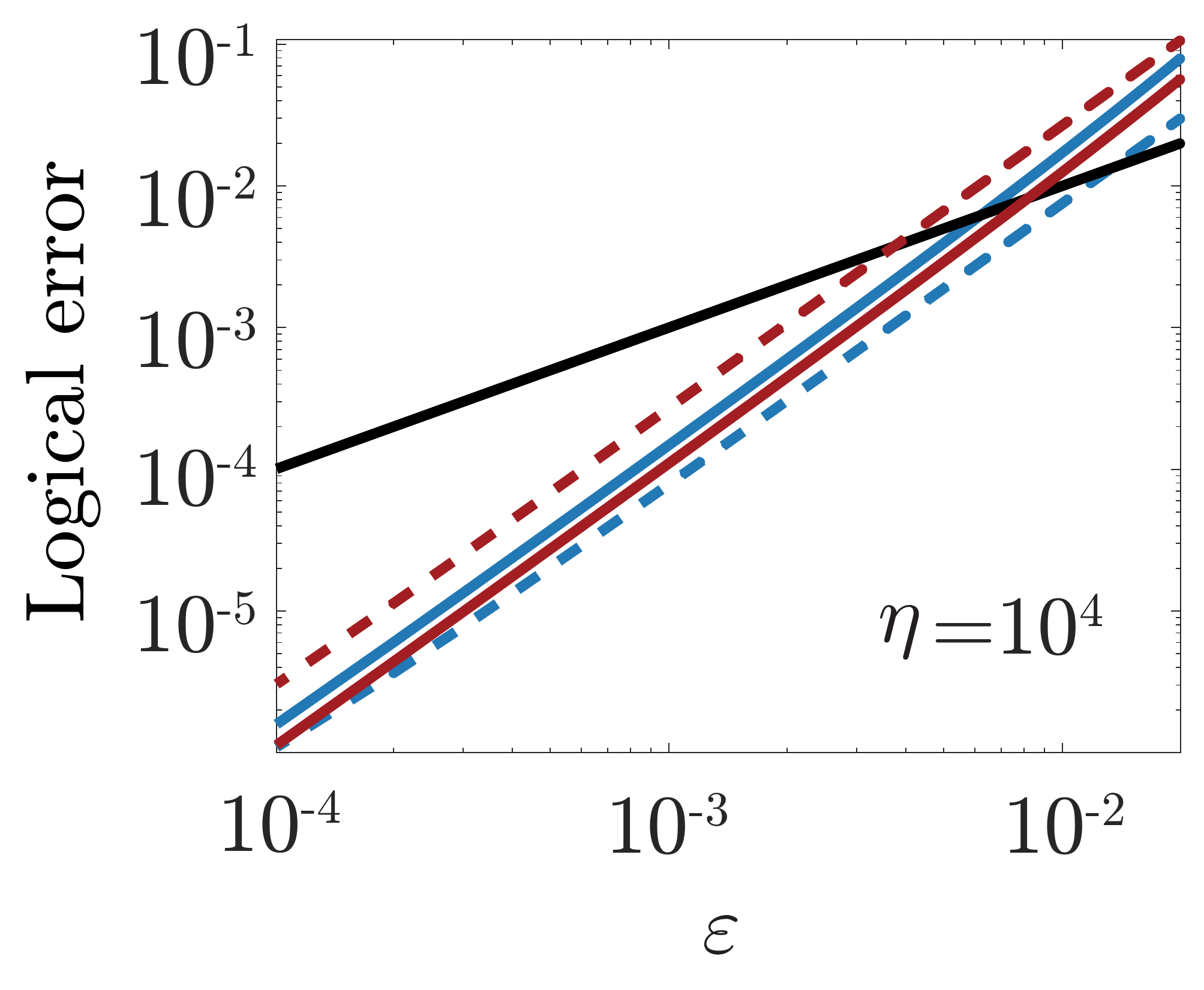}
 \caption{Logical dephasing and non-dephasing error rate for state preparation given in Eq.~\eqref{zT} (solid blue line) and Eq.~\eqref{xT} (dashed blue line) using $n=r=r_\mathrm{z,L}=3$. For comparion the logical dephasing and non-dephasing rates from ref.~\cite{webster2015reducing} are shown using solid and dashed red lines, respectively. The black line with slope=1 is shown for reference.}
 \label{magic2}
 \end{centering}
 \end{figure}

A logical dephasing error in block 2 could result from the failure of the error-correction block to detect any error. This happens if there is a phase-flip error on all the $n$ qubits in block 2. Each qubit $i$ in block 2 can undergo a phase-flip during any one of the $2r$ CX gates, the $ZZ(\theta)$ gate, preparation and when the $i^\mathrm{th}$ qubit in block 1 undergoes a phase flip. In block 1, a phase error on a qubit can occur during preparation, measurement, $ZZ(\theta)$ gate and any one of the $r_\mathrm{z,L}$ CZ gates. Additionally, a bit-flip error in the ancilla for $\overline{\mathcal{M}}_{\hat{Z}}$ measurement can also introduce correlated phase-flip errors in block 1, which then propagates to block 2. To estimate the upper bound, we assume that a non-dephasing fault in any of the $nr_\mathrm{z,L}$ CZ gates results in undetectable error in block 2.
Therefore, the probability of a $\sz{\mathrm{L}}$ error in block 2 due to failure of the error-correction gadget is
\begin{align}
\begin{split}
\varepsilon^{(1)}_\mathrm{z,L}&\leq((2r+2)\varepsilon+(r_\mathrm{z,L}+3)\varepsilon)^n+nr_\mathrm{z,L}\frac{\varepsilon}{\eta}\\
&\leq(2r\varepsilon+r_\mathrm{z,L}\varepsilon+5\varepsilon)^n+nr_\mathrm{z,L}\frac{\varepsilon}{\eta}.
\end{split}
\label{zL1}
\end{align}
In addition, correlated phase errors in block 1 and block 2 is another source of fault. Recall that if the error syndromes obtained from the measurement of the stabilizer generators do not agree with the $\overline{\mathcal{M}}_{\hat{Z}}$ measurement on block 1 then the state is rejected. Correlated phase errors in the $i^{\mathrm{th}}$ qubit in blocks 1 and 2, would however be missed, leading to an incorrect preparation of the magic state. Such correlated errors could occur during the $n$ $ZZ(\theta)$ gates (probability $\varepsilon_\mathrm{zz}$) or due to independent errors $\varepsilon^2$ occurring during preparation, measurement, $ZZ(\theta)$ gates or $r_\mathrm{z,L}$ CZ gates in the $\overline{\mathcal{M}}_{\hat{Z}}$ measurement. Therefore, the probability of an $\varepsilon^{(2)}_\mathrm{z,L}$ error due to correlated phase errors is
\begin{align}
\varepsilon^{(2)}_\mathrm{z,L}&\leq n\varepsilon_\mathrm{zz}+n(r_\mathrm{z,L}\varepsilon+3\varepsilon)(2r\varepsilon+2\varepsilon).
\label{zL2}
\end{align}
A final source of fault is the correlated errors in block 2 and stabilizer measurements themselves. 
If a qubit in block 2 undergoes a phase-flip and if the stabilizers are measured correctly, then a comparison between block 1 and 2 will reveal this error and the state can be discarded. 
However, if the two stabilizer generators involving that qubit are measured incorrectly, then the error in block 2 will be missed. 
An upper bound on the probability for this fault path is
\begin{align}
\varepsilon^{(3)}_\mathrm{z,L}\leq   n(2r\varepsilon+r_\mathrm{z,L}\varepsilon+5\varepsilon)\left(\binom{r}{\frac{r+1}{2}}(4\varepsilon)^{\frac{r+1}{2}}\right)^2.
\label{zL3}
\end{align}
Finally, the total probability of a $\varepsilon_\mathrm{z,L}$ error is obtained by combining Eq.~\eqref{zL1}, Eq.~\eqref{zL2} and Eq.~\eqref{zL3},
\begin{align}
\begin{split}
\varepsilon_{\mathrm{z,L}}\leq &\ nr_\mathrm{z,L}\frac{\varepsilon}{\eta}+ (2r\varepsilon+r_\mathrm{z,L}\varepsilon+5\varepsilon)^n+n\varepsilon_\mathrm{zz}\\
&\ +n(r_\mathrm{z,L}\varepsilon+3\varepsilon)(2r\varepsilon+2\varepsilon)\\
&\ +n(2r\varepsilon+r_\mathrm{z,L}\varepsilon+5\varepsilon)\left(\binom{r}{\frac{r+1}{2}}(4\varepsilon)^{\frac{r+1}{2}}\right)^2.
\end{split}
\label{zT}
\end{align}
We can now compare the expressions for the logical error rates in Eq.~\eqref{xT} and Eq.~\eqref{zT} to those in~\cite{webster2015reducing}. 
Recall that the only difference in the preparation schemes described here is the error correction gadget based on stabilizer measurements using CX gates. 
This does not affect the leading order sourced of error, which is correlated phase-flips in block 1 and 2 ($\propto \varepsilon_\mathrm{zz},\varepsilon^2$). 
The main difference is due to the contribution of correlated errors in block 2 and syndrome measurements (Eq.~\eqref{zL3}) to a logical $\sz{\mathrm{L}}$ error. 
In contrast, faults due to measurement errors in the error-correction gadget used in Ref.~\cite{webster2015reducing} contribute to a logical $\sxx{\mathrm{L}}$ error and are of higher order (see Eq.~(17) in~\cite{webster2015reducing}). Figure~\ref{magic2} shows the logical error rates computed here and that from Ref.~\cite{webster2015reducing} for $n=r=r_\mathrm{z,L}=3$. As expected, the probability of non-dephasing error with the CX based error-correction is slightly lower than in Ref.~\cite{webster2015reducing}, while the probability of dephasing errors is similar.

\subsection{Fault tolerance with a measurement code}~\label{subsec_optimal}
As we discussed in section~\ref{sec_thresh}.\ref{subsec_rep} the naive way to decode by measuring $(n-1)$ stabilizer generators is suboptimal. 
We will now discuss how we can improve decoding by using what we refer to as, a measurement code. 
To construct a measurement code, we desire that our syndrome measurement procedure measures a total of $s$ elements of the stabilizer group (not necessarily the specified generators) by coupling to ancillas and that it can correct any $t =(d-1)/2 $ phase-flip errors on the $n$ qubits. 
That is, we wish to have a classical code with parameters $[n+s,n,d]$. 
However, not every classical code with those parameters is admissible, because the classical parity checks must still be compatible with the stabilizers of the original quantum code, in this case the repetition code. 
In particular, each parity check in the measurement code must have even weight when restricted to the data qubits so that it commutes with the logical $\sz{\mathrm{L}}$ operator of the quantum phase-flip code. 
In fact, consistency with the stabilizer group of the base quantum code is the only constraint on a measurement code. 

The general form of a measurement code can be specified by the parity check matrix $H_M$. 
This in turn is specified as a function of the (generally redundant) parity checks $H_Z$ of the quantum repetition code and an additional set of $s$ ancilla bits that label the measurements. 
Given $H_Z$, the parity check matrix of the measurement code is the block matrix
\begin{align}
    H_M = \bigl( H_Z\ I_s \bigr)
\end{align}
where $I_s$ is the $s\times s$ identity matrix. 
Since there are $s$ ancilla bits for readout $H_M$ is an $s \times (n+s)$ matrix. 
The fact that the rows of $H_Z$ come from the stabilizers of a quantum repetition code is captured by the constraint they must all have even weight. 
The rows are clearly linearly independent, so the associated code has parameters $[n+s,n,d]$ for some $d \le n$. 
The distance is never greater than $n$ since a string of $\szz$ operators on the data qubits, corresponding to $1$'s on exactly the first $n$ bits, is always in the kernel of $H_M$. 

The measurement of the $j$th parity check in the measurement code can be done by a standard choice of circuit. 
We simply apply a CX gate to qubit $i$ if there is a $1$ in column $i$, and target the ancilla labeled in column $n+j$. 
Note that by construction there is always a $1$ in position $(j,n+j)$ of $H_M$. 
The effective error rate of this bare-ancilla measurement gadget will depend on the number of CX gates used, and hence on the weight of the stabilizer being measured. 
Therefore, all other things (such as code distance) being equal, lower weight rows are preferred when designing a measurement code. 

The two examples we consider here are generated from the following choices for $H_Z$, displayed here in transpose to save space:
\begin{align}
\label{eq:HZmats}
\arraycolsep=4pt
\def\arraystretch{0.5}
H_Z^T &= \left(
\begin{array}{ccc}
1 & 1 & 0\\
1 & 0 & 1\\
0 & 1 & 1
\end{array}
\right)\quad ,\\
H_Z^T &= 
\left(
\begin{array}{ccccccccc}
 1 & 0 & 0 & 0 & 1 & 1 & 0 & 0 & 1 \\
 1 & 1 & 0 & 0 & 0 & 0 & 1 & 1 & 0 \\
 0 & 1 & 1 & 0 & 0 & 1 & 0 & 1 & 0 \\
 0 & 0 & 1 & 1 & 0 & 0 & 0 & 1 & 1 \\
 0 & 0 & 0 & 1 & 1 & 0 & 1 & 1 & 0 
\end{array}
\right).
\label{eq:HZmats2}
\end{align}
These codes were chosen to saturate the distance bound, so $d=n$ for each code (so $d=3$ and $d=5$ respectively). 
These were found by guesswork, and no attempt at finding optimal measurement codes was made, although these are the best of the few that were tested. 
To contrast our choices with the choice associated to repeating the measurements of the standard generators $r$ times for $n=r=3$, the measurement code is specified by
\begin{align}
\arraycolsep=4pt
\def\arraystretch{0.5}
H_Z^T = \left(
\begin{array}{cccccc}
1 & 1 & 1 & 0 & 0 & 0\\
1 & 1 & 1 & 1 & 1 & 1\\
0 & 0 & 0 & 1 & 1 & 1
\end{array}
\right).
\end{align}
Both this choice and the $n=3$ choice in Eq.~(\ref{eq:HZmats}) have distance $d=3$ as measurement codes. 
However, our choice corresponds to a $[6,3,3]$ measurement code whereas the naive repeated generator method yields a $[12,3,3]$ measurement code. 
In general, the naive scheme yields an $[n + (n-1)r,n,d(n,r)]$ code, and for smaller $r$ the distance will not yet saturate to $n$. 
For the case $n=5$ case, we need $r=2$ before the measurement code has distance $3$, and $r=4$ before the distance saturates at $d=5$. 
Thus, the naive scheme yields either an $[13,5,3]$ code or a $[21,5,5]$ code, which are inferior in either distance or rate respectively to the $[14,5,5]$ code that results from the choice in Eq.~(\ref{eq:HZmats2}).

These examples also illustrate a counterintuitive feature of measurement codes. 
Consider again the naive repeated generator method with $n=5$ and $r=2$ or $4$. 
If the decoder works by first decoding the syndrome bits individually, then the data are only protected against at most $(r-1)/2 = 0$ or $1$ arbitrary errors respectively. 
However, a decoder that uses the structure of the associated measurement code can correct $1$ or $2$ arbitrary data errors with these respective parameters, which then reduces the leading order behavior of the code failure probability. 

Both of the above codes in Eq.~\eqref{eq:HZmats} and Eq.~\eqref{eq:HZmats2} are small enough that the exact probability of a decoding failure can be computed via an exhaustive lookup table. 
To demonstrate the advantage of the measurement code over naive encoding and decoding, we estimate the probability of a logical error in the $\overline{\mathrm{CX}}$-gadget using the measurement code in Eq.~\eqref{eq:HZmats2} for $n=5$. 
The corresponding threshold is $\sim 6\times 10^{-3}$. 
On the other hand, to reach a similar threshold using the naive decoder requires $n=11$, $r=5$. 
Clearly, the optimal decoder requires fewer resources than the naive decoder.
In general this optimal (maximum likelihood) decoder is infeasible to implement because it requires exponential resources in $n$ and $s$ to compute, so substantially larger codes will need decoding heuristics such as message passing algorithms to approach peak decoding performance. 
The decoder declares failure whenever the data error is not guessed exactly right, even though this is not strictly speaking necessary.
When repeated rounds of error correction occur, it is sufficient to define success as reducing the weight of any correctable error. 
This more relaxed definition is harder to analyze, however, so our stricter definition of failure is used in all of the threshold calculations. 

\begin{figure*}[ht]
\begin{centering}
 \includegraphics[width=1.5\columnwidth]{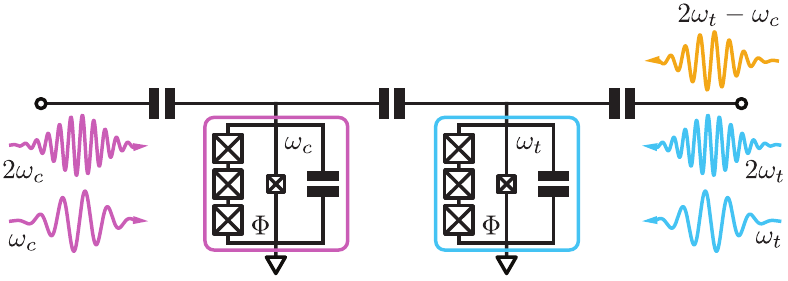}
 \caption{Schematic for implementation of the bias-preserving CX gate with superconducting circuits. Here the Kerr-nonlinear oscillators (of frequencies $\omega_\mathrm{t}$ and $\omega_\mathrm{c}$) are implemented with superconducting nonlinear asymmetric inductive elements or SNAILs~\cite{frattini20173,frattini2018optimizing}. A SNAIL can be biased with an external magnetic field so that it has both three- and four-wave mixing capabilities. It can therefore be used to implement the two-photon driven Kerr nonlinear oscillator and realize a cat-qubit with biased noise channel~\cite{GrimmStabilizing}. The Hamiltonian in Eq.~\eqref{CX_h} can  be simplified as $\hat{H}=-K\hta^{\dag 2}_\mathrm{c}\hta^{2}_\mathrm{c}-K\hta^{\dag 2}_\mathrm{t}\hta^{2}_\mathrm{t}+K\beta^2( \hta^{\dag 2}_\mathrm{c}+\mathrm{h.c.})+K\alpha^2\cos(\phi(t))(e^{i\phi(t)}\hta^{\dag 2}_\mathrm{t}+\mathrm{h.c.})-(Ki\alpha^2\sin(\phi(t))/\beta)(i\hta^{\dag 2}_\mathrm{t}\hat{a}_\mathrm{c}+\mathrm{h.c.})+(K\alpha^4/2\beta)\sin(2\phi(t))(i\hta^\dag_\mathrm{c}+\mathrm{h.c.})-(K\alpha^4\sin^2(\phi(t))/\beta^2)\hta^\dag_\mathrm{c}\hta^\dag_\mathrm{c}-\dot{\phi}(t)\hta^\dag_\mathrm{t}\hta_\mathrm{t}/2+(\dot{\phi}(t)/4\beta))\hta^\dag_\mathrm{t}\hta_\mathrm{t}(\hta^\dag_\mathrm{c}+\mathrm{h.c.})$. By expressing the Hamiltonian in this form, the drives required to realize the Hamiltonian become immediately clear. Firstly, a drive to the control cavity (fixed amplitude and time-dependent phase) centered at $2\omega_\mathrm{c}$ is required for the two-photon term driving the control cavity via three-wave mixing. Next, a drive to the target cavity with time-dependent amplitude at $2\omega_\mathrm{t}$ results in the two-photon term driving the target cavity via three-wave mixing.  An additional drive $2\omega_\mathrm{t}-\omega_\mathrm{c}$
(time-dependent amplitude and phase) is applied to the target cavity to realize the coupling terms $\propto \hta^2_\mathrm{t}\hta_\mathrm{c}$ in Eq.~\eqref{CX_h}. A drive applied directly to the control cavity centered at $\omega_\mathrm{c}$ with time-dependent phase and amplitude realizes the single-photon drive to the control cavity.  
A final drive to the target cavity at $\omega_\mathrm{c}$ with time-dependent amplitude and phase realizes the last term in the Hamiltonian~\cite{touzard2019gated}. }
 \label{ckt}
 \end{centering}
 \end{figure*}

\section{Discussion}
In this paper we have presented a driven cat-qubit with highly biased noise channel and shown how to perform a CX gate which preserves the error bias.  A bias-preserving CX gate with strictly two-dimensional systems is impossible~\cite{aliferis2008fault,guillaud2019repetition}. We are able to circumvent this no-go conjecture by exploiting the phase space topology of the underlying continuous variable system. 

The physical realization of the CX gate requires a three-wave mixing between the oscillators. 
The natural coupling between two oscillators is, however, beam-splitter type. Fortunately, the oscillators are themselves fourth-order, Kerr nonlinear. Thus, the required three-wave mixing can be generated by parametrically driving the target oscillator at a frequency $\omega_\mathrm{d}$ such that $\omega_\mathrm{d}=2\omega_\mathrm{t}-\omega_\mathrm{c}$. Here $\omega_\mathrm{t}$ and $\omega_\mathrm{c}$ are the frequencies of the target and control oscillators respectively. When this condition is satisfied, the fourth order nonlinearity converts a photon in the drive and a photon in the control to two photons in the target. Thereby, an effective three-wave mixing is realized between the control and target. Importantly, the Kerr nonlinearity of the oscillators themselves is sufficient to realize the CX interaction Hamiltonian and no additional coupling elements are necessary. Moreover, because of the parametric nature, the coupling is controllable. A possible realization of the CX gate Hamiltonian in superconducting circuits is shown in Fig~\ref{ckt}. It is feasible to extend the scheme for the CX gate to implement a bias-preserving CCX gate between three cat qubits. A naive circuit would however require a controllable four-wave mixing between the oscillators which is typically much weaker. Remarkably, as described in Appendix~\ref{app}.\ref{cCX}, it is  possible to implement a bias preserving CCX gate by using only three wave-mixing and four cat-qubits. To summarize, the bias preserving set of unitaries discussed in this paper, which are also physically implementable with three-wave mixing (or less) are $\{\mathrm{CX},\mathrm{CCX},ZZ(\theta),Z(\theta),\mathrm{CCZ}\}$. These can be supplemented with state preparations $\mathcal{P}_{\ket{\pm}}$ and measurements $\mathcal{M}_{\hat{X},\hat{Z}}$ for universal fault-tolerant quantum computation.  
 
Furthermore, by adapting the scheme for concatenated error correction in~\cite{aliferis2008fault} we have demonstrated that having bias-preserving CX gates leads to significant improvements in fault-tolerance thresholds and overheads. At the level of repetition code, the estimated bound for fault-tolerance thresholds with naive decoding and experimentally reasonable biases of $\sim 10^3-10^4$ is $\sim 0.55\%=0.75\%$.
Consequently, high quality oscillators will still be required so that the phase-flip error remain small enough. One way to improve the threshold is by using better decoding techniques as discussed in section~\ref{sec_thresh}.\ref{subsec_optimal}. More importantly, the approach based on concatenating a repetition code to another CSS code is not necessary or ideal. A more efficient technique would be to directly implement a code tailored to asymmetric noise such as the surface code~\cite{tuckett2018ultrahigh,tuckett2018tailoring} or cyclic code~\cite{robertson2017tailored} with the cat-qubit. An analysis of these codes tailored to the cat qubits will be carried out in future work. 

\begin{acknowledgements}
We thank Arne L. Grimsmo, Andrew Darmawan and Mazyar Mirrahimi for discussions. 
This work was supported by the National Science Foundation grant number DMR-1609326, by the Canada First Research Excellence Fund and NSERC, Fonds de Recherche du Québec-Nature et technologies,
by the US Army Research Office grant numbers W911NF-18-1-0212, W911NF-14-1-0098, and W911NF-14-1-0103, and by the Australian Research Council Centre of Excellence for Engineered Quantum Systems grant number CE170100009. 
STF thanks the Yale Quantum Institute for its hospitality while this research was carried out. 
\end{acknowledgements}

\section{Appendix}\label{app}
\subsection{Error channel from simulations}
\label{app_ch}
Here we describe how the error channel in the main text is extracted from master equation simulations. The dimension of the system of $s$ cat qubits is $d=2^s$ and the elements of the Pauli transfer matrix $R$ are 
\be
R_{ij}=\frac{1}{d}\mathrm{Tr}[\hat{P}_i\mathcal{E}(\hat{P}_j)].
\ee
In the above expression $\mathcal{E}(\cdot)$ is the error channel and $\hat{P}_i$ are the $d^2$ Pauli operators. The Pauli transfer matrix at time $t$ is extracted by simulating the master equation, using the software package QuTiP~\cite{johansson2012qutip}, 
with the Pauli operators as initial state at $t=0$. Once the $d^2\times d^2$ elements of the Pauli transfer matrix are obtained, the above equation is inverted to obtain the error channel. 


\subsection{White frequency noise }\label{sec_fr}

Coupling with a white dephasing channel leads to the Lindbladian master equation,
\be
\dot{\hat{\rho}}=-i[\hat{H}_0(\phi),\hat{\rho}]+\kappa_\phi \mathcal{D}[\hta^\dag a]\hat{\rho}.
\ee
In the above expression, $\kappa_\phi$ is the dephasing rate. 
Following the approach used in section~\ref{sec_err}.\ref{sec_th}, we describe the dynamics of the oscillator in the quantum-jump formalism. 
In this approach evolution under a non-Hermitian Hamiltonian $\hat{{H}}=\hat{H}_0(\phi)-i\kappa_\phi\hta^{\dag}\hta\hta^\dag\hta/2$ is interrupted by stochastic quantum jumps corresponding to the operators $\hta^\dag\hta$. Again, for $\kappa_\phi\ll |\Delta\omega_\mathrm{gap}|$, the leading order effect of the non-Hermitian terms in $\hat{{H}}$ is to broaden the linewidths of the eigenstates of the oscillator. 
Action of $\hta^\dag\hta$ on a state in the cat subspace causes leakage to $\mathcal{C}_\perp$. 
In the limit of large $\alpha$, $\hta^\dag\hta\catpm\sim\alpha^2\catpm+\alpha\ket{\psi^\pm_{\mathrm{e},1}}$ (note that the parity does not change). 
Therefore, a dephasing event excites the first excited subspace at rate $\sim\kappa_\phi \alpha^2$. 
As we saw in the last section, if a two-photon dissipation channel is introduced then the leakage will be corrected at a rate $4\kappa_\mathrm{2ph}\alpha^2$. 
As a result, leakage errors are suppressed by $\sim \kappa_\phi /4\kappa_\mathrm{2ph}$.  

\begin{figure}
\begin{centering}
 \includegraphics[width=\columnwidth]{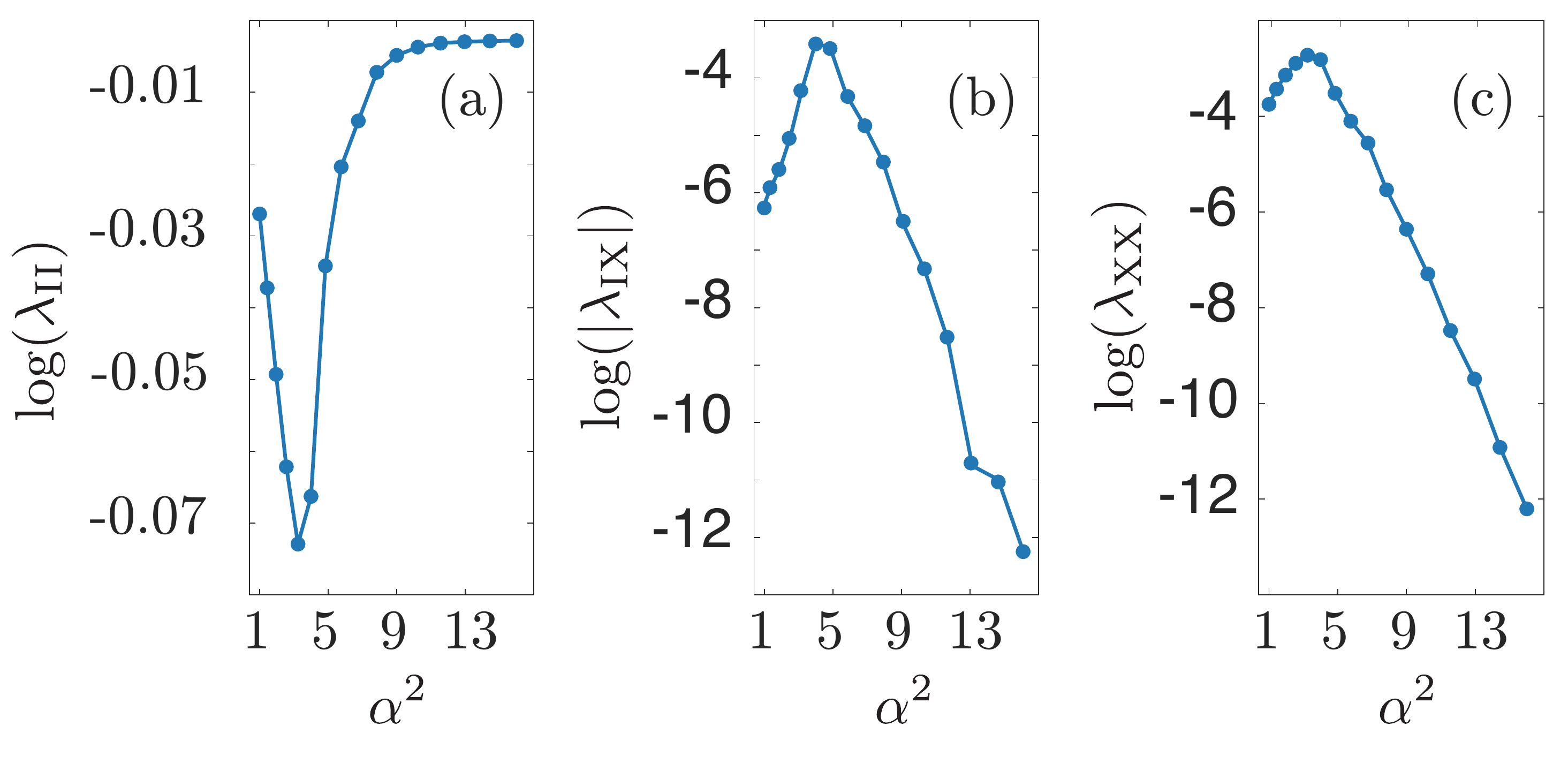}
 \caption{Natural logarithm of the coefficients of the error channel Eq.~\eqref{ch_dephase} of an idle cat qubit in the presence of white thermal noise and two-photon dissipation. As expected the amount of non-phase errors decreases exponentially with $\alpha^2$. The parameters for the simulations are $\kappa_\phi=K/1000$ and $\kappa_\mathrm{2ph}=K/10$ . The coefficients are evaluated at $t=50/K$. }
 \label{dephase_i}
 \end{centering}
 \end{figure}

\begin{figure}
\begin{centering}
 \includegraphics[width=.8\columnwidth]{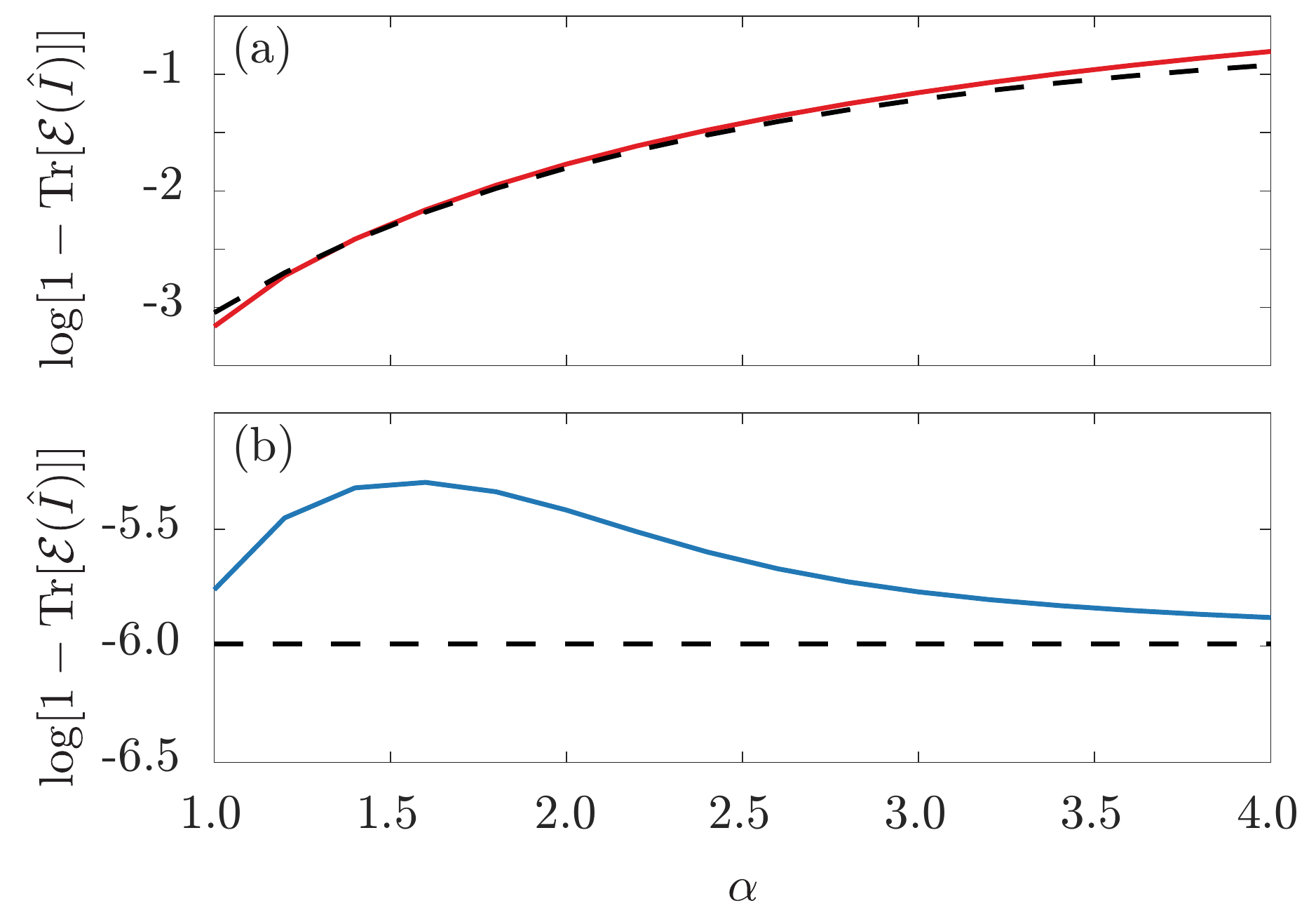}
 \caption{Natural logarithm of the amount of leakage, $\log[1-\mathrm{Tr}[\mathcal{E}(\hat{I})]]$ in the presence of white dephasing noise without two-photon dissipation (red solid line in (a)) and with it (blue solid line in (b)). As expected, the two-photon dissipation autonomously corrects for leakage. The dashed black line shows the leakage predicted by the theoretical expressions for the rates of out-of-subspace excitations ($\sim\kappa_\phi\alpha^2$) and correction due to two-photon loss $\sim4\kappa_\mathrm{2ph}\alpha^2$. These expression are only approximations which become more and more exact as $\alpha$ increases. 
 }
 \label{dephase_oos}
 \end{centering}
 \end{figure}

Observe that both dephasing and two-photon loss events cause transitions within the same parity subspace. Therefore, two-photon loss immediately after a dephasing event does not result in phase-flips. As a result, no phase-flips are introduced but bit-flips can arise because of the energy difference between the states $\ket{\psi^\pm_{\mathrm{e},1}}$. However, the energy difference and consequently the probability of a bit-flip, decreases exponentially with $\alpha^2$ and the noise bias is preserved. The analysis above is confirmed by numerically simulating the master equation,
\begin{align}
\dot{\hat{\rho}}&=-i[\hat{H}_0(\phi_0),\hat{\rho}]+\kappa_\phi \mathcal{D}[\hta^\dag\hta]\hat{\rho}+\kappa_\mathrm{2ph} \mathcal{D}[\hta^2]\hat{\rho}.
\end{align}
The dissipation rates $\kappa_\mathrm{2ph}$ and  $\kappa_\phi$ are fixed, while $\alpha$ is varied by changing $\hat{H}_0(\phi)$ as in Eq.~\eqref{H0_2ph}.
The  error channel is of the form
\begin{align}
\mathcal{E}(\hat{\rho})&=\lambda_\mathrm{II}\hat{I}\hat{\rho}\hat{I}+\lambda_{\mathrm{IX}}\hat{I}\hat{\rho}\hat{X}+\lambda^*_\mathrm{IX}\hat{X}\hat{\rho}\hat{I}+\lambda_\mathrm{XX}\hat{X}\hat{\rho}\hat{X}.
\label{ch_dephase}
\end{align}
Figure~\ref{dephase_i} shows the numerically evaluated coefficients $\lambda_\mathrm{II}$, $\lambda_{\mathrm{IX}}$, and $\lambda_{\mathrm{XX}}$ at time $t=50/K$ for $\kappa_\phi=K/1000$ and $\kappa_\mathrm{2ph}=K/10$. All other coefficients cause a change in parity and hence are zero. As expected for large $\alpha$, the bit-flips and consequently $|\lambda_\mathrm{IX}|$ and $\lambda_\mathrm{XX}$ decrease exponentially with $\alpha^2$. Figure~\ref{dephase_oos} shows the amount of leakage $1-\mathrm{Tr}[\mathcal{E}(\hat{I})]$ with and without the two-photon dissipation. When $\alpha\rightarrow 0$, the cat states $\ket{\mathcal{C}^\pm_0}$ are the vacuum and single-photon Fock state respectively. In this case, frequency jumps do not cause any leakage because the vacuum and single-photon Fock states are eigenstates of $\hta^\dag\hta$. Therefore, in the limit of small $\alpha$, the leakage and consequently $|\lambda_{\mathrm{IX}}|$ and $\lambda_{\mathrm{XX}}$ increase as $\alpha$ increases. Once $\alpha$ becomes sufficiently large, the exponential suppression of bit-flips with increase in $\alpha$ begins. The behavior in the limits of small and large $\alpha$ explains the trend in $|\lambda_{\mathrm{IX}}|$ and $\lambda_{\mathrm{XX}}$ (and hence also $\lambda_\mathrm{II}$) shown in Fig.~~\ref{dephase_i}. Figure~\ref{dephase_oos} shows the numerically and theoretically estimated leakage. The theoretical estimation uses the expression $\kappa_\phi\alpha^2$ for the rate of leakage and $4\kappa_\mathrm{2ph}\alpha^2$ for the rate of autonomous correction. We find that the theoretically predicted leakage is a good approximation for the numerically simulated leakage.  


\subsection{Threshold for imperfect rotation}\label{app_thresh}
In this section, we will first provide a qualitative estimate for the threshold of rotation error $\Delta_\mathrm{th}\sim0.5$ in the CX gate (section~\ref{sec-CX}.\ref{subsec-CX_overview}).
We then provide a numerical estimate for the threshold which is in good agreement with the qualitative result. 

To begin with, consider a single cat-qubit in a two-photon driven nonlinear oscillator. We work in the limit of large $\alpha$, so that the computational states $\ket{0,1}$ are well approximated by the coherent states $\ket{\pm\alpha}$. Now suppose that the oscillator is initialized in a coherent state $e^{i\Delta\hta^\dag\hta}\ket{\alpha}=\ket{e^{i\Delta}\alpha}$ or $e^{i\Delta\hta^\dag\hta}\ket{-\alpha}=\ket{-e^{i\Delta}\alpha}$. These are not computational states, but as we have seen before, addition of two-photon dissipation will bring them back into the computational basis. Our task is to find the threshold $\Delta_\mathrm{th}$ so that for $\Delta<\Delta_\mathrm{th}$ the initial state $e^{i\Delta\hta^\dag\hta}\ket{\alpha}$ evolves preferably to $\ket{\alpha}$ instead of $\ket{-\alpha}$, while the state $e^{i\Delta\hta^\dag\hta}\ket{-\alpha}$ evolves (exponentially) preferably to $\ket{-\alpha}$ instead of $\ket{\alpha}$. Consequently, for $\Delta<\Delta_\mathrm{th}$ phase-flips will be exponentially suppressed. It is reasonable to assume that the threshold obtained in this manner is identical to that for the rotation error in the CX gate. 

\begin{figure}
\begin{centering}
 \includegraphics[width=\columnwidth]{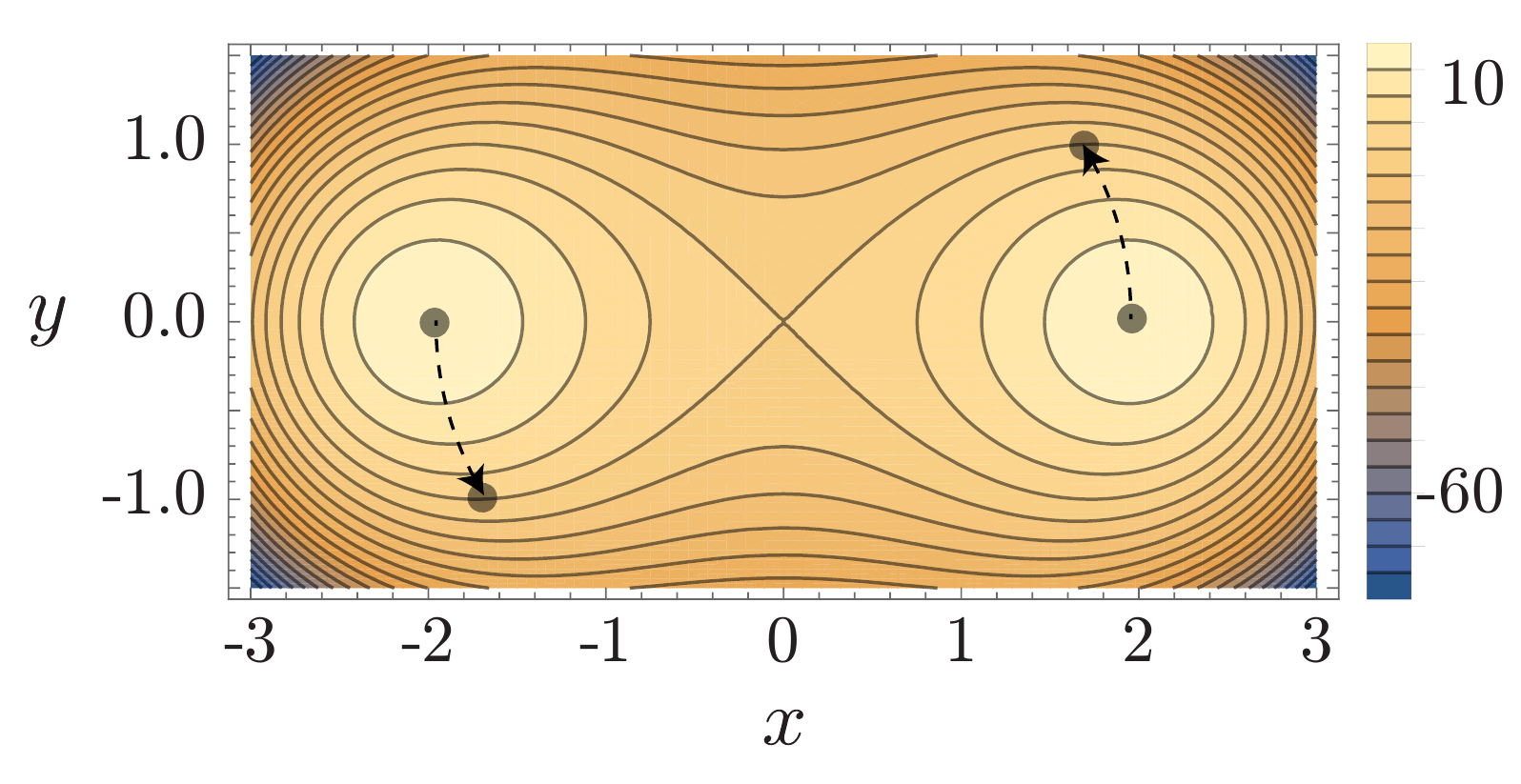}
 \caption{The meta potential of the Hamiltonian on the two-photon driven nonlinear oscillator, $f(x,y)=-K(x^2+y^2)^2+2K\alpha^2(x^2-y^2)$. The dark lines are the lines of equi-potential $f(x,y)=$ const. or the trajectories along which a classical particle moves. Since the nonlinearity if negative, the motion of the particle is clockwise. The trajectories for which $f(x,y)<0$ are closed around $\pm \alpha$, while they are delocalized for $f(x,y)\geq 0$. At equilibrium the particle is at one of the two semiclassical stable points $(x,y)=(\pm\alpha,0)$, shown in grey circles.
 A rotation error with $\Delta=\pi/3$ places the particle on the first delocalized path $f(x,y)=0$.}
 \label{meta}
 \end{centering}
 \end{figure}

Consider the master equation of such an oscillator in the presence of two-photon dissipation,
\be
\dot{\hat{\rho}}=-i[\hat{H}_0(\phi_0),\hat{\rho}]+\kappa_{2\mathrm{ph}}\mathcal{D}[\hta^{ 2}]\hat{\rho}.
\ee
Using $\langle\hta\rangle=x+i y$, in the semiclassical approximation
\begin{align}
\dot{x}&=2Ky[\alpha^2+(x^2+y^2)]-\kappa_2x[\alpha^2-(x^2+y^2)]\nonumber\\
\dot{y}&=2Kx[\alpha^2-(x^2+y^2)]+\kappa_2y[\alpha^2+(x^2+y^2)].
\end{align}
The equations of motion have the form of a particle moving on a metapotential or phase-space potential. For $K\gg\kappa_2$, the metapotential is $f(x,y)=-K(x^2+y^2)^2+2K\alpha^2(x^2-y^2)$, which is an inverted double well in the phase-space with peaks at $\pm\alpha$~\cite{dykman2012fluctuating,puri2017quantum}. Figure~\ref{meta} shows the contour plot of the metapotential. In semiclassical approximation, these contours are the trajectories along which a particle moves (clockwise because the nonlinearity is attractive.) The trajectories corresponding to $f(x,y)<0$ are closed around $\pm\alpha$ and for $f(x,y)\geq 0$ they are delocalized. The state of the oscillator $\ket{e^{i\Delta}\alpha}$ or $\ket{-e^{i\Delta}\alpha}$ is equivalent to the particle being at $(x_0,y_0)=(\alpha\cos(\Delta),\alpha\sin(\Delta))$ or $(x_0,y_0)=(-\alpha\cos(\Delta),-\alpha\sin(\Delta))$ in the semiclassical approximation. If $\Delta$ is small, then $(x_0,y_0)$ lies on one of the closed trajectories around $\pm\alpha$. Whereas, for large $\Delta$, $(x_0,y_0)$ lies on one of the delocalized trajectories. 
If the semiclassical trajectory is closed, then under the influence of quantum fluctuations and two-photon jumps, the initial states $\ket{\pm e^{i\Delta}\alpha}$ evolve preferably to $\ket{\pm\alpha}$ respectively. However, if the phase-space path is delocalized, then the initial states $\ket{\pm e^{i\Delta}\alpha}$ will evolve equally likely to either $\ket{\alpha}$ or $\ket{-\alpha}$. The threshold $\Delta_\mathrm{th}$ is then the rotation angle such that $\alpha e^{i\Delta}$ lies on the first of delocalized trajectory $f(x,y)=0$. Thus, solving the equation $f(\alpha\cos(\Delta),\alpha\sin(\Delta))=0$ gives the threshold $\Delta_\mathrm{th}=\pi/6\sim0.52$. This semiclassical estimate of the threshold is valid in the regime $K\gg\kappa_2$. In the opposite regime $\kappa_2\gg K$ the threshold is $\sim \pi/2$~\cite{leghtas2015confining}. In the intermediate regime where $\kappa_2$ and $K$ are of similar magnitudes, we expect the threshold to be between $\pi/6$ and $\pi/2$.

\begin{figure}
\begin{centering}
 \includegraphics[width=.8\columnwidth]{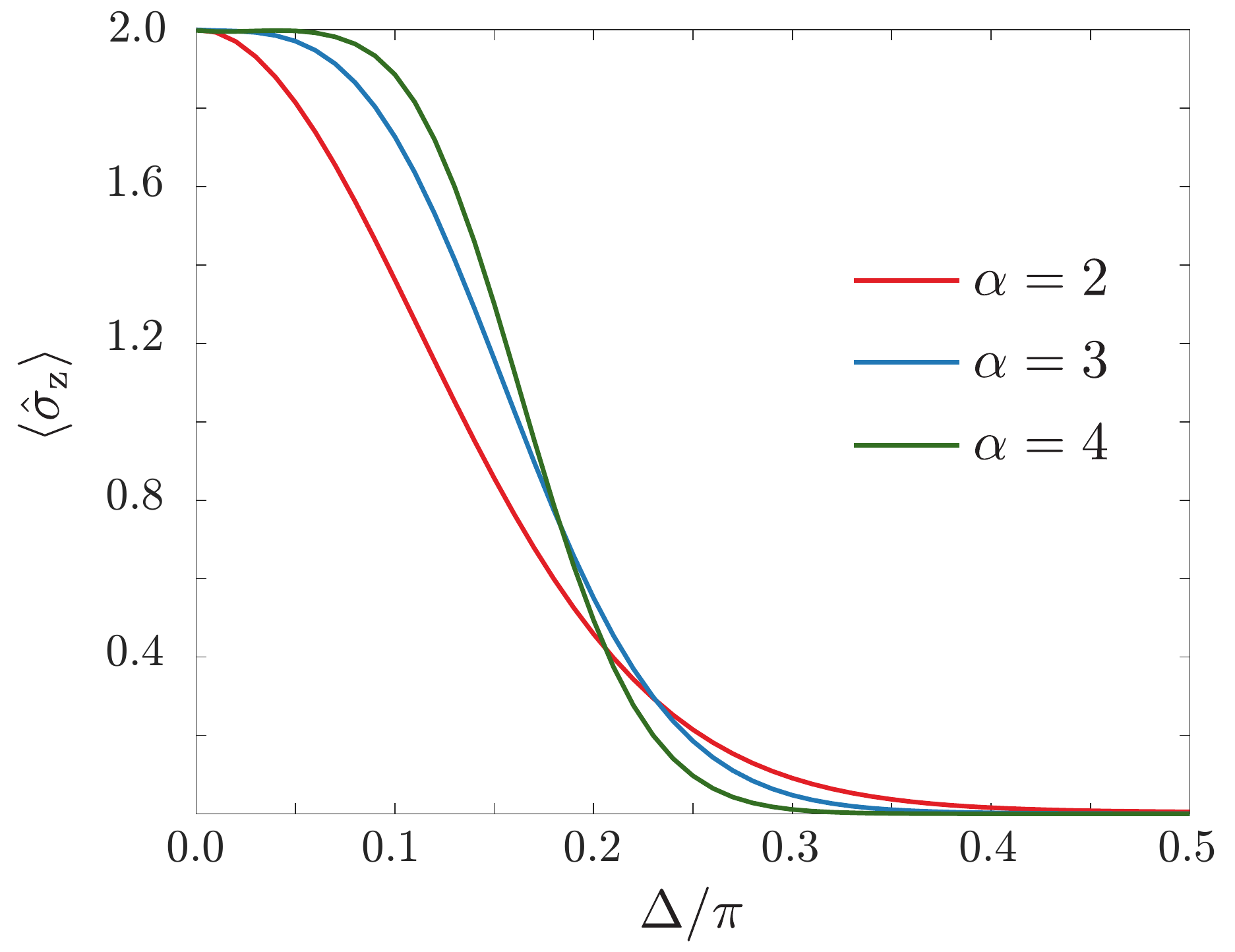}
 \caption{$\langle\szz\rangle$ as a function of the rotation error $\Delta$ for different $\alpha$ at $t=2/\kappa_2$ for $\kappa_2=K/100$. The initial state of the oscillator is rotated by $\Delta$, that is $\szz\rightarrow e^{i\hta^\dag\hta}\szz e^{-i\hta^\dag\hta}$. Two photon dissipation is added to refocus the rotated state back into the cat basis. Non-dephasing errors can be introduced during this correction process which would decrease $\langle\szz\rangle$. However, as expected from analytics, the numerical simulations show a thresholding behavior. That is, the $\szz$ decreases less rapidly for larger $\alpha$ for small $\Delta$. The plot suggests a thresholding behaviour around $\Delta\sim 0.5$.}
 \label{thresh}
 \end{centering}
 \end{figure}

To justify this simple analysis of we numerically simulate the master equation $\dot{\hat{\rho}}=-i[\hat{H}(\phi_0),\hat{\rho}]+\kappa_2\mathcal{D}[\hta^2]\hat{\rho}$ with  $\hat{H}(\phi_0)=-K\hta^{\dag 2}\hta^2+P(\hta^{\dag 2}e^{2i\phi_0}+\hta^2e^{-2i\phi_0})$, $P=\alpha^2\sqrt{K^2+\kappa_2^2/4}$ and $2\phi_0=\tan^{-1}(\kappa_2/2K)$. The input state at $t=0$ is $\exp(i\Delta\hta^\dag\hta)\szz\exp(-i\Delta\hta^\dag\hta)$, where $\szz$ is the Pauli operator $\szz=\ket{0}\bra{0}-\ket{1}\bra{1}=\catp\catmb+\catm\catpb$. From the simulations we estimate $\langle\hat{Z}\rangle$ at time $t=2/\kappa_2$, $\kappa_2=K/100$ for different $\alpha$ and $\Delta$. As shown in Figure~\ref{thresh}, the dependence of $\langle\hat{Z}\rangle$ as a function of $\Delta$ for different $\alpha$ suggests a thresholding behaviour around $\Delta\sim 0.5$. The figure shows that the numerical result does agree qualitatively with the analytic prediction. The large Hilbert space size limits the numerical simulations for larger $\alpha$. 

\subsection{Geometric and topological phases during the CX gate}\label{appen_phases}

As described in the main text, the rotation of cat states in phase space along a closed loop gives rise to both geometric and topological phase. The topological phase arises simply because the state $|\mathcal{C}^-_{\beta}\rangle$ is $2\pi$ periodic in the phase of $\beta$, while the state $|\mathcal{C}^+_{\beta}\rangle$ has the periodicity of $\pi$. The geometric or Berry phase~\cite{berry1984quantal} depends only on the path in the phase space. The difference in this phase for the two cat states is related to the mean particle number difference. Near $\beta=0$ the state $|\mathcal{C}^+_{\beta}\rangle$ becomes the vacuum and the state $|\mathcal{C}^-_{\beta}\rangle$ becomes the single-photon Fock state, so that the photon number difference is 1. However for large $\beta$, the mean photon numbers for the two cat states become exponentially close. 
In phase space where $x=(\beta+\beta^*)/2$ and $x=(\beta-\beta^*)/2$, the Berry phase for the states $\catpm$ can be defined using the corresponding Berry connection as~\cite{chaturvedi1987berry,girvin2019modern},
\begin{align}
    \Phi^\pm_\mathrm{g}=\int A^\pm_x dx+A^\pm_{y} dy
\end{align}
where,
\be
    A^\pm_x&=\frac{1}{2}\left(\frac{1\mp e^{-2(x^2 +y^2)}}{1\pm e^{-2(x^2 +y^2)}}\right)y,\nonumber\\
    A^\pm_{y}&=-\frac{1}{2}\left(\frac{1\mp e^{-2(x^2 +y^2)}}{1\pm e^{-2(x^2 +y^2)}}\right)x\nonumber\\
\ee
The corresponding Berry curvatures are,
\begin{align}
\Omega^\pm=&\frac{\partial A^\pm_{y}}{\partial x}-\frac{\partial A^\pm_x}{\partial y}\nonumber\\
=&-\frac{1\mp e^{-2(x^2+y^2)}}{1\pm e^{-2(x^2+y^2)}}\mp \frac{4(x^2+y^2)e^{-2(x^2+y^2)}}{(1\pm e^{-2(x^2+y^2)})^2}
\end{align}
The Berry curvature and connection can be thought of as a magnetic flux and an electromagnetic vector potential respectively.
The difference in the Berry curvature or flux $\Omega^+-\Omega^-$ is small near the origin and at large $x,y$ or $|\beta|^2$. The difference in Berry connection or vector potential is large near the origin and decreases exponentially away from it. Therefore, as long as the $|\beta|^2$ is large enough so that the closed path in phase space is not too close to the origin, then the geometric phase difference between the $\catp$ and $\catm$ will be exponentially small. In this case, topology is the only source of phase difference between the $\catp$ and $\catm$ state. Consequently, for the topological implementation of the CX gate, the size of the cat state must be large.

\subsection{CX gate in the presence of thermal noise}\label{appen_thm}
The error channel of the CX gate in the presence of thermal noise was given in section~\ref{sec-CX}~\ref{CX_num}. The channel is obtained from the transfer matrix, which itself is obtained in two steps. First, the master equation for the CX, $\dot{\hat{\rho}}=-i[\hat{H}_\mathrm{CX},\hat{\rho}]+\sum_{i}\mathrm{c,t}\kappa(1+n_\mathrm{th})\mathcal{D}[\hta_i]\hat{\rho}+\kappa n_\mathrm{th}\mathcal{D}[\hta^\dag_i]\hat{\rho}$, is simulated for time $T=10/K$ with $\phi(t)=\pi t/T$, $\alpha=\beta=2$ and with Pauli matrices as the input. Here, $i=\mathrm{c,t}$. Next, all the interactions between the control and target oscillators are removed and the Hamiltonian of the system is set to that two uncoupled oscillators $\hat{H}'=-K[\sum_i\hta^{\dag 2}_i\hta^2_i+\alpha^2(\hta^{\dag 2}_i+\hta^2_i)]$. Now the master equation with two-photon dissipation is simulated, $\dot{\hat{\rho}}=-i[\hat{H}',\hat{\rho}]+\sum_{i}\mathrm{c,t}\kappa(1+n_\mathrm{th})\mathcal{D}[\hta_i]\hat{\rho}+\kappa n_\mathrm{th}\mathcal{D}[\hta^\dag_i]\hat{\rho}+\kappa_2\mathcal{D}[\hta^2_i]\hat{\rho}$, for time $T'=2/\kappa_2$ and using the density matrix from the output of the CX simulation as the input. The transfer matrix from the second simulation is inverted to obtain the error channel. 

As shown by Eq.~\eqref{CX_et}, two-photon dissipation on the target oscillator $\mathcal{D}[\hta^2_\mathrm{t}]$ during the CX gate introduces additional phase-flip errors in the control oscillator. 
This implies that, although the two-photon dissipation corrects leakage, it will also reduce the gate fidelity. It is possible to overcome this problem by adding time-dependent, correlated dissipation between the control and target oscillators. However, the numerical simulations become significantly harder. To avoid this, we use the two-step process discussed above.

\subsection{Impossibility of universal set of bias-preserving gates at the level of physical qubits}
\label{appen_univ}
Consider physical qubits with a biased noise channel such that dephasing errors are dominant over non-dephasing fault. Suppose that the set of gates $(U_1,U_2,...,U_k)$ are universal and preserve the bias. This set of operations can be single- or multi-qubit gates. Since each $U_i$ preserves the bias, $U_i\sz{m}=f(\sz{1},\sz{2},...)U_i$, where $\sz{m}$ is the Pauli matrix for the $m^\mathrm{th}$ qubit and $f(\szz)$ is some function of $\szz$-matrices of some number of qubits. Since $(U_1,U_2,...,U_k)$ is a set of universal gates, it should be possible to construct a Hadamard gate on the $m^\mathrm{th}$ qubit, $H=\Pi_i U_i$. If this were true, then $H\sz{m}=(\Pi_i U_i)\sz{m}=f'(\sz{1},\sz{2},...)\Pi_i U_i=f'(\sz{1},\sz{2},...)H$. However, we know that $H\sz{m}=\sxx{m}H\neq f'(\sz{1},\sz{2},...)H$ for any function $f'$. Clearly, this a contradiction showing that it is not possible to have a universal set of bias-preserving gates to begin with.

\subsection{Clifford gadgets for the second-level encoding}
\label{appen_cliff}
\begin{figure}
\begin{centering}
 \includegraphics[width=.8\columnwidth]{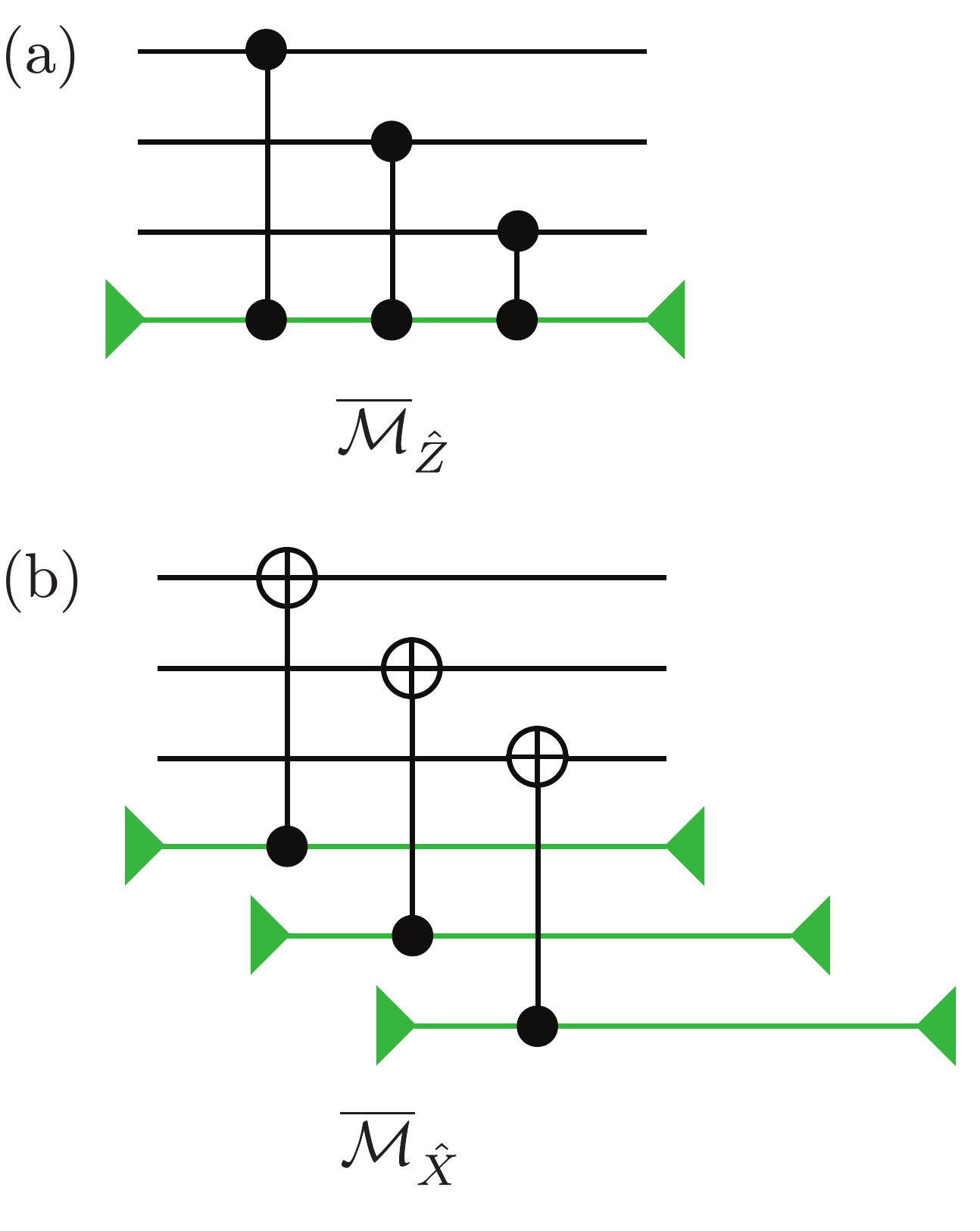}
 \caption{Fault-tolerant gadgets for non-destructive measurements (a)$\overline{\mathcal{M}}_\mathrm{\hat{Z}}$ and (b) $\overline{\mathcal{M}}_\mathrm{\hat{X}}$.
 }
 \label{gadget}
 \end{centering}
 \end{figure}

The $\mathcal{C}_1$ protected $\mathcal{C}_2$ Clifford gadgets are $\{\overline{\mathrm{CX}},\overline{\mathcal{P}}_{\ket{+}},\overline{\mathcal{P}}_{\ket{0}},\overline{\mathcal{M}}_{\hat{Z}},\overline{\mathcal{M}}_{\hat{X}}\}$. The $\overline{\mathrm{CX}}$ gadget has been discussed in the main text. 
The rest of the gadgets are similar to ref.~\cite{aliferis2008fault}. 
The scheme for non-destructive measurement $\overline{\mathcal{M}}_\mathrm{\hat{Z}}$ is shown in Fig.~\ref{gadget}(a). 
The ancilla is prepared in state $\ket{+}$ and CZ gates are implemented between the ancilla and each qubit in the repetition code. 
Finally, the ancilla is measured along the $X$-axis. 
For fault-tolerance measurement with ancilla is repeated $r$ times and $\sz{\mathrm{L}}$ is determined by a majority vote on the measurement outcome. Figure~\ref{gadget}(b) shows the gadget $\overline{\mathcal{M}}_\mathrm{\hat{X}}$. 
The non-destructive measurement of $\sz{\mathrm{L}}$ is carried out by performing transversal measurement of $\szz$ on each qubit in the repetition code. For measuring $\szz$ on each qubit, a CX gate is realized between the qubit and an ancilla prepared in $\ket{+}$. 
The ancilla is then measured along $\sx$. For fault-tolerance, each $\szz$ measurement is repeated $r$ times and taking the majority of the outcome. 
The state $\ket{+}_\mathrm{L}$ can be prepared transversally by preparing each cat qubit in the $\catp$ state. 
The preparation $\ket{0}_\mathrm{L}$ proceeds by first preparing $\ket{+}_\mathrm{L}$ and then performing non-destructive measurement of $\sz{\mathrm{L}}$.

\subsection{Bias-preserving CCX using only three-wave mixing}\label{cCX}

One can generalize the CX Hamiltonian, Eq.~\eqref{CX_h}, to a Hamiltonian
for a controlled-controlled-NOT (CCX) gate---also known as a Toffoli gate---by
adding a term to stabilize the second control oscillator and adapting the terms
$(\beta\pm\hta_\mathrm{c})/2\beta$ (and their conjugates) to more involved
expressions
\begin{align}
  \frac{\beta-\hta_\mathrm{c}}{2\beta}
  &\to
  \frac{1}{4}\left(1-\frac{\hta_{\mathrm{c}1}}{\beta}
  -\frac{\hta_{\mathrm{c}2}}{\gamma}
  +\frac{\hta_{\mathrm{c}1}\hta_{\mathrm{c}2}}{\beta\gamma}\right)
  \\
  \frac{\beta+\hta_\mathrm{c}}{2\beta}
  &\to
  \frac{1}{4}\left(3+\frac{\hta_{\mathrm{c}1}}{\beta}
  +\frac{\hta_{\mathrm{c}2}}{\gamma}
  -\frac{\hta_{\mathrm{c}1}\hta_{\mathrm{c}2}}{\beta\gamma}\right)\,,
\end{align}
where $\hta_{\mathrm{c}1}$ and $\hta_{\mathrm{c}2}$ are the annihilation
operators for the two control oscillators, with stabilized subspaces defined by
the coherent-state amplitudes $\beta$ and $\gamma$.  This direct implementation
of a CCX is undesireable, though, because it requires interaction terms that
are quartic in the creation and annihilation operators arising from the cross
terms between the above expressions and the two-photon drive of the target
oscillator.  We instead propose to implement a CCX in a less direct manner
that only requires interaction terms cubic in the creation and annihilation
operator.  We accomplish this through the use of an ancillary oscillator and a
three-body entangling gate that only requires cubic interaction terms and using
the ancillary oscillator as the control for a final CX.

\begin{figure}
  \begin{centering}
    \includegraphics[width=\columnwidth]{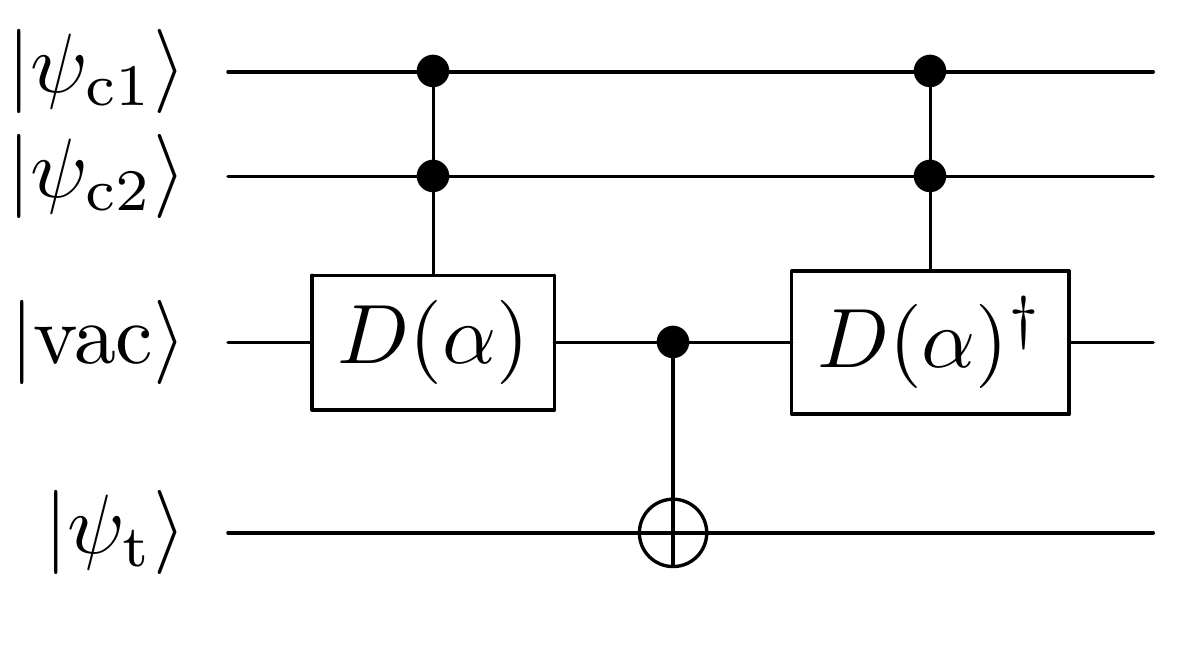}
    \caption{Implementation of CCX using the CCD gate.}
    \label{ccd_ckt}
  \end{centering}
\end{figure}

The three-body gate we propose is the controlled-controlled displacement (CCD)
gate.  The Hamiltonian for the gate is
\begin{widetext}
\begin{align}
  \hat{H}_\mathrm{CCD}(t)
  &=
  -K\left(\hta_{\mathrm{c}1}^{\dag 2}-\beta^2\right)
  \left(\hta_{\mathrm{c}1}^{ 2}-\beta^2\right)
  -K\left(\hta_{\mathrm{c}2}^{\dag 2}-\gamma^2\right)
  \left(\hta_{\mathrm{c}2}^{ 2}-\gamma^2\right)
  -K\left(\hta^{\dag 2}_\mathrm{t}-\alpha(t)^2\right)
  \left(\hta^2_\mathrm{t}-\alpha(t)^2\right)
  \nonumber\\
  &\quad
  -\delta\left[\hta_\mathrm{t}^\dag
  -\alpha(t)\frac{1}{2}\left(
  1+\frac{\hta_{\mathrm{c}1}^\dag}{\beta}
  +\frac{\hta_{\mathrm{c}2}^\dag}{\gamma}
  -\frac{\hta_{\mathrm{c}1}^\dag\hta_{\mathrm{c}2}^\dag}{\beta\gamma}
  \right)\right]
  \left[\hta_\mathrm{t}
  -\alpha(t)\frac{1}{2}\left(
  1+\frac{\hta_{\mathrm{c}1}}{\beta}
  +\frac{\hta_{\mathrm{c}2}}{\gamma}
  -\frac{\hta_{\mathrm{c}1}\hta_{\mathrm{c}2}}{\beta\gamma}
  \right)\right]
  \nonumber\\
  &\quad
  +i\dot{\alpha}(t)\left(\hta_\mathrm{t}^\dag-\hta_\mathrm{t}\right)
  \frac{1}{2}\left(
  1+\frac{\hta_{c1}+\hta_{c1}^\dag}{2\beta}
  +\frac{\hta_{c2}+\hta_{c2}^\dag}{2\gamma}
  -\frac{(\hta_{c1}+\hta_{c1}^\dag)(\hta_{c2}+\hta_{c2}^\dag)}{2\beta\gamma}
  \right)\,,
\label{ccd_h}
\end{align}
\end{widetext}
where $\alpha(t)$ adiabatically transitions from $\alpha(0)=0$ at the beginning
of the gate to $\alpha(T)=\alpha$ by the end of the gate.

This gate displaces the target oscillator from vacuum at $t=0$ to
$|{-\alpha}\rangle\approx|1\rangle$ at $t=T$ if both control oscillators are in
$|1\rangle$, otherwise it displaces the target to $|\alpha\rangle$.  By applying
this gate from two controls to an ancilla prepared in vacuum, applying a CX
from the ancilla to the ultimate target, and unentangling the ancilla by
applying the inverse of the displacement gate (see Fig~\ref{ccd_ckt}), an
effective CCX is applied from the two controls to the ultimate target without
the need for any quartic interaction terms.

In the computational subspace the gate effects the isometry
\begin{align}
  \mathrm{CCD}:
  &\,
  |00\rangle\mapsto|000\rangle
  \\
  &\,
  |01\rangle\mapsto|001\rangle
  \\
  &\,
  |10\rangle\mapsto|010\rangle
  \\
  &\,
  |11\rangle\mapsto|111\rangle\,.
  \label{ccd_isometry}
\end{align}

To show that the evolution given by Eq.~\eqref{ccd_h} results in a bias-preserving CCD, we perform simulations equivalent to those described in Sec.~\ref{CX_num} for the CX gate.
A Schr\"{o}dinger-equation simulation of the three oscillators, with $\alpha=\beta=\gamma=2$, $\delta=K/2$, $\alpha(t)=t/T$, and $T=10/K$, yields an average-gate infidelity with the ideal CCD gate of $1.42\times10^{-5}$.

To simulate the master equation evolution in the presence of loss, we simplify the bath for the two control oscillators such that the spectral density of its thermal photons are narrow, as discussed in Sec.~\ref{sec_narrow_therm}, allowing us to approximate them as qubits with effective Lindblad operators given by Eqs.~\eqref{narrow_therm_loss_op} and~\eqref{narrow_therm_gain_op}.
The target oscillator is simulated as evolving in the presence of a thermal bath with a white-noise spectrum as discussed in Sec.~\ref{sec_th}.

Using the same Hamiltonian parameters as for the Schr\"{o}dinger-equation simulation and setting the error-channel parameters to $\kappa=K/4000$ and $n_\mathrm{th}=1\%$ yields an average-gate fidelity of $98\%$. 
The leakage due to thermal photons in this case is $\sim1.5\times10^{-5}$, and the bias is $\eta_\mathrm{CCD}\sim2500$. 
We also simulate adding two-photon dissipation for the gate time $T$ after the gate has completed, with $\kappa_2=K/5$, as a means of reducing the leakage. 
In this case, the leakage is reduced by almost an order of magnitude $\sim2\times10^{-6}$, at the cost of reducing the average-gate fidelity to $96\%$ and the bias to $\eta_\mathrm{CCD}\sim2300$.

In calculating leakage and bias for the CCD gate there are some subtleties due
to the isometric nature of the gate. 
The process matrix is most naturally constructed as a map from operators on the domain of the gate (the subspace of inputs to the gate) to operators on the range of the gate (the subspace spanned by all outputs of the gate), restricted to the computational subspace of the oscillators. 
For all processes, the coefficient for the mapping from identity on the domain to identity on the codomain (the vector space the gate maps into, which can generally be larger than the range) tells how much of the trace is preserved when restricting to the computational subspace. 
For unitaries, this is the same as the coefficient mapping from identity on the domain to identity on the range, since the range equals the codomain. 
For isometries, this is not the case, and the coefficient mapping identity on the domain to identity on the range does not capture how much trace is retained in the computational subspace.  
In the particular case of the CCD gate, when restricting to the computational
subspaces of the oscillators, the range is the 4-dimensional subspace spanned by
$\{|000\rangle,|001\rangle,|010\rangle,|111\rangle\}$, while the codomain is the
8-dimensional subspace spanned by all 3-bit strings. 
Errors in the gate implementation can map input operators into operators with support outside of the range but still within the codomain, and we must be careful to not count these errors as leakage. 
It is important then to explicitly take the image of the computational-identity input in the whole oscillator space and project on the computational subspace (the codomain) in order to properly account for how much trace is preserved, and therefore how much leakage occurs.

The bias calculation is complicated because one does not probe all inputs to a 3-qubit error channel. 
This is because the states in the range of an ideal CCD cannot distinguish between certain distinct error processes in the computational subspace. 
It is cumbersome to express the indistinguishable error processes for the CCD gate, so for clarity consider the simpler CD gate
\begin{align}
    \mathrm{CD}:
    &\,
    |0\rangle\mapsto|00\rangle
    \\
    &\,
    |1\rangle\mapsto|11\rangle\,.
\end{align}
This gate is also an isometry, mapping to a range that is a 2-dimensional subspace of a 4-dimensional codomain. 
The range consists of states of the form $c_0|00\rangle+c_1|11\rangle$. 
Because of the symmetry of these states, $IZ$ and $ZI$ errors have exactly the same effect, so as far as the CD gate is concerned these error processes are equivalent. 
To characterize the inequivalent errors in our implementation, it is useful to look at the images of errors on the input states. 
For the CD gate, a $Z$ error on the input has the same effect as either an $IZ$ or a $ZI$ error on the output. 
So for the CD gate, one can determine the probability of a dephasing-type error occurring on the output by computing the probability of a dephasing error occurring at the input.
This ensures that a $ZZ$ error on the output is correctly ignored, since on the range $ZZ$ acts like the identity.

Similarly, for the CCD gate, one determines that the probability of a dephasing error occurring on the output is equal to the probability of a $IZ$, $ZI$, or $ZZ$ error occurring on the input.
The probability of other errors occurring is the probability of staying in the computational subspace, minus the probability of the $II$ action and the dephasing errors. 
The bias is then calculated as the ratio of these two probabilities.

\bibliographystyle{apsrev4_2}
\bibliography{Gatesbib}{}

\end{document}